\documentclass[onecolumn]{emulateapj}
\usepackage{mathptmx}
\begin{document}

\title{Spectral Signatures of Photon-Particle Oscillations from Celestial Objects}
\author{Doron Chelouche\altaffilmark{1,2,3}, Ra\'ul Rabad\'an\altaffilmark{2}, Sergey S. Pavlov\altaffilmark{4}, and Francisco Castej\'on\altaffilmark{5}}
\altaffiltext{1}{Canadian Institute for Theoretical Astrophysics, University of Toronto, 60 St. George st., Toronto, Canada; doron@cita.utoronto.ca} 
\altaffiltext{2}{School of Natural Sciences, Institute for Advanced Study, Einstein Drive, Princeton 08540, USA} 
\altaffiltext{3}{Chandra Fellow} 
\altaffiltext{4} {Institute of Plasma Physics, Kharkov Institute of Physics and Technology, Kharkov, Ukraine} 
\altaffiltext{5} {Laboratorio Nacional de Fusi\'on por ConÞnamiento Magn\'etico, Asociaci\'on Euratom-Ciemat para Fusi\'on, 28040 Madrid, Spain}

\shortauthors{Chelouche D. et al.}
\shorttitle{Photon-Particle Oscillations}

\begin{abstract}

We give detailed predictions for the spectral signatures arising from photon-particle oscillations in astrophysical objects. The calculations include quantum electrodynamic effects as well as those due to active relativistic plasma. We show that, by studying the spectra of compact sources, it may be possible to {\it directly} detect  (pseudo-)scalar particles, such as the axion, with much greater sensitivity, by roughly three orders of magnitude, than is currently achievable by other methods. In particular, if such particles exist with masses $m_a<10^{-2}$\,eV and coupling constant to the electromagnetic field, $g>10^{-13}\,{\rm GeV}^{-1}$, then their oscillation signatures are likely to be lurking in the spectra of magnetars, pulsars, and quasars. 
\end{abstract}

\keywords{dark matter --- elementary particles --- plasmas --- pulsars: general --- magnetic fields --- quasars: general }

\section{Introduction}

It is well known that the combined product of the charge-conjugation (C) and parity (P) symmetry holds for electromagnetic interactions. Nevertheless, for some weak interactions it is known to be violated; so called CP-violation. This near (but incomplete) symmetry of nature in weak interactions is contrasted by the fact that the strong interaction, i.e., quantum chromodynamics (QCD), seems to be consistent with preserving CP symmetry to a very high precision since, otherwise, the electric dipole moment of the neutron would have been $\sim 10$ orders of magnitude higher than current observed limits. In particular, the lack of broken symmetry requires fine tuning of QCD so that one of its parameters (treated as an effective angle in the Lagrangian) is very small instead of being of order unity, as may be a-priori expected. A way to overcome the fine tuning problem is via the Peccei-Quinn mechanism (Peccei \& Quinn 1977) in which the angle becomes a dynamical field. This field, dubbed the axion (Wilczek 1978), gets a mass from QCD instanton effects, leading to a spontaneous relaxation of the angle  whereby the field energy is minimized (an insightful and entertaining review is given by Sikivie 1996). The axion is a (charge-less) pseudoscalar field that interacts very weakly with matter (e.g., Dine et al. 1981, Kim 1979, Sikivie 1983). Interestingly enough, pseudoscalar particles with axion-like properties and can be found in string theory (see among others, Banks \& Dine 1997, Bar 1985, Choi \& Kim 1985 and Green et al. 1987). Such particles also serve as cold dark matter candidates (e.g., Kolb \& Turner 1990) forming as a Bose condensate  in the very early stages of our universe evolution, when Peccei-Quinn symmetry breaking takes place (e.g., Turner 1987 and references therein).  We note that besides pseudo-scalar particles, scalar particles may exist such that  their properties do not change sign under reflection. Neither pseudoscalar particles nor scalar particles have been detected to date  (a possible detection was reported recently by the PVLAS collaboration but the claim was later withdrawn; see below and also Asztalos et al. 2006 for a general review).

The properties of the pseudo-scalar particle (hereafter referred to sometimes as axions) are very loosely predicted by current theory. In particular, the axion mass cannot be predicted from first principles though various cosmological and astrophysical constraints, as well as experimental data, suggest it is probably $<1$\,eV (Turner 1987). (For the axion to be a viable candidate for dark matter, its mass should be of order $10^{-5}$\,eV.) In addition, the value of the coupling constant, $g$,  between axions and the electromagnetic field is unknown although quantum chromodynamics (QCD) models predict a relation between the (unknown) particle mass and the coupling amplitude such that more massive axions couple more strongly. Axions can, in principal, couple to photons, leptons, and baryons with the strength of the interaction depending on the Peccei-Quinn charges of the $u$ and $d$ quarks and the electron's. Different axion models predict different interaction strengths depending on the charges' values which are also not known from first principles.

As mentioned before, the axion mass and the coupling constant are essentially unknown and, in principal, may span a very large volume of the parameter space. Realistically, however, a part of the parameter space is less favored by astrophysical and cosmological considerations: axions whose mass is $<10^{-6}$\,eV are likely to (but not necessarily) result in $\Omega_M>0.3$ thereby closing the universe, in contrast to observations. Axions whose mass is in the eV range and which constitute a fair fraction of the mass of our galaxy (as any viable dark matter particle candidate would) are more likely to decay to photons - via their stronger coupling constants - and be observable as optical light glow over the sky, in contrast to observations (Turner 1987). These considerations suggest that the more relevant mass range for the axion is $10^{-6}<m_a<10^{-3}$\,eV (though a more extended range cannot be firmly rejected). Additional constraints come from the coupling of  axions to photons in the interiors of stars which, for strong enough coupling, would enhance the cooling rate of those objects thereby shortening their lifetime. Measurements of old stellar populations in globular clusters [horizontal branch (HB) and red giant stars] suggest that the coupling constant is limited to $g<10^{-10}\,{\rm GeV}^{-1}$ (see Raffelt 1996). This agrees with current mass limits for standard axion models. 

Apart from general theoretical arguments that indirectly constrain the physical parameters of axions, there is a considerable ongoing effort to directly detect these particles. We make no attempt to fully cover the numerous experimental methods  especially devised for this purpose but note a few general classes of them and that most utilize coupling of axions to photons. Microwave cavity experiments (Sikivie et al. 1983) are most relevant for detecting axions as dark matter candidates  and use the proposition that the number of axions through any given surface is large if they are to constitute dark matter hence, in the presence of a magnetic field, some of them would convert to photons. For axion mass of $\sim 10^{-5}$\,eV and energies typical of the virialized halos of galaxies, the photons with which such particles mix are the microwave band and a cavity is built with appropriate resonances so as to enhance their conversion. These experiments are sensitive to very low $g-$values yet scanning the entire mass range quickly and effectively is yet to be fulfilled (e.g., Duffy et al. 2006, Hagmann et al. 1990, Wuensch et al. 1989). That said, if axions exist but their density is low (and so do not constitute dark matter) then they cannot be easily detected by such experiments. The currently operating Cern Axion Solar Telescope (CAST; see Lazarus et al. 1992 and references therein for the general concept and Andriamonje et al. 2007 for recent findings) utilizes the fact that $\sim$\,keV photons in the sun's core convert to axions that reach the earth. Applying a $\sim 10^5$\,G magnetic field, this experiment attempts to reconvert solar axions to photons and observe them. This approach has yielded limits on the coupling constant of axions to photons which are comparable to indirect astrophysical arguments constraints suggesting that  $g<10^{-10}\,{\rm GeV}^{-1}$ (e.g., Raffelt 2007 and references therein). Photon regeneration experiments ("light shining through walls") are based on a similar concept and use a light source and a magnetic field to convert a fraction of the photons to axions (e.g.,  van Bibber et al. 1987, Cameron et al. 1993, Sikivie 1983, and also Adler et al. 2008). A mounted wall blocks the light ray from propagating but not the axion ray which is later reconverted to photons by a similar magnetic field (e.g., Rabad\'an et al. 2006 and references therein). Using such a method, upper limits on the coupling constant, $g$, which are about 4 orders of magnitude higher than current astrophysical constraints are obtained (e.g., Robilliard et al. 2008; but see improvements recently suggested  by Sikivie et al. 2007). One such ongoing experiment is the PVLAS experiment (e.g., Zavattini et al. 2006). This experiment announced a tentative detection of a signal that may be interpreted as due to the elusive  axion. Nevertheless, a more thorough investigation  of the measurements suggested a problem with the experimental setup and the claim was later withdrawn by the same group (Zavattini et al. 2007).   We note that the coupling of the axion to other fields have also been investigated. In particular, a different type of experiment was initiated by Youdin et al. (1996) and utilizes a gravitational interaction potential between spin and matter (Moody \& Wilczek 1984). Much heavier, axion-like particles that are predicted by some theories are probed by the DAMA experiment (e.g., Bernabei et al. 2008a,b and references therein). To date, the axion has not been detected and the current limits are about 10 orders of magnitude higher than most theoretical predictions.  

A major limitation of most terrestrial experiments for the detection of photon-particle mixing has to do with  the fact that the expected axion-photon conversion probability $P_{\gamma \rightarrow a} \propto g^2 B^2R^2$ is small (where $B$ is the magnetic field and $R$ the size of the system; see \S4 for the accurate expression for the conversion probability in the general non-linear case, and also Raffelt \& Stodolsky 1988 and Sikivie 1983). The small  probabilities require high signal-to-noise (S/N) data which render secure detection challenging. It is worthwhile to compare laboratory expected probabilities to those which may be expected from astrophysical objects under the assumption that the magnetic field is in equipartition with gravity i.e., $GM^2/R \sim B^2R^3$; here, $G$ is Newton's constant and $M$ the object's mass (this seems to be a fair approximation for many astrophysical systems and seems to be supported by recent numerical simulations; e.g., Igumenshchev \& Narayan 2002). In this case we obtain that, in the limit of small conversion probabilities and in the linear regime holds (see however \S 4.1), the ratio between the conversion probabilities is
\begin{equation}
\frac{P_{\gamma \rightarrow a}^{\rm laboratory~experiments}}{P_{\gamma \rightarrow a}^{\rm celestial~objects}} \simeq \left ( \frac{B_{\rm lab}}{B} \right)^2 \left ( \frac{R_{\rm lab}}{R} \right )^2 \simeq  \left ( \frac{R_{\rm obj}}{R_s} \right )^{2} \frac{B_{\rm lab}^2 R_{\rm lab}^2}{c^4/8G} \ll 1,
\end{equation}
where $R_s\equiv 2GM/c^2$ is the Schwartzschild radius. For the particular case of a celestial object whose size is of the order of its Schwarzschild radius, and taking $B_{\rm lab}=10^5$\,G and $R_{\rm lab}=10$\,m, one obtains a ratio  $<10^{-30}$ (non-linear effects are treated in \S 4).  This demonstrates the potentially greater sensitivity that may be achieved in the case of  astrophysical objects compared to laboratory experiments. We note that several other works have already taken an astrophysical approach for constraining the axion properties. An incomplete list includes: Deffayet et al. (2002) who considered photon-particle oscillations as an  explanation to supernovae dimming, M\"ortsell \& Goobar (2003) who investigated the physical properties of very light axions (whose mass, $m_a\sim 10^{-16}$\,eV) using {\it Sloan digital sky survey} quasar spectra,  Brockway et al. (1996) who deduced an upper limit on the coupling constant for very light axions of $\sim 10^{-11}\,{\rm GeV}^{-1}$ from supernovae data, Rubbia \& Sakharov (2008) who put more stringent constraints on heavy, $m_a>10^{-4}$\,eV axions from polarization studies of the prompt emission in a $\gamma$-ray burst, and Hochmuth \& Sigl (2007) who investigated the observational implications of the (recently withdrawn) PVLAS experiment results. More relevant to our study is the recent work by Lai \& Heyl (2006) who explored the possibility for axion detection in the case of magnetars.

In this work we wish to see whether, by studying at the spectra of various astrophysical objects, one can hope to observe the signatures of photon-particle conversion down to low values of the coupling constant and extend the physical parameter space accessible to us. This approach has been suggested in the past and was qualitatively treated in several works (e.g., Lai \& Heyl 2006 and references therein). Nevertheless, the application of such methods is more complicated and requires that we have good understanding of the astrophysical object and can distinguish between photon-particle spectral oscillation features and other spectral imprints such as atomic lines, edges, and continuum features. In particular, detailed predictions for the spectral signatures of photon-particle mixing are crucial for correctly interpreting the observations. Unlike terrestrial experiments whose setup may be controlled and the results verified or refuted (e.g., PVLAS), an astrophysical experiment cannot be controlled and other corroborative means are required to assess the validity of the results. As we shall see, the observable properties of the photon-particle spectral  oscillation features depend rather sensitively on the physical properties of the object and can therefore be used to our advantage. 

This paper is organized as follows: we start with a general layout of the formalism used in this work to calculate the conversion probabilities (\S 2).  Various relevant physical processes leading to light refraction  are discussed in \S 3 among which (cold and relativistic) plasma and quantum electro-dynamics (QED) effects. Section 4 focuses on the general properties of the predicted photon-particle oscillation features and study their dependence on the physical parameters of the problem (e.g., on medium stratification). Readers who are less interested in the technical aspect of the problem but wish to better understand the observational consequences for specific astrophysical systems (e.g., magnetars, pulsars, quasars, X-ray binaries, Ap stars, white dwarfs, and cataclysmic variables) may skip to section 5. The discussion follows in \S 6 where a short "user's manual" is provided. Summary follows in \S 7.

\section{The Formalism}

There are two equivalent approaches to the problem of particle oscillations: a Schr\"odinger-like approach (e.g., Sikivie 1983) or a Heisenberg-like approach (Raffelt and Stodolsky 1988). Here we choose the latter approach.  The Lagrangian density of the electromagnetic and a (pseudo-)scalar particle  fields, and their interaction, can be written as:
\begin{equation}
\mathfrak{L}=-\frac{1}{4}F_{\mu\nu}F^{\mu\nu}-j_\mu A^\mu  +\frac{1}{2}\left [  \partial^\mu a \partial_\mu a + m_a^2a^2 \right ] + \frac{1}{4}g F_{\mu \nu} \tilde{F}^{\mu \nu} a+\frac{\alpha^2}{90m_e^4}\left [ \left ( F_{\mu \nu}F^{\mu \nu} \right )^2 + \frac{7}{4} \left ( F_{\mu \nu} \tilde{F}^{\mu \nu} \right )^2 \right ]
\end{equation}
[we work in rationalized units where $\hbar=c=1$ and the electromagnetic fields (charges) have been reduced  (multiplied) by a factor $\sqrt{4\pi}$]. The Lagrangian terms are as follows:
\begin{itemize}
\item The first and second terms comprise the Lagrangian for the electromagnetic field in the presence of sources (e.g., Jackson 1975). In our calculations we shall assume that all charges and currents reside outside the propagation region of the photon (in particular, the currents responsible for the magnetic field threading space lay outside the region relevant to our study). Charges and currents in the plasma are small and are treated perturbatively whereby the free electromagnetic field Lagrangian is considered but with a different phase velocity for the photons (i.e., $c \rightarrow c/n$ where $n$ is the refractive index).  This is justified for the cases considered here. An additional source for the electromagnetic field is that of the axion field, which we discuss below.
\item
The third term in the Lagrangian is the Klein-Gordon equation for the spin-$0$ axion field. Here $a$ is the particle field. We assume there are no sources or sinks for axions within the calculation's domain (e.g., Raffelt \& Stodolsky 1988).
\item
The fourth term is the photon-axion interaction term which must be a scalar. For pseudo-scalar particles, $a\rightarrow -a$ upon $x_\mu \rightarrow -x_\mu$ ($x_\mu$ is the 4-coordinate) hence the interaction term must consist of another pseudo-scalar multiplying the axion field. The simplest one involving the electromagnetic field is $F_{\mu \nu}\tilde{F}^{\mu \nu}$ with $\tilde{F}^{\mu \nu}=\frac{1}{2}\epsilon_{\mu \nu \rho \sigma} g^{\rho \eta} g^{\sigma \zeta} F_{\eta \zeta}$ ($\epsilon$ is the Levi-Civita anti-symmetric tensor). In terms of the electric field, $E$, and magnetic field, $B$, $F_{\mu \nu}\tilde{F}^{\mu \nu}\propto {\bf E \cdot B}$. (For scalar particles the interaction terms reads $F_{\mu \nu}F^{\mu \nu}\propto {\bf B}^2 - {\bf E}^2$.) $g$ is the coupling constant whose value is not predicted by current theory (see \S 1). The process by which such axions are generated is called the Primakoff process and is depicted in figure 1.
\item
The last term is a QED extension to free electromagnetic field Lagrangian density and stands for photon-photon interactions via scattering off virtual electrons (Heisenberg \& Euler 1936). This term may be obtained by Taylor expanding the full Euler-Heisenberg action (see equation 10 in Adler 1971) including only quadratic terms in the electro-magnetic field. It is valid as long as the magnetic field strength, $B=\vert {\bf B} \vert$ is smaller than the critical value of $B_c=m_e^2/e\simeq 4.41\times 10^{13}$\,gauss (in CGS units where $e$ is the electron charge; in rationalized units one obtains $B_ c\simeq 1.24\times 10^{13}$\,gauss). This term gives rise to vacuum birefringence (see Fig. 1a) and may be included in the calculation in a perturbative manner via its effect on the phase velocity of photons, similarly to the case of plasma refraction (see Eq. 7 below). Expressions for the effective refractive index due to vacuum birefringence are given in Adler (1971) who calculated this effect for any values of $B$ (using the full Euler-Heisenberg Lagrangian; see his equation 10) and, in particular, for $B>B_c$ which are relevant to the case of magnetars. A diagram showing a general dispersive graph is shown in figure 1 and we note the even number of interactions with the external field.
\end{itemize}

The equations of motion for the fields may be obtained by requiring that the action $S=\int d^4x \sqrt{-\mathfrak{g}}\mathfrak{L}$ is minimal with $\mathfrak{g}$ being the metric. This is accomplished by deriving the well known Euler-Lagrange equations. In what follows we shall assume flat space-time so that $\sqrt{-\mathfrak{g}}=1$ which is justified for photons emitted far from the Schwarzschild radius and is applicable for most cases considered here (see \S3.2.4). For the electromagnetic and axion fields one derives (neglecting currents and charges as well as the Euler-Heisenberg term for the moment) 
\begin{equation}
\partial_\mu F^{\mu \nu}+g\tilde{F}^{\mu \nu} \partial_\mu a =0 ~~~{\rm and}~~~ \left ( \partial^\mu \partial_\mu  + m_a^2 \right ) a-\frac{1}{4}gF_{\mu \nu}\tilde{F}^{\mu \nu}=0
\end{equation}
(where we recall that $\partial_\mu \tilde{F}^{\mu \nu}=0$). In terms of the electric (${\bf E}$) and magnetic (${\bf B}$) fields these equations translate to
\begin{equation}
\begin{array}{c}
\displaystyle \nabla \cdot {\bf E}-g{\bf B} \cdot \nabla a=0,~~\nabla \times {\bf B} -\frac{\partial {\bf E}}{\partial t}-g\left ( {\bf E} \times \nabla a- {\bf B}\frac{\partial a}{\partial t} \right )=0,~~{\rm and}~ \left (\square +m_a^2 \right )a +g{\bf E} \cdot {\bf B}=0 \\
\displaystyle \nabla \cdot {\bf B}=0,~~\nabla \times {\bf E} -\frac{\partial {\bf B}}{\partial t}=0
\end{array}
\end{equation}
where the last two expressions result from $\partial_\mu \tilde{F}^{\mu \nu}=0$. The equations for the electro-magnetic field are essentially Maxwell's equations for free fields with source terms associated with the particle (axion) field. The above expressions are rather general and difficult to solve. Below we consider the more specific case in which a photon propagates in a medium threaded by a strong external magnetic field, ${\bf B}^{\rm ext}$ whose amplitude $B^{\rm ext}$ is much larger than that of the radiation field (i.e., the photon). We further assume that there is no external (static) electric field which is a good approximation for plasma in which the charges can quickly rearrange to shield the enclosed volume where the photons propagate. Using the following decomposition for the external and radiation electromagnetic fields,
\begin{equation}
F_{\mu \nu} = F_{\mu \nu}^{\rm ext}+F_{\mu \nu}^{\rm rad},
\end{equation}
we may approximate all source terms in equation 3 such that $F_{\mu \nu}\simeq F_{\mu \nu}^{\rm ext}$ (and likewise for the dual field: $\tilde{F}_{\mu \nu}\simeq\tilde{F}_{\mu \nu}^{\rm ext}$).  In this case, the equations reduce to the form 
\begin{equation}
\partial_\mu F^{\mu \nu,\,{\rm rad}}+g\tilde{F}^{\mu \nu,\,{\rm ext}} \partial_\mu a =0~~~{\rm and}~~~ \left ( \partial^\mu \partial_\mu  + m_a^2 \right ) a-\frac{1}{4}gF_{\mu \nu}^{\rm ext}\tilde{F}^{\mu \nu,\,{\rm ext}}=0.
\end{equation}
Higher order corrections related to the back-reaction of the field on itself would be of order $B^{\rm rad}/B^{\rm ext} \ll 1$ and are therefore negligible.

\begin{figure}
\plottwo{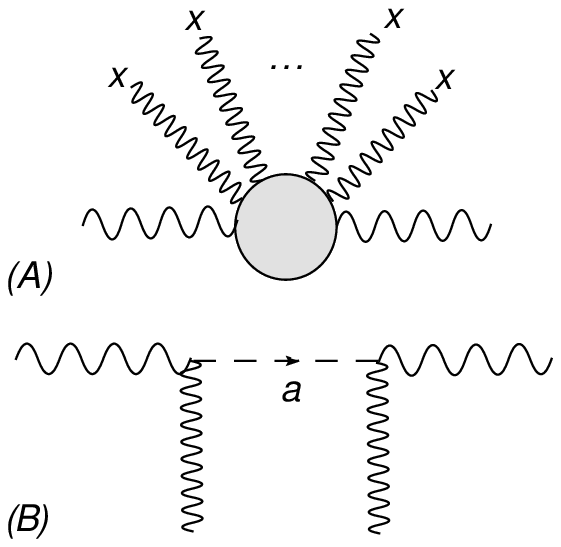}{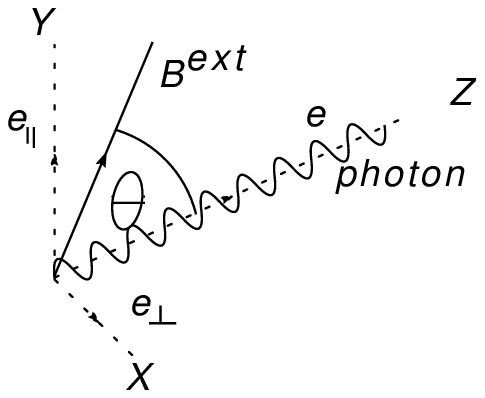}
\caption{{\it Left:} The physical processes associated with photon propagation include (A) refraction whose general dispersive graph is shown and involves an even number of interactions with the external field (denoted by $x$), (B) photon-axion-photon conversion through the Primakoff process involving energy-conserving oscillations where excess momentum is transmitted to the external field.  {\it Right:} The coordinate system adopted in this work. The magnetic field is in the plane defined by the propagation direction of the photon and ${\bf e}_\|$ (with ${\bf e}_\bot$ being orthogonal to both). We note that this definition is different than the one used in Adler (1971).}
\end{figure}

We consider a coordinate system in which the propagation direction of the photon is denoted by ${\bf e}_\gamma$, the direction parallel to the external magnetic field component which is orthogonal to ${\bf e}$ by ${\bf e}_\|$, and the third direction which is orthogonal to both, ${\bf e}_\bot$. This forms the orthonormal set shown in figure 1. We note that although the photon propagates in plasma and not in vacuum, its electric and magnetic field may be considered orthogonal for all practical purposes provided its frequency, $\omega$, is greater than the plasma frequency, $\omega_p$. For this reason, the vector potential for the radiation field may be chosen to be entirely in the plane spanned by $({\bf e}_\|,{\bf e}_\bot)$ while its properties depend on the spatial coordinate along the photon's direction of propagation, ${\bf e}_\gamma$, which we denote by $\gamma$ and on time, $t$, so that ${\bf A}=[0,A_\| (\gamma),A_\bot (\gamma)] e^{i \omega t}$. The equations of motions for the fields with a photon energy $\omega$ can then be concisely written as
\begin{equation}
\left ( \square-m_a^2 \right ) a=gB^{\rm ext}_\| \omega A_\| {\rm ~~~~~and~~~~~~} \left ( \square-m_{\gamma,\lambda}(\omega)^2 \right ) A_\lambda = gB^{\rm ext}_\| \omega \delta_{\lambda \|}a
\end{equation}
where $\lambda$ is an index denoting the polarization ($\lambda=\bot,\|$). The effective photon mass, $m_{\gamma,\lambda}(\omega)$, results from  light refraction induced by the plasma effects and the vacuum birefringence terms (i.e., it is here where the contribution of plasma charges and currents as well as the Euler-Heisenberg QED terms in Eq. 2 come in). Basically, because of coherent scattering, the phase velocity of light in the medium depends on the photon frequency and polarization. This phase velocity is directly related to the photon acquiring an effective mass 
\begin{equation}
m_{\gamma,\lambda}^2=\omega^2-k^2\simeq 2\omega^2(n_{\lambda}-1)
\end{equation} 
($\vert n_\lambda-1 \vert \ll 1$ applies to all the cases considered here). There is no restriction on whether the mass is real or imaginary (depending on whether the phase velocity is $<c$ or $>c$ ; the latter case does not contradict special relativity since information travels at the group velocity). One should note that, generally, the refractive index and the effective mass are tensors so that equation 7 may be concisely written in matrix form as
\begin{equation}
\left ( \omega^2+\partial_\gamma^2+2\omega\Delta \right ) \mathfrak{A}=0 
\end{equation}
where 
\begin{equation}
\Delta \equiv \left \vert 
\begin{array}{ccc}
\Delta_{\bot \bot} & \Delta_{\bot \|} & 0 \\
\Delta_{\bot \|}^\star & \Delta_{\| \|}  & \Delta_{\| a} \\
0 & \Delta_{\| a} & \Delta_{a a}
\end{array}
\right \vert~~~~{\rm and}~~~~
\mathfrak{A}\equiv \left [
\begin{array}{c}
A_\bot \\
A_\| \\
a 
\end{array}
\right ].
\end{equation}
The terms are self-explanatory with $\Delta_{\bot \bot},\Delta_{\| \|}$ being related to the refractive indices (or effective masses) of each polarization. $\Delta_{\bot \|}$ stands for the  Faraday rotation and Cotton-Motton effects in optically active plasma (and its complex conjugate $\Delta_{\bot \|}^\star$). $\Delta_{\| a}=gB_\|^{\rm ext}/2$ is the (real) photon-axion conversion term, and $\Delta_{a a}=-m_a^2/2\omega$ stands for  the axion mass term.  We discuss the relevant contributions to the refractive indices  in the following section.

It is possible to further simplify the equation of motion by noting that the photon wavelength is by far the shortest length scale in the problem in which case
\begin{equation}
\square\equiv\partial_t^2+\partial_\gamma^2=\omega^2+\partial_\gamma^2=(\omega+i\partial_\gamma)(\omega-i\partial_\gamma)\simeq 2\omega(\omega-i\partial_\gamma)
\end{equation}
where the last step requires that the refractive index is close to unity and that the particles are relativistic, i.e., that the photon undergoing oscillations satisfy $\omega \gg m_a,~\omega_p$ (where $\omega_p$ is the plasma frequency, see \S3). As we shall see below, these conditions are, generally, satisfied for all cases considered here. In particular, cases where this approximation breaks down (e.g., at cyclotron line frequencies and below the particle frequency) are irrelevant to this work since the photon source becomes optically thick to radiation making it unsuitable for our purpose. Therefore, for all cases considered here, the equation of motion takes a Schr\"odinger-like form:
\begin{equation}
\left (\omega-i\partial_\gamma+\Delta \right ) \mathfrak{A}=0 \rightarrow i\partial_\gamma \mathfrak{A}=\mathfrak{H} \mathfrak{A}
\end{equation}
The great advantage of this form for the equation of motion of the photon-particle system is the fact that probabilities may now be calculated using the well-developed quantum mechanical formalism in its various representations. One difference is that the time coordinate is replaced here by the space coordinate along the photon propagation direction. In particular, the evolution of some initial state, $\left \vert \mathfrak{A} (\gamma=0) \right > $ with distance is
\begin{equation}
\left \vert \mathfrak{A}(\gamma) \right >=e^{i\mathfrak{H}\gamma} \left \vert \mathfrak{A}(\gamma=0) \right >
\end{equation}
Furthermore, the probability that some final state $\left \vert \mathfrak{A}' \right > $ is obtained after the system has propagated a  finite distance $\gamma$  along the (original) photon propagation direction  is simply
\begin{equation}
P_{i\rightarrow f}=\left \vert   \left <  \mathfrak{A}'  \vert \mathfrak{A}(\gamma) \right>    \right \vert^2=\left \vert   \left <  \mathfrak{A}' \right \vert e^{i\mathfrak{H}\gamma}  \left \vert \mathfrak{A} (\gamma=0)\right>    \right \vert^2 = \left \vert \sum_{i=1}^3 e^{iE_i\gamma} \left < \mathfrak{A}' \vert \tilde{\mathfrak{A}}_i \right >   \left < \tilde{\mathfrak{A}}_i \vert \mathfrak{A}(\gamma=0) \right >  \right \vert^2
\end{equation}
where $E_i$ are the eigenvalues corresponding to the eigenvectors of the Hamiltonian $\left \vert \tilde{\mathfrak{A}}_i \right >$. Clearly, any global phase is immaterial and may be discarded so that only relative phases determine the interference pattern. 

The orthonormal states that we define are:
\begin{equation}
\bot \equiv
\left [ \begin{array}{c}
1 \\
0 \\
0
\end{array} \right ], ~~\|\equiv
\left [ \begin{array}{c}
0 \\
1 \\
0
\end{array} \right ], ~~{\rm and}~~a\equiv
\left [ \begin{array}{c}
0 \\
0 \\
1
\end{array} \right ],
\end{equation}
The first two states are (pure) photon polarization states and the third one is a pure particle state.  Clearly, any initial photon state in our coordinate system may be constructed from the first two eigenvectors. We emphasize that the above vectors are {\it not} eigenvectors of the Hamiltonian but are the ones which characterize the system at the creation point of the photon and are those which can later be measured by us. We note that if photon scattering occurs at some point along the photon propagation direction then the wave-function collapses to a pure photon state from which it continues to propagate according to the above equation of motion (until the next scattering or photon measurement by an observer takes place).  

The above formulation of the problem is completely analogous to the Mikheyev-Smirnov-Wolfenstein (MSW) effect of neutrino oscillations and the solution is alike. Specifically, there is a resonance conversion where the probability is highest when $\Delta_{\| \|}(\omega)=\Delta_{a a}$. Using the definitions for $\Delta_{ij}$ we find that resonance occurs at photon energy $\omega_0$ so that the momentum transfer to the field, $q$, satisfies
\begin{equation}
q=n(\omega_0)\omega_0-\sqrt{\omega_0^2-m_a^2}=0;
\end{equation}
(for a specific example see below). Put differently, when the particle mass and the photon's effective mass equal, no momentum transfer takes place during the oscillation process and the probability for conversion is maximized. Basically, the problem is now reduced to finding the photon frequency (or frequencies) that solves this equation which would correspond to region(s) in the  spectrum where photon-particle oscillations are most likely to occur (calculating the exact conversion probabilities is a different matter which would be dealt with in \S4). It is important to note that, for this equation to have a solution, it is necessary that $n(\omega_0)<1$. As we shall later show, this requires the presence of plasma without which no resonance occurs and the conversion probability is lower. Non-resonance conversion, while generally having lower probability than resonance conversion, may still be observed  in cases where the coupling constant, $g$, is large enough. For probing the regime of small $g$, most relevant to particle searches, resonance conversion should be sought after.

Thus far we have assumed that the Hamiltonian itself is independent of location; i.e., $\gamma$. Nevertheless, the density and magnetic field of astronomical objects vary with distance and allowance should be made for their effect. The formalism developed above still holds locally but the evolution of the initial state is now given by 
\begin{equation}
\left \vert \mathfrak{A}(\gamma) \right >=e^{i\int_0^\gamma d\gamma' \mathfrak{H}(\gamma')} \left \vert \mathfrak{A}(\gamma=0) \right > .
\end{equation}
Here, care should be taken when applying the operators on the initial condition since the local Hamiltonians are, in general, non-commutative. Clearly, to solve for the evolution of the initial photon state one requires better knowledge of the dielectric tensor (refractive indices) of the medium through which the photon propagates. Characterizing those is the main objective of the following section. The means by which equation 17 may be numerically solved are discussed in \S4.2.

\section{Refractive processes}

To calculate the effective photon mass, we must first identify the relevant processes governing the phase velocity of photons, namely the polarization-dependent refractive index $n$. In general, the problem is very complex and requires knowledge of the material in which the photon propagates such as its ionization level, density structure, magnetic field configuration threading it, etc.  To simplify the problem yet maintain a reasonable description of the relevant astrophysical systems, we shall work in the weak dispersion regime where the refraction index $n$ is close to unity. This hold for photon frequencies above the plasma frequency (below that frequency the plasma is optically thick) and far from cyclotron lines which occupy only a negligibly small energy range of the spectrum (we discuss their effect in greater detail in \S3.2.2). In this weak dispersion limit, contributions to the refraction index from various processes are decoupled and can be linearly added to obtain their combined effect on the photon's phase velocity or its effective mass. At any photon frequency, $\omega$, and polarization, $\lambda$, one can therefore define
\begin{equation}
n_\lambda(\omega)=1+\sum_i \delta n_\lambda^i(\omega)
\end{equation}
where $\delta n^i_\lambda$ is the individual contribution to the refractive index from physical processes pertaining to plasma and QED effect which we now discuss in more detail.

\subsection{Vacuum Birefringence}

\begin{figure}
\center \includegraphics[width=8cm]{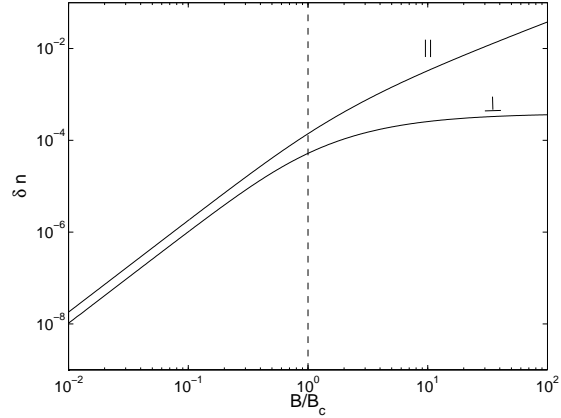}
\caption{The refractive index due to vacuum birefringence for the two photon polarizations and taking into account the full Euler-Heisenberg Lagrangian (Adler 1971). For field strengths below the critical value, $B_c$, the dependence follows from equation 21 and the two photon polarization have similar (bit not identical) refractive indices.  However, for large enough magnetic fields ($B>B_c\simeq 4\times 10^{13}$\,G; applicable to magnetars), deviations from a $B^2$ dependence are evident. Note also the large difference in refractive index between different light polarizations for high values of $B$.  The weak dispersion limit is fully justified for field strengths relevant to all known astrophysical objects.}
\end{figure}

QED predicts that photon propagation in vacuum is accompanied by the constant creation of electron-positron virtual pairs. When a magnetic field is present, it affects each member of the pair differently resulting in an effective birefringence of the vacuum - i.e., different photon polarizations travel with different  velocities. The magnitude of this process has been calculated by several people and here we follow the results of Adler (1971) who calculated the dielectric tensor of the vacuum in relation to the photon-splitting effect and finds:
\begin{equation}
\delta n_{\|,\bot}= -\frac{\alpha}{4\pi}\int_0^\infty \frac{du}{u} e^{-u} K_{\|,\bot}(uB/B_c)
\end{equation}
and
\begin{equation}
K_\|(\zeta)=-\frac{2}{3} \zeta {\rm coth}\, \zeta +\zeta^{-1}{\rm coth}\, \zeta -{\rm sinh}^{-2}\, \zeta,~~~
K_\bot(\zeta)=-\zeta^{-1}{\rm coth}\, \zeta - (1-2\zeta{\rm coth}\,\zeta) {\rm sinh}^{-2}\, \zeta
\end{equation}
(here $\zeta \equiv uB/B_c$). Note that our definition of polarization is different from Adler's which interchanges $\bot$ and $\|$. In the limit of weak magnetic fields, $B/B_c \ll 1$, one encounters the more familiar expressions involving two interactions with the magnetic field (the box diagram contribution) which is quadratic in the magnetic field so that the refraction indices are given by
\begin{equation}
\delta n_\bot \simeq \frac{4}{2} \frac{\alpha}{45\pi} \left ( \frac{\vert {\bf B}^{\rm ext}\cdot {\bf e} \vert }{B_c} \right )^2 ,~~\delta n_\| \simeq \frac{7}{2} \frac{\alpha}{45\pi} \left ( \frac{\vert {\bf B}^{\rm ext}\cdot {\bf e} \vert }{B_c} \right )^2
\end{equation}
At super-critical magnetic fields, the refraction depends linearly on the magnetic field for the parallel polarization and independent of it for the perpendicular one. The results are shown in figure 3. The weak dispersion limit for the QED effect requires that $B \ll 10^{17}$\,gauss which is justified for all known astronomical objects, including the most magnetized magnetars detected to date. Note that the phase velocity due to QED effects is always smaller than the speed of light, i.e.,  $n>1$. As discussed above, the presence of resonances requires $n<1$ and so efficient conversion cannot occur in pure vacuum. We emphasize that this does {\it not} preclude the detection of photon-particle oscillations and that this will occur if the value of the coupling constant $g$ is high enough. As we shall see below, the presence of plasma makes efficient (resonance) conversion possible.

\subsection{Plasma effects}

We are interested in the propagation of photons in space which is threaded by large magnetic fields. As there is no pure vacuo in astrophysical systems, nor in the universe at large, plasma terms must be included. This turns out to have a crucial effect on photon-particle oscillations, as we show below. 

Gaseous regions in astrophysics may be fully or partially ionized. The propagation of photons can be very different in each case since atomic transitions, such as lines and edges, can drastically change the propagation of photons via absorption and dissipation of their energy. These, in turn, depend on the plasma composition, density, ionization level, temperature, and even plasma kinematics. To keep the problem tractable, we shall neglect all atomic transitions in our calculations of photon-particle oscillations (but briefly mention their effect in \S4.3). This simplification is not unreasonable since, as we shall show below, effective conversion takes place in regions threaded by a strong magnetic fields. Such regions are found close to compact objects (such as black holes and neutron stars) where the surrounding gas is likely to be fully ionized. Furthermore, in some astrophysical objects it is possible that the gas contains a considerable fraction of relativistic pair plasma, in which case atomic transitions are likely to be unimportant. In addition to neglecting atomic processes, we neglect inverse Bremmstrahlung (free-free) opacity and synchrotron self-absorption which might become an issue at long photon wavelengths. 

Without loss of generality we can write
\begin{equation}
\delta n_{ij}= \frac{1}{2} \left ( \epsilon_{ij} -1 \right)=-\frac{1}{2}\frac{\omega_p^2}{\omega^2}F_{ij}(\omega; \rho,T,B,\theta)
\end{equation}
where $F$ is a function of photon energy, plasma properties, and the magnetic field configuration [hereafter we denote this function by $F(\omega)$ for short]. $\epsilon$ is the dielectric tensor and is linearly related to $\delta n$ so long as $\delta n \ll 1$. The light refraction index in plasma, $\delta n_{\| \|}$, may be either positive or negative (see below), and so resonance conversion is, in principal, possible which would not have otherwise occurred. For magnetic fields below the critical value, equation 21 holds and the photon energy where resonance occurs may be found from equation 16, in the limit $(1-n) \ll 1$, to be
\begin{equation}
\omega_0\simeq \frac{m_a}{\sqrt{2(1-n)}} =  \omega_p \frac{B_c}{B} \sqrt{ \frac{F_{\| \|}(\omega_0)-m_a^2/\omega_p^2}{7\alpha/45\pi}}
\end{equation}
This equation is non-linear but becomes simple for non-active cold plasma (where $F_{\| \|}=1$; see below). Clearly, one requires that $F(\omega_0)>m_a^2/\omega_p^2$ for resonance photon-particle conversion to take place. For deductive purposes we treat three types of plasma in this paper, in this order: 1) non-active cold plasma, 2) active cold plasma, and 3) active hot plasma.  Each type of plasma in the above order adds a further complication to the problem whose effect is extensively discussed in the following sub-sections. 

\subsubsection{Inactive cold plasma}

\begin{figure*}
\plotone{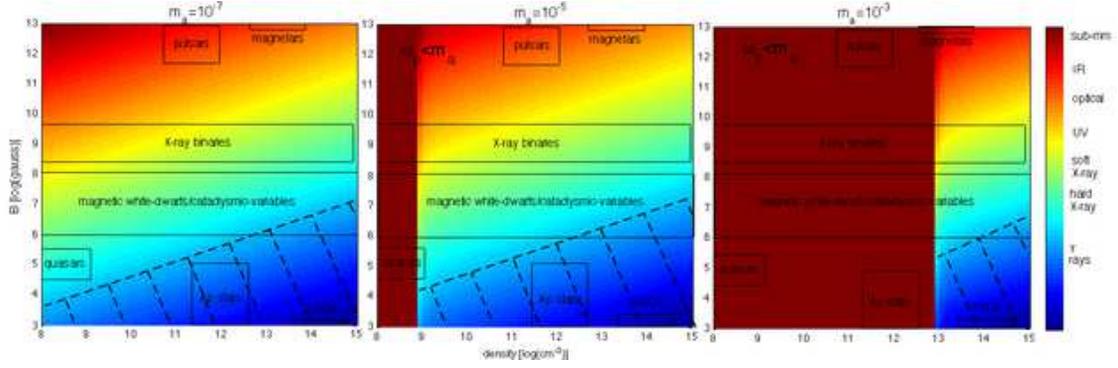}
\caption{The photon energy range where photon-particle resonance occurs as a function of the cold plasma density and the magnetic field intensity for cold non-active uniform (within some volume) plasma. Color shade bands are denoted next to the color bar on the right.  Three particle masses are considered: $10^{-7},~10^{-5},10^{-3}$\,eV. The position of various astrophysical sources is shown including the estimated uncertainties on their physical properties (see also  \S5). The brown region in the two right panels indicates a region where resonance conversion does not occur since the plasma frequency is lower than the particle mass (equation 23). Clearly, photon-particle resonance conversion is expected to occur over a broad range of photon energies. For example,  in the low mass end ($m_a<10^{-6}$\,eV), resonance conversion would occur in the X-ray energies for AGN and in the infrared for pulsars. The results for magnetic fields in excess of the critical value are not shown since the frequency dependence is no longer given by equation 23. The hatched region indicates frequencies above the threshold for pair production, at $\gamma$-ray energies. Evidently, astrophysical sources can be useful probes for particle masses $<10^{-3}$\,eV. Diagrams for $m_a<10^{-7}$\,eV are quantitatively similar to the left panel.}
\label{param_space}
\end{figure*}

This case, while not very realistic, does provide insight to the general features of the photon-axion conversion process and is therefore treated first. In this case $F_{ij}=\delta_{ij}$ where $\delta_{ij}$ is Kronecker's identity matrix so that
\begin{equation}
\delta n_{ij}=-\frac{1}{2}\frac{\omega_p^2}{\omega^2}\delta_{ij}.
\end{equation}
Clearly, the medium is inactive in this case, having no off-diagonal terms which would mix the photon polarizations. 

The resonance frequency for the case of sub-critical magnetic fields is given by equation 23 (with $F=1$) and is shown in figure \ref{param_space} for a range of plasma densities and magnetic field strengths (assumed to be constant within the relevant volume). Three different axion masses are considered and their effect on the resonance energy investigated. Evidently, compact astrophysical objects are potentially good probes for axion masses in the mass range of interest. In particular, most astrophysical sources are likely to be able to probe the low mass end with $m_a<10^{-5}$\,eV. Judging by figure \ref{param_space}, the resonance feature may be detected over a broad range of photon energies depending on the object's properties and the axion mass (dependence on the latter is weak if $m_a\ll \omega_p$). Specifically, for $m_a=10^{-7}$ the conversion feature would be observed in the hard X-ray spectra of AGN and at far infrared energies for pulsars. Magnetars are potentially good probes yet they have super-critical fields  and so deviate from the predictions of equation 23 and are not shown here (but are extensively discussed in \S4.1). For axion masses exceeding $10^{-3}$\,eV, the plasma density in most objects is too low for a resonance feature to be detected unless relativistic plasma is present (see below). We note that uncertainties in the density and magnetic field of some objects result in the energy position of the resonance conversion being somewhat uncertain. In particular, for X-ray binaries, the feature may be observed from the UV to X-ray energies for a plausible range of object's plasma density.

Whether or not a conversion feature may be observationally detected depends not only on whether resonance energy exists but also on the strength of the feature (i.e., the peak conversion probability), the width of the spectral feature, and the resolution of the instrument. These important issues are discussed in the next section.

\subsubsection{Active cold plasma}

\begin{figure*}
\plottwo{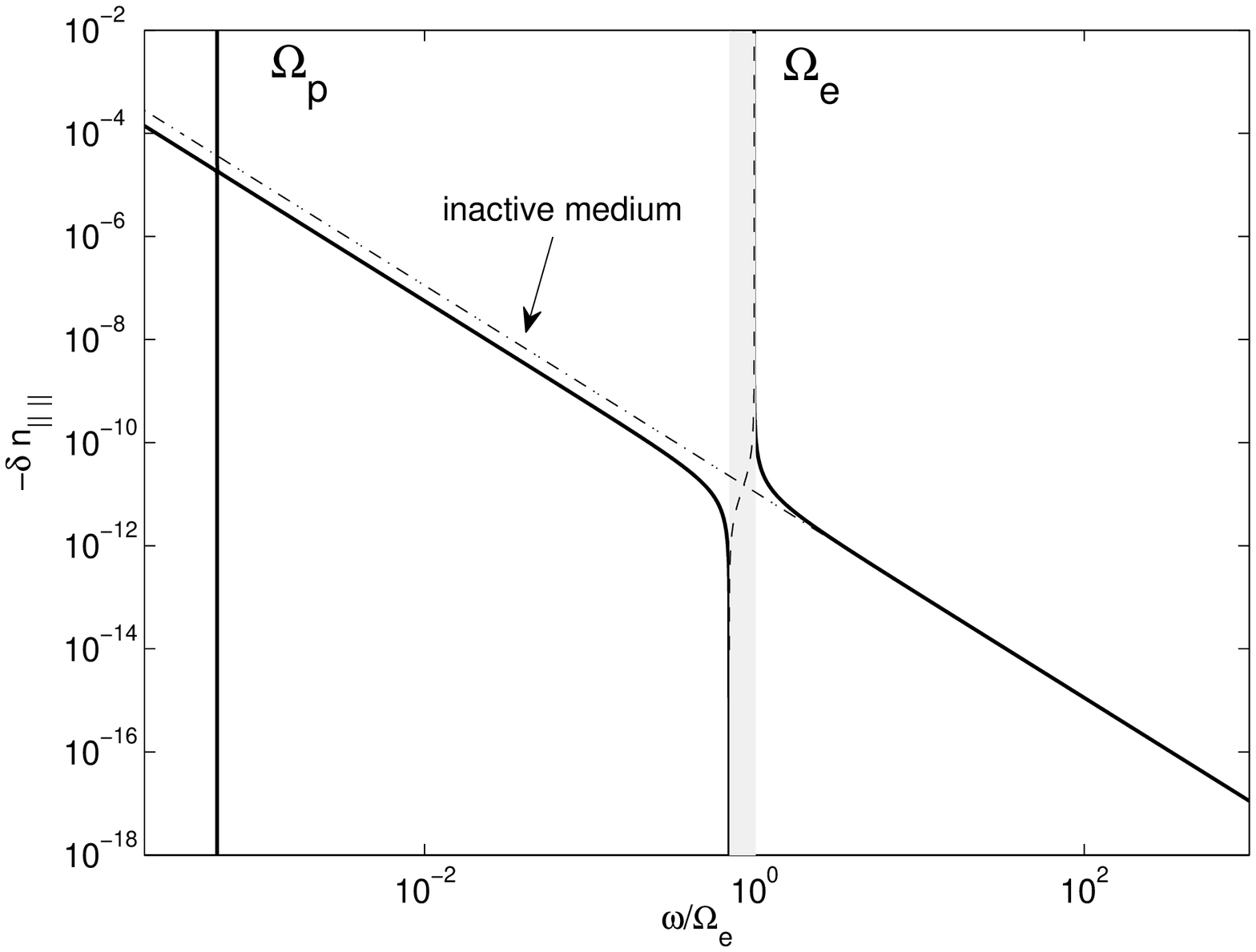}{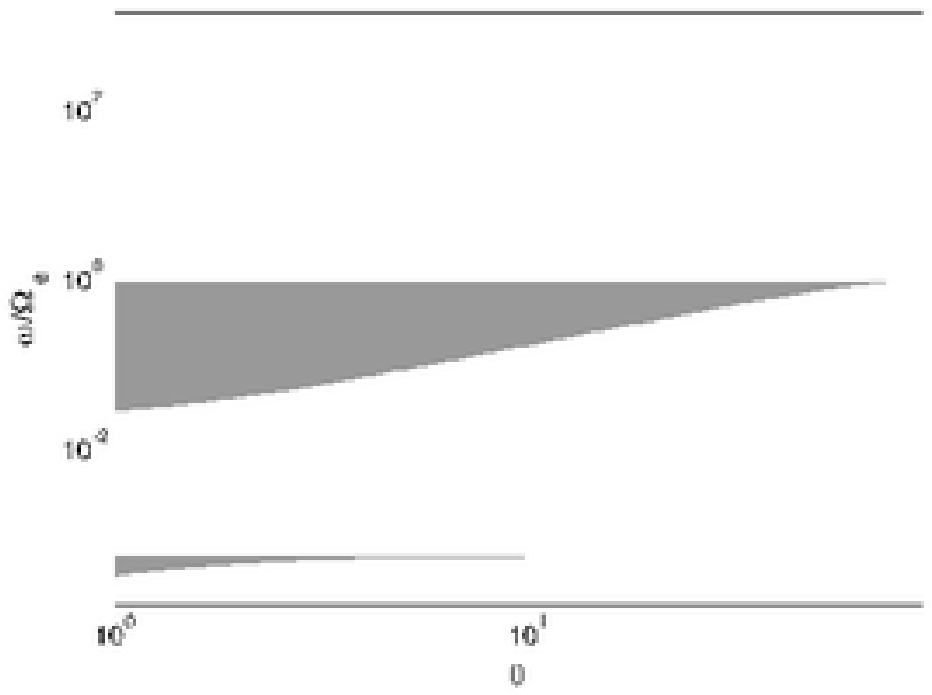}
\caption{{\it Left:} The refractive index, $\delta n_{\| \|}$, for cold active plasma as a function of photon frequency, normalized to the electron cyclotron frequency (for $\rho/m_p\sim 10^{14}~{\rm cm^{-3}}$ and $B\sim 10^{10}$\,gauss then $\Omega_e\sim 0.1$\,keV). The case for proton-electron plasma is shown and the results for electron-positron plasma are similar (apart from the lack of proton cyclotron lines and a slightly elevated values - note however that electron-positron plasma is inactive; dash-dotted line). The energy range where $\delta n>0$ (dashed line regions) cannot result in resonance conversion and are seen to occur just below the electron/proton cyclotron frequencies. The width of the region depends on $\theta$. {\it Right:} The energy range where resonance conversion is forbidden (grey region) as a function of $\theta$. Clearly, for intermediate values of $\theta$ the forbidden energy range is only a small fraction of the entire spectral band. Whether or not photon-particle resonance conversion is possible in specific objects requires knowledge of their emission mechanisms and the particular energy at which the resonance feature is expected to occur.}
\label{active_cold}
\end{figure*}

When a magnetic field is threading the plasma, symmetry is broken and the medium becomes active, i.e.,  different photon polarizations travel with different phase velocities. It is most simple to calculate the dielectric tensor, $\epsilon'_{ij}$, in a coordinate system in which one axis (the third axis in this case) is parallel to the magnetic field. This coordinate system is different than the one adopted above and a simple transformation exists between the two; see below. This classical result appears in text books and is included here for 
completeness purposes: 
\begin{equation}
\epsilon_{11}'=\epsilon_{22}'=1-\sum_s\frac{\omega_{p,s}^2}{\omega^2-\Omega_s^2},~~\epsilon_{12}'=-i\sum_s \frac{\Omega_s}{\omega} \frac{\omega_{p,s}^2}{\omega^2-\Omega_s^2},~~\epsilon_{33}'=1-\sum_s\frac{\omega_{p,s}^2}{\omega^2}, ~~{\rm and}~~\epsilon_{13}'=\epsilon_{23}'=0.
\end{equation}
Here $s$ denotes a sum over all relevant charges (e.g., electrons, positrons, protons, ions). $\Omega_s$ is the cyclotron frequency ($\Omega_s=e_sB/m_sc$). Optical activity is clearly evident since, generally, $\epsilon'_{12} \ne 0$. The dielectric tensor in  the coordinate system of our choice (Fig. 1) is given by
\begin{equation}
\epsilon= {\bf A}^T \epsilon' {\bf A}~~~{\rm where}~~~ {\bf A}= \left (
\begin{array}{ccc}
1 & 0 & 0 \\
0 & {\rm cos}\theta & {\rm sin}\theta \\
0 & -{\rm sin}\theta & {\rm cos}\theta
\end{array} \right )
\end{equation}
where only $\epsilon_{ij}$ with $i,j=1,2$ is of relevance to our study (the electric field along the
propagation direction of the photon is negligible). The resulting expressions are cumbersome and we do not give them here (see Lai \& Heyl 2006). We do plot however the $\delta n_{\| \|}=(\epsilon_{\| \|}-1)/2$ component of the dielectric tensor which is most relevant to the photon-particle conversion in figure 4a. Clearly, the picture is qualitatively different than the case of cold, non-active plasma in the sense that there is a region where $\delta n_{\| \|}>0$ hence resonance conversion does not occur. The range where this occurs depends on $\theta$ as shown in figure 4b. For photons traveling at small angles to the magnetic field, the region where resonance conversion is forbidden is large and an energy range spanning two orders of magnitude may be less suitable  for photon-particle conversion detection. At present, our understanding of the photon emission mechanisms in many astrophysical environments is limited and so $\theta$ is not known. Due to the lack of evidence suggesting otherwise, and assuming  intermediate values for $\theta$, we obtain that the energy range where resonance cannot occur is, generally, only a small fraction of the observable spectrum. The effect of protons and the heavier ions is always small. To conclude, the basic effect of active plasma is to open energy gaps within which resonance conversion does not occur (yet photons may still propagate).

Another effect relates to the fact that resonances occur as roots of equation 23 and, for $F_{\| \|}$ having a non-trivial dependence on $\omega$, more than one root is possible. In particular, $F_{\| \|}(\omega)$ may attain a large range of values  near cyclotron lines where additional solutions may occur (depending on whether the the main resonance feature is located below or above the cyclotron lines; see also \S5.3).  Nevertheless, observationally, it may be difficult to tell cyclotron line absorption from photon-particle conversion and we do not treat these pathological cases in this work. We note, however, that, such features would be narrow with respect to the main resonance feature (see \S5.3.1).

\subsubsection{Active hot plasma}

When solving for the dielectric tensor of cold active plasma it is assumed that charges are basically at rest and that their motion is set by the local phase of the impinging electric field of the propagating photon. However, when the thermal motion of the electrons is considerable and, specifically, when their kinetic energy is a non-negligible fraction of their rest mass, electrons travel with appreciable velocities in all directions while reacting to the electric field. In particular, for hot enough plasma, the path travelled by the electron over one period of the electric field may be comparable to the photon wavelength. This effect is considerably more complicated to account for and extends beyond the reach of simple analytic calculations. Nevertheless,  gas temperatures of order $\sim 40$\,keV have been measured from pulsar cyclotron lines and pair produced plasma in the magnetospheres of pulsars and magnetars is thought to exceed those temperatures resulting in the electrons having sub-relativistic velocities. As such, the effect of particle motion needs to be included. Under a wide range of plasma conditions (e.g., in thermodynamic equilibrium) the protons, if present, may be considered relatively cold.

\begin{figure*}
\plottwo{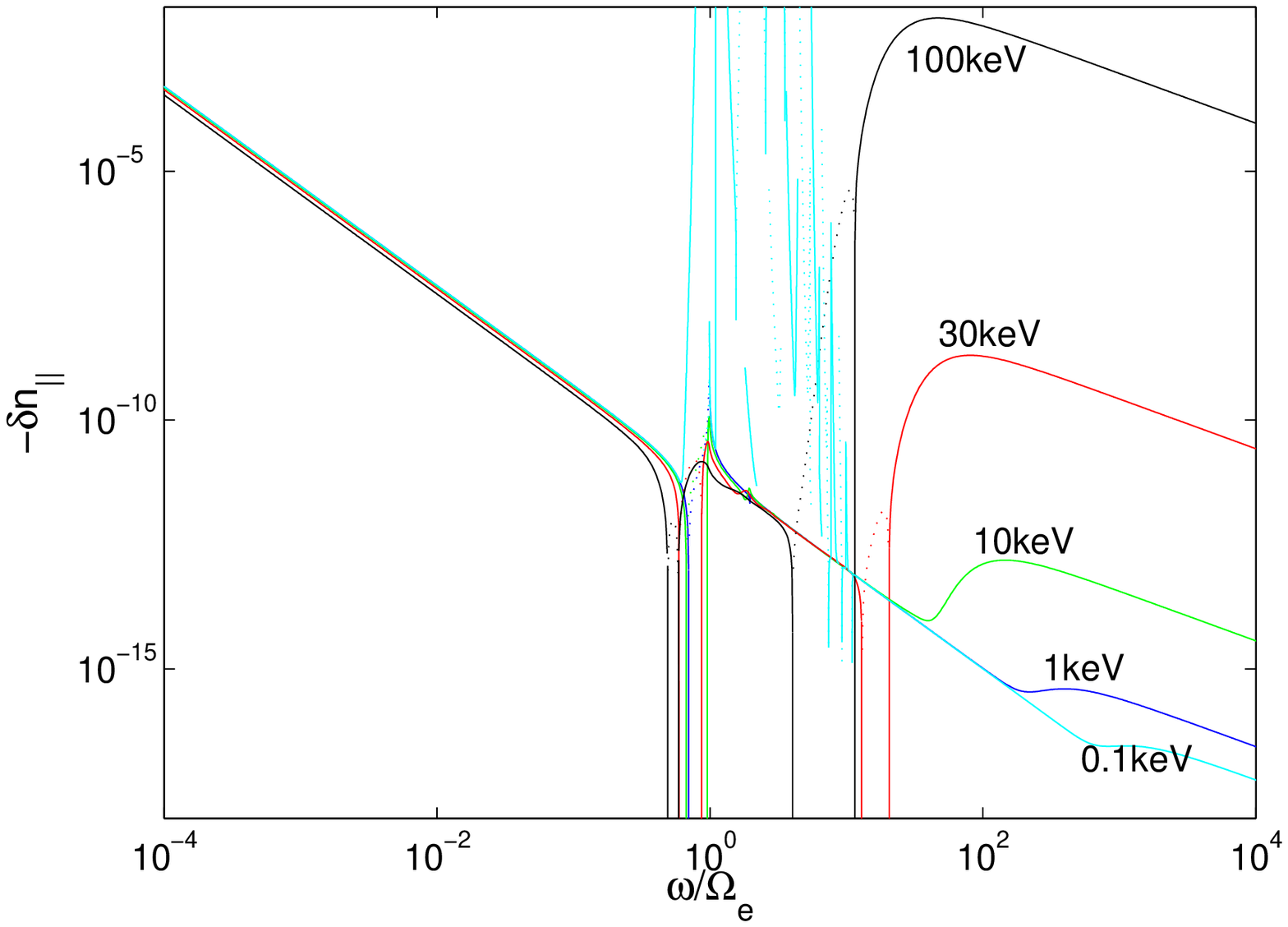}{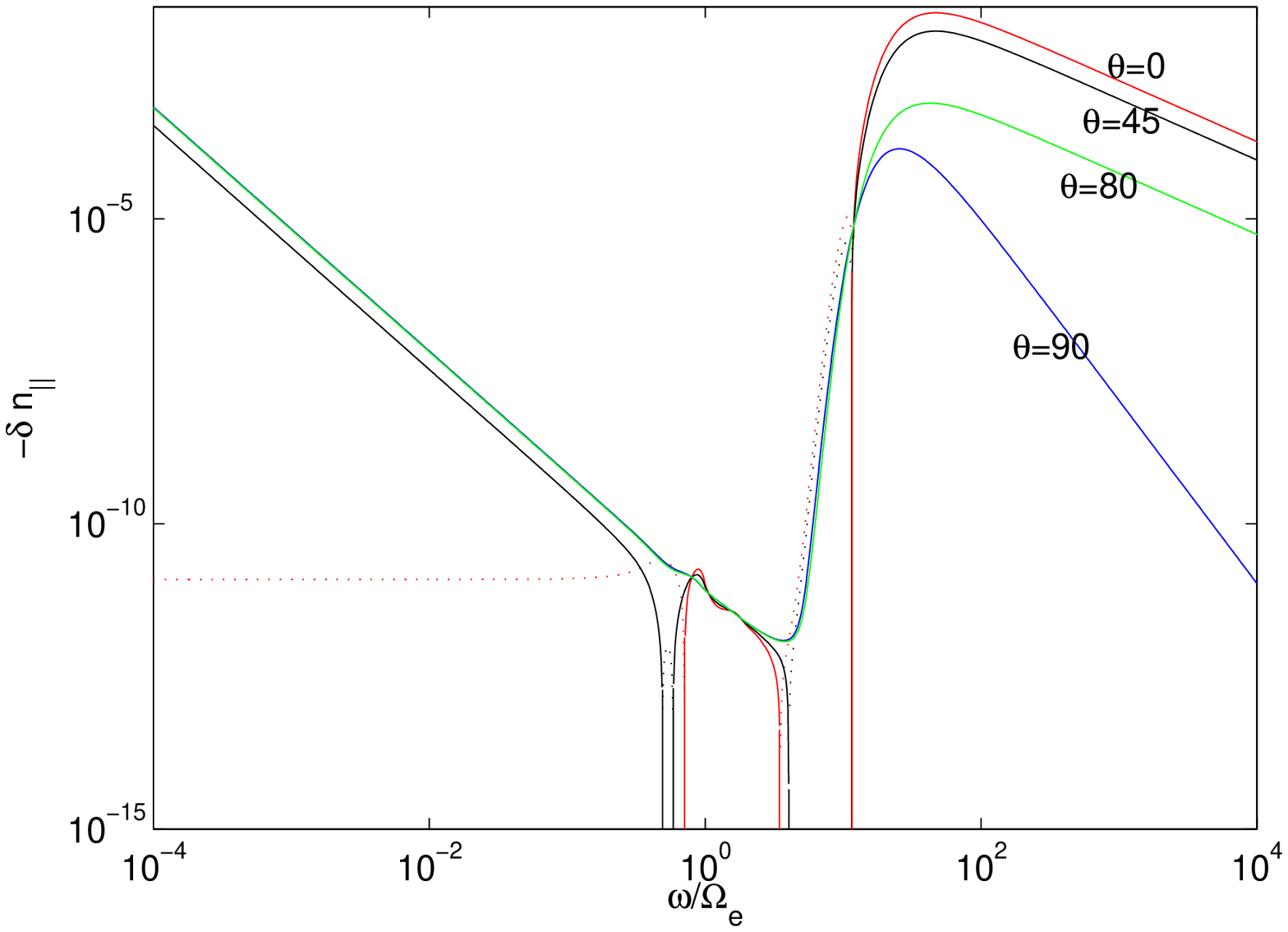}
\center \includegraphics[width=8cm]{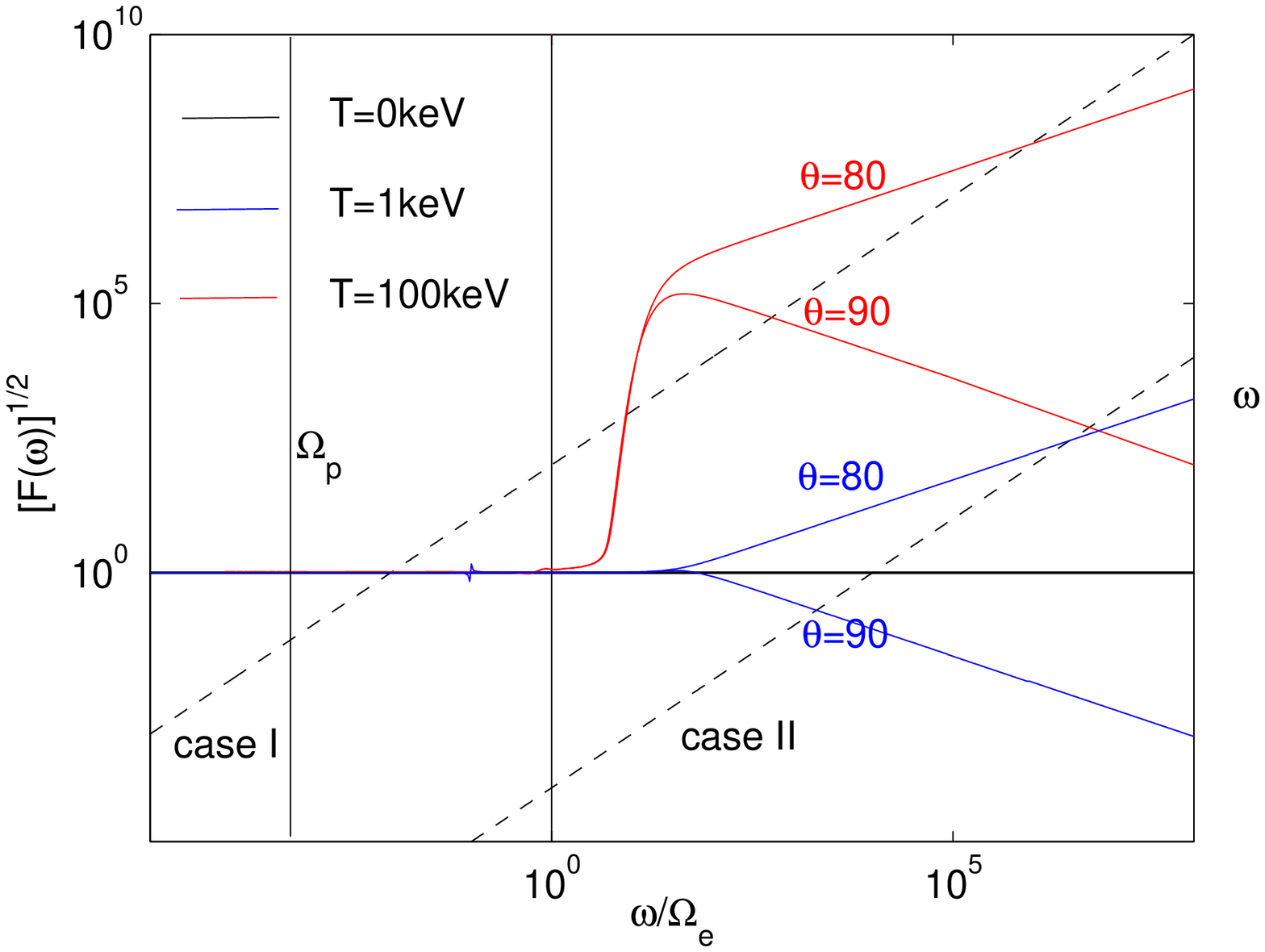}
\caption{{\it Left:} The refractive index, $\delta n_{\| \|}$, for relativistic active plasma (for $\rho/m_p\sim 10^{14}~{\rm cm^{-3}}$, $B\sim 10^{10}$\,gauss then $\Omega_e\sim 1$\,keV and $\theta=45^\circ$) with temperatures denoted next to each curve (proton contribution is not shown). Dotted sections of the curve denote positive values for $\delta n_{\| \|}$ whence resonance conversion does not occur. Clearly, 
the results are quantitatively different than those of cold plasma (Fig. 4a) and a prominent hump develops at high energies (see text). {\it Right:} The dependence of $\delta n_{\| \|}$ on the angle $\theta$ between the photon and the magnetic field (100\,keV  plasma is assumed). Clearly, large gaps form for $\theta \sim 0$ (dotted lines) where resonance conversion does not occur.  These gaps narrow as $\theta$ increases. {\it Bottom:} The solutions to equation 23 where sub-relativistic (red line) and weakly relativistic (black line) are considered ($\sqrt{F(\omega)}$ in solid lines and $\omega$ in dashed lines). Two cases may be distinguished: case I - where the resonance frequency in the case of cold plasma lies below the electron cyclotron frequency, $\Omega_e$. Case II - where that frequency is above $\Omega_e$. Clearly, the effect of finite plasma temperature does not effect the resonance below $\Omega_e$ (unless that one falls in a range where $\delta n_{\| \|}>0$ - not shown here) but could add more resonance features at higher energies. Case II resonances may be significantly affected by finite temperature plasma to the point where the might become undetectable (see text). The dependence on $\theta$ is significant for small $\theta$ for both weakly and sub-relativistic plasma.}
\label{active_rel}
\end{figure*}

In approaching this problem we assume that the electron gas follows the (relativistic) Maxwell velocity
distribution. While this may not be entirely justified in all cases, it is the only reasonable assumption to make given our limited understanding of those systems. The general form of the dielectric tensor for relativistic electron gas was first presented by Trubnikov (1959) as a steady-state solution to the linearized Vlasov equation for particles with a relativistic Maxwell velocity distribution. Nevertheless, evaluating the tensor for arbitrary plasma and wave parameters proved to be challenging since the Hermitian parts contain complicated Cauchy-type singularities (e.g., Bornatici \& Ruffina 1985). These difficulties gave rise to various perturbative methods for evaluating the tensor. In particular, the weakly relativistic regime has been calculated using appropriate plasma dispersion function (PDF) expansion series which is appropriate for low temperature plasma. More recently, a general method for calculating the fully relativistic tensor using exact PDFs was introduced by Castej\'on \& Pavlov (2006) on the basis of the theory of Cauchy-type integrals. In this method, the tensor is presented in a compact closed form as a finite Larmor radius expansion in terms of the exact PDF and all is required is the evaluation of the latter (see appendix).

The fully relativistic tensor has a rather slow convergence with the Larmor radius expansion for high temperatures,  and we use the mildly relativistic approximation which is fully justified for plasma temperatures below 100\,keV, as is likely to be the case for all astrophysical environments considered here. In this case the tensor is given by
\begin{equation}
\begin{array}{l}
\displaystyle \epsilon_{11}=1-\mu \left ( \frac{\omega_p}{\omega} \right )^2 \sum_{n=1}^N \left [ Z_{n+3/2}(z_n)+Z_{n+3/2}(z_{-n}) \right ] n^2 \frac{A_n(\lambda)}{\lambda} \\
\displaystyle \epsilon_{12}=-\epsilon_{21}=i\mu \left ( \frac{\omega_p}{\omega} \right )^2 \sum_{n=1}^N  \left [ Z_{n+3/2}(z_n)-Z_{n+3/2}(z_{-n}) \right ] n \frac{dA_n(\lambda)}{d\lambda} \\
\displaystyle \epsilon_{22}=\epsilon_{11}+\mu \left ( \frac{\omega_p}{\omega} \right )^2 2\lambda \left \{ Z_{7/2}(z_0)\frac{dA_0(\lambda)}{d\lambda}+ \sum_{n=1}^N  \left [ Z_{n+5/2}(z_n)-Z_{n+5/2}(z_{-n}) \right ] \frac{dA_n(\lambda)}{d\lambda} \right \} \\
\displaystyle \epsilon_{33}=1-\mu \left ( \frac{\omega_p}{\omega} \right )^2 \left \{ \left [ Z_{5/2}(z_0)+\frac{\partial ^2 Z_{7/2}(z_0)}{2\partial z_0^2} \right ] A_0(\lambda) +    \sum_{n=1}^N \left [ Z_{n+5/2}(z_n)+\frac{\partial^2 Z_{n+7/2}(z_n)}{2\partial z_n^2}+Z_{n+5/2}(z_{-n})+\frac{\partial^2 Z_{n+7/2}(z_{-n})}{2\partial z_{-n}^2} \right ] A_n(\lambda) \right \} \\
\epsilon_{13}=\epsilon_{31}=0 \\
\epsilon_{23}=-\epsilon_{32}=0
\end{array}
\end{equation}
where $\mu$ is the thermal velocity of the gas, e.g., electrons (in units of the speed of light), $z_n=\mu (1- n\Omega_e/\omega)/\sqrt{4a}$ ($a=\mu N_3^2/2$ where $N_3=k_{3}c/ \omega$ and $k_3$ is the photon wave vector along the axis defined in \S3.2.2) and $n$ is an integer. $A_n(\lambda) = 2\int_0^\infty dx J_n^2(\sqrt{2\lambda} x){\rm exp}(-x^2)$ and $J_n$ is the Bessel function. $\lambda=(k_\bot r_L)^2$ where $r_L$ is the Larmor radius of the relevant particle and $k_\bot$ is photon wave vector which is perpendicular to $k_3$ hence to the magnetic field. $Z$ are the exact plasma dispersion functions (PDF) whose definition in  further discussed in the appendix. We find that, for temperatures of up to 100\,keV, $N<30$ suffices to converge to a solution.

Our calculation of the component of the dielectric tensor term most important for photon-particle conversion calculations, $\epsilon_{\| \|}$  is shown in figure 5a as a function the photon wavelength and for different  electron temperatures (for $\theta=45^\circ$). The details of the dielectric tensor and the related refractive plasma index, $\delta n_{\| \|}$ change considerably as the electron temperature increases. In particular, several zones where $\delta n{\| \|}>0$ are created within which photon-particle resonance oscillations cannot occur. These zones become numerous (yet still do not cover a very broad energy range) for mildly relativistic gas, as shown for the case of gas with $T\sim 0.1$\,keV. Most importantly, there is a hump at high energies where the dielectric tensor resumes high (negative) values whose magnitude is an increasing function of the electron temperature. This results from taking into account the transverse relativistic dispersion of the plasma.  The hump in the refraction index becomes significant in the framework of photon-particle oscillations when the location of resonances is concerned since  $\epsilon_{\| \|}$ becomes a multi-valued function (and not only near cyclotron lines) so that detectable spectral oscillation features may occur in several places in the spectrum (see below and also \S5.3).  As shown, finite temperature plasma gives qualitatively different results than cold plasma models. While the differences are the largest for the more relativistic case, even small changes in the slope (as shown at high frequencies for low temperature plasma; Fig. 5a) may result in considerable shifts and shape changes of resonance features. We return to these effects in \S5.2,5.3.

As noted before, our knowledge of the photon emitting mechanisms from compact astrophysical objects in general, and in pulsars and magnetars in particular, is limited hence the angle $\theta$ between the propagating photon and the magnetic field is not known. It is therefore worthwhile to explore the dependence of the relativistic dielectric tensor on $\theta$ as we show in Fig. 5b. The results for the relativistic case is similar to the non-relativistic case whereby smaller angles result in broader energy regions where resonance conversion will not occur (since the refractive index, $n>1$). In addition, there could be several energy bands where $n>1$ further reducing the wavelength range over which the sensitivity to the detection of photon-particle conversion features. We note that similar slope changes at high frequencies as a function of $\theta$  occur also for weakly relativistic plasma such that for $\theta \lesssim 90^\circ$ the slope is steep (with a powerlaw index $<-2$) while for $ \theta \gtrsim 0$ it is generally shallower (with  a powerlaw index $>-2$). 

The frequency dependence of the refraction index at high photon energies can have a crucial effect on the resonance energy (and overall shape; see below) of the photon-particle spectral feature. Broadly, two cases may be distinguished: case I in which the resonance feature occurs at energies lower than the electron cyclotron frequency, $\Omega_e$, and the opposite case, which we denote as case II (see Fig. 5c). Uncertainties in the plasma temperature have little effect on case I. In particular, the energy of the feature is not shifted unless the temperature and angle $\theta$ conspire to have the forbidden energy range (i.e., where $\delta n_{\| \|}>0$) fall at those energies. That said, additional resonance features may occur where equation 23 has a non-trivial solution (these correspond to the intersections of the dashed case I line with the solid lines). The additional resonance frequencies occur at high energies. For weakly relativistic plasma another resonance feature will occur around the electron cyclotron frequency yet  more may be observed at even higher energies. Case II is qualitatively different. In this case, the resonance frequency lies above $\Omega_e$ and its location is very sensitive to the energy dependence of the refraction index which is temperature and $\theta$-dependent. As a result, case-II resonances may appear or disappear from any given spectral band (whose energy $>\Omega_e$) once the plasma properties and the propagation direction of the photon with respect to the magnetic field changes. As we shall later see (\S 5), different astrophysical objects can be associated with different cases.

\subsubsection{Plasma kinematics and gravitational redshift}

In the above calculations we have assumed plasma at rest. Nevertheless, in the general case, the plasma may be moving with respect to the observer with potentially high velocities (reflecting, perhaps, the escape velocity from the magnetosphere/accretion disk). In this case, a dielectric permittivity tensor should be boosted from the plasma frame (where the above calculations hold) to the observer's frame (e.g., Tamor 1978 and references therein). Clearly, the problem is quite rich since besides the angle between the magnetic field and the propagation direction of the photons, there are additional angles corresponding to the velocity direction of the plasma system with respect to the adopted coordinate system.  

In addition to the motion of the plasma, gravitational redshifting near compact objects may occur. Both of these effects combine to shift the frequency at which conversion occurs. At present, however, plasma kinematics near compact objects is poorly understood and the precise position of the conversion feature is uncertain (even if all other parameters are well determined). We note, however, that, unless the plasma is moving with relativistic velocities or the region where resonance conversion occurs is close to the event horizon, the observable effects  are likely to be modest and probably much smaller than other uncertainties in the problem (e.g., concerning the magnetic field intensity and plasma properties). For these reasons, we choose to work under the assumption of no gravitational redshift and plasma at rest in the remainder of this work.

\section{Spectral Properties}

In this section we focus on the following questions: how does a photon-particle conversion feature look like and how may it be distinguished from other spectral features such as atomic lines and edges? Calculating the spectral shape goes beyond identifying the energy where resonance occurs (as we qualitatively did for the case of uniform cold plasma in \S3.2.1) and requires the solution of equation 14 as a function of photon energy over the entire spectral range. It is also a way by which non-resonant conversion may be predicted and its spectral signatures derived.   We first treat the case of uniform plasma and magnetic field configuration and then discuss the more realistic case where both these quantities are location dependent.  For the sake of simplicity, we shall work in the  cold plasma regime treating relativistic plasma in later sections when specific astronomical objects are discussed. 

\subsection{Uniform Conditions}

\begin{figure}
\center \includegraphics[width=8cm]{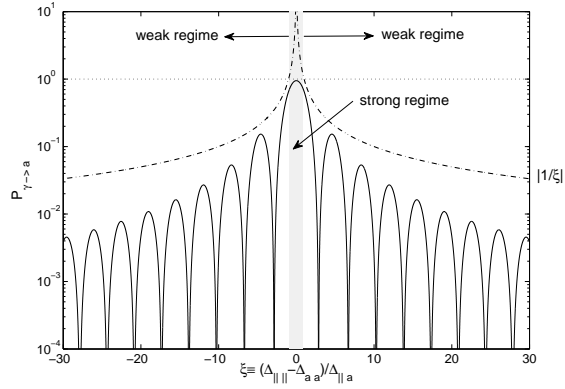}
\caption{The probability for photon-particle conversion under uniform conditions (magnetic field and density) as a function of  $\xi\equiv \Delta _{\| a}^{-1}(\Delta _{\| \|} -\Delta_{aa} )$ ($\xi=0$ corresponds to the resonance). $\xi$ is some measure of the photon energy (in particular, for $\xi<0$, $\xi \propto \omega^{-1}$ while $\xi \propto \omega^2$ for $\xi>0$  in cold plasma).  The strong, non resonance, regime is defined as the energy range where $\xi^{-1}>1$ (shown in dashed line corresponding to the energy range gray area). The weak regime is defined here as the energy range where $\xi^{-1}<1$. Clearly, most of the energy range corresponds to the weak regime. Evidently, the highest conversion probability is obtained at resonance, $\xi=q=0$. As such, it is crucial for detecting particles down to the lowest possible values of the coupling constant, $g$.}
\label{strong_weak}
\end{figure}

Solving for the probability of conversion from a pure photon state to an axion state is relatively straightforward when all quantities are constant in space and when optical activity is neglected (e.g., Raffelt \& Stodolsky 1989)
\begin{equation}
P_{\gamma \rightarrow a}=\left [ \Delta_{\| a} \gamma  \frac{{\rm sin}\left ( 2\pi \gamma/ \gamma_{\rm osc} \right)}{2\pi \gamma/\gamma_{\rm osc}} \right ]^2=\left [ {\rm sin}(2\phi){\rm sin}(2\pi \tilde{\gamma}) \right ]^2
\end{equation}
where $\tilde{\gamma}\equiv \gamma/\gamma_{\rm osc}$ and 
\begin{equation}
\gamma_{\rm osc}\equiv \frac{2\pi}{\Delta_{\| a}} {\rm sin} 2\phi~~~{\rm and}~~~\phi\equiv\frac{1}{2}{\rm atan} \left [ 2\Delta_{\| a}/ (\Delta_{\| \|} - \Delta_{a a}) \right ].
\end{equation}
Here $\gamma_{\rm osc}$ is the physical length-scale over which photon-axion-photon conversion occurs. System for which $\tilde{\gamma} \ll 1$ would show little conversion since photons have not had time to oscillate into particles (see below). The observed flux from an object whose emission is given by $F_0(\omega)$ would therefore be
\begin{equation}
F(\omega)=F_0(\omega) \left [ 1-P_{\gamma \rightarrow a} (\omega) \right ].
\end{equation}
These simple expressions for the conversion probability result in a rich spectrum of oscillation signatures. 

There are three distinct regimes of interest in the problem which are defined by the value obtained by $\xi \equiv \Delta_{\| a}^{-1}(\Delta_{\| \|}-\Delta_{a a})$:
\begin{itemize}
\item
{\it Strong resonance conversion} which occurs when $\xi \propto \Delta_{\| \|}-\Delta_{a a}=0$, i.e.,  $q=0$ (equation 16). In this case $\phi=\pi/4$ and $\gamma_{\rm osc}=2\pi/\Delta_{\| a}$ so that
\begin{equation}
P_{\gamma \rightarrow a}=\left [  {\rm sin} \left ( \Delta_{\| a} \gamma \right ) \right ] ^2 \simeq \frac{1}{4}g^2B^2\gamma^2
\end{equation}
where the last  expression on the right-hand side holds if $\Delta_{\| a} \gamma \ll 1$ and the conversion probability is small (see \S 1). Clearly, the conversion probability is very small for $\gamma \ll \gamma_{\rm osc}$ simply because not enough phase has been accumulated along the photon's path to considerably mix with particles. Considerable conversion does take place for $\gamma \simeq \gamma_{\rm osc}$. For $\gamma \gg \gamma_{\rm osc}$, considerable conversion occurs when $\gamma/\gamma_{\rm osc} \ne (N+1/2)\pi/2$ where $N$ is an integer.  Clearly, even small change of system dimensions could result in a significant change to the conversion probability. Realistically, averages are better defined quantities in these cases. 
\item
{\it Strong non-resonance conversion}: This cases arise when $\xi^{-1} = \Delta_{\| a}/(\Delta_{\| \|}-\Delta_{a a})\gg 1$ yet the denominator $\Delta_{\| \|}-\Delta_{a a} \ne 0$ (c.f. above). Here too, $\phi \simeq \pi/4$ and equation 30 holds. This regime occurs naturally just off the resonance peak where $\Delta_{\| a}/(\Delta_{\| \|}-\Delta_{a a})\sim 0$ but could also occur independently of it provided that either the coupling constant $g$ is large or the magnetic fields are high enough. 
\item
{\it Weak conversion}: In this case $\xi^{-1}\ll 1$ so that $\gamma_{\rm osc}\simeq  4\pi/(\Delta_{\| \|}-\Delta_{a a} )$ and 
\begin{equation}
P_{\gamma \rightarrow a}=\left \{ 2\frac{\Delta_{\| a}}{\Delta_{\| \|}-\Delta_{a a}} {\rm sin} \left [ \frac{1}{2}  (\Delta_{\| \|}-\Delta_{a a} ) \gamma \right ] \right \} ^2 \ll 1.
\end{equation}
This case is of limited observational use as the conversion probability is very small. It does not allow us to probe down to low values of the coupling constant, $g$, which are of particular interest. That said, accounting for this regime may well be important for the correct interpretation of a high S/N spectrum where prominent oscillation features are observed.
\end{itemize}

An example for the photon-particle probability distribution is shown in figure \ref{strong_weak}. The different regimes are designated. In particular, resonance occurs at $\xi=0$ (see caption) with strong, non-resonance conversion taking place at its wings (remember though this such conversion may, in principal, occur without the existence of a resonance yet this requires stronger coupling), and weak conversion extending to large values of $\vert \xi \vert$. In particular, far from resonance equation 32 holds, and the signal shows rapid oscillations with an overall envelope tracing a $\vert \xi^{-2} \vert$ curve.

It is clear from the above that the most observationally interesting regime is that of strong resonance conversion since this allows us to probe down to the lowest possible values of the coupling constant $g$. In particular, assuming that one can detected conversion probabilities down to some threshold, $P_{\rm th}$ (here we assume a fiducial number of 5\%), then the minimum coupling constant, $g_{\rm min}$, that may be probed under the assumption of {\it uniform} conditions in an object up to radius $R$, is (Eq. 31)
\begin{equation}
g_{\rm min} \simeq 5\times 10^{-16} \left ( \frac{10^{15}\,{\rm G}}{B} \right ) \left ( \frac{10^{6}\,{\rm cm}}{R} \right ) \left ( \frac{P_{\rm th}}{5\%} \right )^{1/2},
\end{equation}
where we have taken values characteristic of magnetars (see \S 5.1). Clearly, these limits on $g$ are much lower, by almost six orders of magnitude(!), than those attainable by most current detection means. We note, however, that deviation from uniformity may have a crucial effect, as will be discussed in \S4.2 and that it is likely that $g_{\rm lim}$ is somewhat higher than that obtained above. 

An estimate for the width of a resonance feature centered on $\Delta_{\| \|}-\Delta_{a a} =0$ is where the probability, shown in figure \ref{strong_weak}, is first nullified. This implies that significant conversion occurs for 
\begin{equation}
\left [ \Delta_{\| \|}(\omega)-\Delta_{a a}(\omega) \right ] \gamma < 2\pi.
\end{equation}
This is completely analogous to the requirement that the momentum transfer in the conversion process corresponds to wavelengths larger than the size of the system  (i.e., $q\gamma <2\pi$). One can analytically estimate the relative width of the lines for the case of cold plasma in a sub-critical magnetic fields,
\begin{equation}
\frac{\delta \omega}{\omega_0} = \sqrt{ \frac{720 \pi^3}{7\alpha}} \frac{ B_c/B}{\gamma \omega_p}.
\end{equation}
It is worthwhile to estimate the required variation in $B$  which would shift the conversion feature by more than its width,
\begin{equation}
\frac{\delta B}{B} = \frac{\delta \omega}{B} \left \vert \frac{\partial \omega_0}{\partial B} \right \vert ^{-1}=\frac{\delta \omega}{\omega_0}
\end{equation}
To put some numbers, we find that $\delta \omega/\omega_0$ may be as low as $10^{-4}$ for some pulsars and even lower for magnetars (see below but note that the above expression does not hold for super-critical fields). As realistic astrophysical objects are unlikely to maintain their properties fixed with such high precisions over large volumes and over time, the observed features are likely to be different than those considered here, and the effect of stratification in the magnetic field and plasma density (as well as other properties) should be considered. This is the main topic of \S4.2.

\begin{figure*}
\plottwo{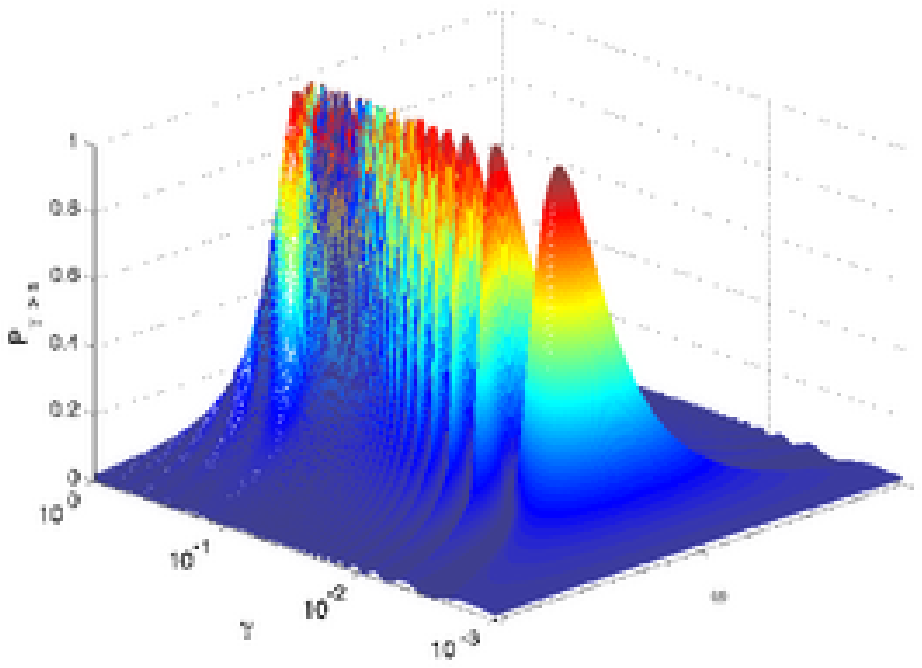}{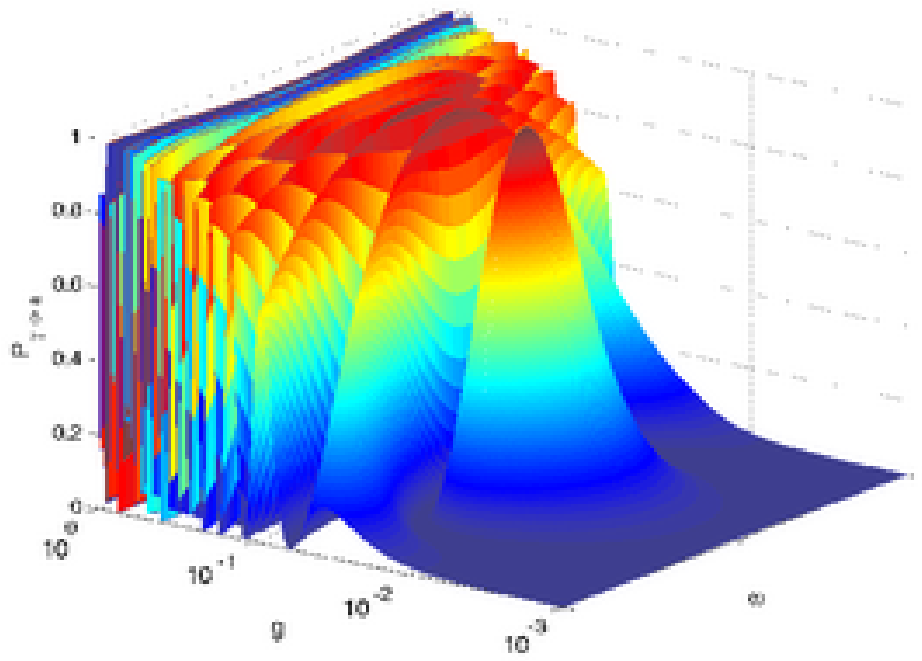}
\caption{{\it Left:} The dependence of the energy-dependent probability, $P_{\gamma \to a}(\omega)$, on the path length through the plasma, $\gamma$. For low values of $\gamma$, conversion is negligible across the spectrum. For larger $\gamma$, a resonance feature appears whose envelope is conserved for much larger $\gamma$. Large values of $\gamma$ result in more rapid variations in the conversion probability over narrow energy intervals yet the overall shape of the envelope is conserved. {\it Right:} $P_{\gamma \to a}(\omega)$ as a function of the coupling constant $g$ with otherwise fixed parameters. Here the envelope shape is not conserved and the feature becomes wider with increasing $g$. In particular, for large enough $g$, the feature may extend over several decades of energy (this excludes the broadening effect of spatial stratification in the magnetic field intensity and plasma density discussed in \S4.2. }
\label{profiles}
\end{figure*}

In figure \ref{profiles} we show the dependence of the conversion probability  as a function of the photon energy     near resonances on the path length of the photon through the medium, $\gamma$, and the coupling constant, $g$. In the former case, the mean overall shape of the conversion probability profile is conserved. Nevertheless, taking slices at constant $\gamma$, which corresponds to an observed spectrum at some location, yields a larger number of narrower components within the envelope with increasing $\gamma$, as indicated by equation 35. In particular, for large enough $\gamma$ the details of the conversion probability profile may vary considerably over very short $\delta \gamma$ intervals. This implies that in such regimes, and for all practical purposes, only the shape of the outer envelope is well defined. In particular, in systems which correspond to such regimes and whose dynamical timescales are short compared to the exposure time of the observation, only the outer envelope could be observed. Furthermore, as all spectrographs have finite resolutions, instrumental smoothing of the signal will occur. We emphasize that a major differences between atomic features and oscillation features is that while optical depth is monotonically increases with path length through the plasma (neglecting contribution from emission to the overall flux), larger values of $\gamma$ may result in particles oscillating back to photons. This aspect is generic and is not related to the specific assumptions employed here.

The dependence of the conversion probability profile on the coupling constant is quite different (Fig.\ref{profiles}). In particular, increasing $g$ broadens the line considerably: for large enough $g$, photons over a very broad (the entire observable)  spectral range may be converted to axions and vice-versa. 

\subsubsection{Comparison with atomic lines}

\begin{figure}
\center{\includegraphics[width=8cm]{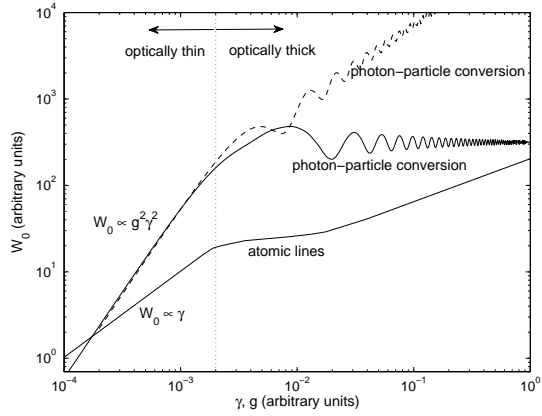}}
\caption{A "curve-of-growth" for photon-particle spectral oscillation features compared to that of atomic lines. Clearly, the behavior is qualitatively different. Most importantly, for small values of $\gamma$ the dependence is strong compared to that of atomic lines. For large $\gamma$, the rest equivalent width, $W_0$, oscillates around a saturation value which is very different from the monotonic increase of $W_0(\gamma)$ for atomic lines. The dependence of $W_0$ on the coupling constant, $g$, is also shown (see text). The non-linear regime at smaller $g\gamma$ values has a significant effect on the detectability of oscillation features in stratified plasma.}
\label{cog}
\end{figure}

It is interesting to compare some of the properties of photon-axion conversion spectral resonance features to atomic absorption lines.  In the field of atomic spectroscopy, it is customary to define the rest-equivalent width of a line so that $W_0\equiv \int_0^\infty d\omega (1-{\rm exp}[-\tau(\omega)])$ where the optical depth $\tau(\omega)=\tau(\omega_l)\chi (\omega-\omega_l)$ and $\chi$ is the line profile (whether Gaussian, Lorentzian, or other shapes, depending on properties of the gas; e.g., Rybicki \& Lightman 1985) centered around the resonance line frequency $\omega_l$. $\tau(\omega_l)$ depends on the product of the line cross-section and the distance along the line-of-sight.  The rest equivalent width for photon-particle conversion may be analogously defined as
\begin{equation}
W_0= \int_0^\infty d\omega P_{\gamma \rightarrow a}(\omega).
\end{equation}
We note that the above definitions assume that one knows the shape of the underlying spectrum. This is rarely an issue for narrow atomic lines but could be less straightforward to determine once broad (conversion) features are involved. The curve of growth for photon-particle conversion features and for atomic lines is qualitatively compared in figure \ref{cog}.

In the case of atomic lines, increasing the path of the photon through the plasma, $\gamma$, results in a monotonic increase in $W_0$. This is clearly not the case for photon-particle conversion features where individual components become narrower as $\gamma$ increases yet the overall shape is conserved.  This is physically related to the fact that, unlike absorption processes, in conversion processes photons are not lost but  are merely converted. In particular, for large enough $\gamma$, particles will convert back and forth between pure photon and pure particle states maintaing the overall envelope of the conversion feature while changing only its inner structure.  Furthermore, the curve-of-growth is non-monotonic although it reaches an asymptotic value for very large $\gamma$ when numerous oscillations average out.

Atomic lines and photon-particle conversion features are very different also in the "optically-thin" limit. While $W_0\propto \gamma$ for atomic lines, $W_0\propto \gamma^2$ for photon-particle conversion features. The non-linearity in the latter case has profound observational consequences especially in cases where spatial dispersion is important or the effective size of the system is reduced by e.g., photon scattering. Basically, the combined {\it incoherent} contribution from $N$ small regions of size $L/N$ to the oscillation amplitude does not equal that of one region of size $L$.

Studying the dependence of $W_0$ on the coupling constant $g$ reveals a behavior different from that of spectral lines: While $\gamma$ and $g$ are interchangeable as far as pure atomic line-absorption is concerned (since $W_0$ depends only on their combination, $\tau$), these parameters have a very different effect in the case of photon-particle oscillations. In particular, in the latter case $W_0$ is very sensitive to $g$ over the entire range, increasing at first as $g^2$ and roughly as $g$ in the "optically thick" limit which is much steeper than the behavior for atomic features. Specifically, for large enough $g$, the conversion feature could span a very broad energy range and, as such, resembling more a continuum feature rather than a line-like signature. [We note that for large enough $g$ our analytic predictions for the width of individual  conversion components (Eq. 34-36) may not hold since these were estimated in the weak conversion regime while in this case the strong non-resonant regime may be more appropriate.] The sharp dependence of $W_0$ on $g$ raises the interesting possibility that the spectra of known astrophysical objects can be used to set stringent limits (or even detect!) photon-particle oscillation feature (see \S4).

\subsection{Varying Conditions}

For real astrophysical objects it is  unlikely that their properties, such as the magnetic field and plasma density, are kept constant with  high precision within a given volume and are zero outside it. More realistically, the magnetic field and density are smoothly varying functions of position. As indicated by the dependence of the resonance energy $\omega_0$ on the magnetic field and plasma density, narrow spectral conversion features are expected to be considerably affected by the spatial stratification and/or time dependence  in the medium properties. In particular, any attempt to give quantitative spectral predictions should take such effects  into account.

We adopt the following numerical scheme in solving the problem of photon-particle oscillation in a stratified medium: we divide space into $N$ zones of size $\delta \gamma_j$ where $j=1...N$. Over each zone the conditions are considered to be (quasi-)uniform. In particular, for small enough $\delta \gamma_j$, there is a well defined local Hamiltonian, $\mathfrak{H}(\gamma_j)$ at location $\gamma_j$.  In this case, the initial state for the first zone is that given by the initial condition of the problem while the initial state for the $j$'th zone ($1<j\le N$) is the evolved state through $j-1$ zones. More formally,  equation 17 now takes the form
\begin{equation}
\left \vert \mathfrak{A}(\gamma) \right >=e^{i\sum_i \mathfrak{H(\gamma_i)}\delta \gamma} \left \vert \mathfrak{A}(\gamma=0) \right > = \sum_{i_1,i_2,...i_N}  \left (  \prod_{j=N...2}   \left < \mathfrak{A}_{i_j} \left \vert e^{i\mathfrak{H}(\gamma_{j}) \delta \gamma_{j}} \right \vert  \mathfrak{A}_{i_{j-1}} \right >  \right )
\left < \mathfrak{A}_{i_1} \vert \mathfrak{A}(\gamma=0) \right >  \left \vert \mathfrak{A}_{i_N} \right >
\end{equation}
where $i_j$ goes over the number of eigenstates of the Hamiltonian $\mathfrak{H}_j$ which, in our case, means $i_j=1,2,3$ for all $j$'s. The probability for photon-particle conversion given the initial and final states  is then given by $\left \vert \left < a \vert \mathfrak{A}(\gamma_{N}) \right > \right \vert^2$ (c.f. equation 14). In the following calculations we have made sure that proper convergence is obtained by repeating the calculation using a finer grid (larger $N$) and requiring that the relative probability differences does not exceed 5\% at any energy bin. For very large $\gamma$, hence very narrow features, differences may still be encountered, even for very  large $N$, and we then require that the relative difference of the outer envelopes agree to within 5\% accuracy. The value of $N$ depends among other things on how fast the conditions of the system vary with radius. We find that, for most relevant applications, $N\sim 10^3-10^4$ suffices. We used spatial logarithmic  spacing in our calculations.

\begin{figure}
\plotone{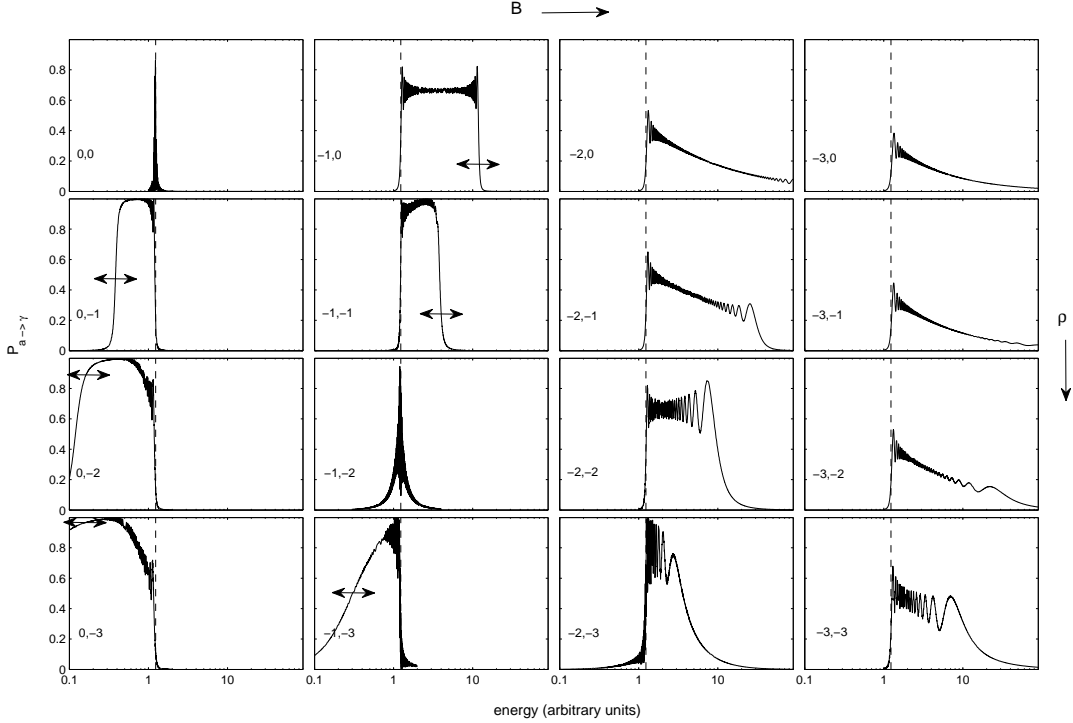}
\caption{The probability for photon-particle conversion, $P_{\gamma \rightarrow a}$, as a function of the energy for several powerlaw profiles for the magnetic field and density. The $\alpha,\beta$ powerlaws are designated in each panel in that order. Clearly, the probability distributions depend sensitively on the nature of the stratified magnetic field and plasma. In particular, cases in which the magnetic field and plasma density decrease rapidly with distance tend to have extended blue wings toward higher photon energies with oscillations showing up at the high energy tail since the relative width of features is $\propto (B\sqrt{\rho})^{-1}$ (Eq. 35). The probability distribution is narrow in cases where $\alpha=\beta/2$ (see text). Cases in which the plasma density changes relatively little with distance (so that it is higher at large $r$), show more rapid oscillations. Whether these may be observed, depends on the instrumental resolution and time-dependent issues (see text). Double arrow indicate that the feature is rather sensitive to the outer radial cutoff assumed (e.g., increasing the radial cutoff would broaden the feature and vice-versa).}
\label{bn_var}
\end{figure}

At present, our understanding of many astrophysical objects is rather qualitative and most models rely on self-similarity arguments or general physical principles to obtain rough scalings of the object's properties with the spatial coordinates (e.g., angle, radial distance). Somewhat reassuring is the fact is that in several cases where more realistic numerical simulations have been conducted, a qualitative agreement was found with the analytical approach. We shall therefore treat cases in which the density and magnetic field depend only on the radial coordinate from the photon emitting source. While probably an oversimplification, we are interested here mainly in the qualitative difference between spatially varying configurations and uniform ones. Further refinements await better understanding of the physics of particular environments. 

In  the following examples we consider density and magnetic field variations with the radial coordinate alone so that they take the following form
\begin{equation}
B=B(r_\star)\left ( \frac{r}{r_\star} \right )^\alpha~~~~{\rm and}~~~~\rho =\rho(r_\star) \left ( \frac{r}{r_\star} \right )^\beta.
\end{equation}
In particular, we neglect possible variations of the plasma temperature and composition with distance (these will be shown in \S 5 to have a smaller effect on the shape of the spectral feature) and do not consider variations in $\theta$. Although an over-simplification, our understanding of these  properties  is poor and so any assumption is as good as others. We also limit our discussion to the case of non-active cold plasma which suffices to show the main differences with the uniform case (see \S 5 for a proper treatment of those effects in the context of specific astrophysical systems).

Qualitative understanding of what might be expected can be gained by looking at equation 23 for the value of the resonance frequency. Clearly, as the magnetic field strength decreases (increases)  the resonance frequency is pushed to higher (lower) energies where the plasma refraction index is lower so that cancellation of the vacuum birefringence term may occur. Generally (including the case of active cold/hot plasmas) one expects the resonance frequency to shift with location due to the varying $B-$field.  In addition, the width of the feature would also vary (e.g., increasing with decreasing field; equation 35). Varying the plasma density will likewise change the location and width of the resonance feature: the resonance frequency is proportional to $\omega_p\propto \sqrt{\rho}$ and the fractional width $\delta \omega/\omega_0\propto \rho^{-0.5}$.  

A few calculated examples for sub-critical magnetic fields are shown in figure \ref{bn_var} as a function of $\alpha,\beta$ (all other, arbitrarily chosen, parameters are similar). Clearly, the effect of spatial stratification in either the magnetic field or plasma density results in a considerable change in the properties of the conversion features. Generally, spatial dispersion tends to broaden the feature since resonance conversion occurs at a continuous range of photon energies. For many astrophysical objects (see below), the magnetic field intensity falls off as a powerlaw with index $-3<\alpha<-2$ (with $\alpha=-3$ corresponds to a dipole field while $\alpha=-2$ describes a field in equipartition with the plasma in singular isothermal sphere models). As shown in figure \ref{bn_var}, such values for $\alpha$ result in the probability (and the spectral feature) having a blue wing which may extend up  to several orders of magnitude in energy (if strong conversion is at all possible). In particular, the slope of the blue wing depends on $\alpha$: smaller $\alpha$ result in a steeper blue wing. For $\alpha\sim -1$ the resulting probability is flattened which results from the fact that $rB(r)$ has only a weak dependence on the radial coordinate.  The dependence on the density profile (i.e., $\beta$) is weaker and its main effect is to control the width of the spectral feature and less so its shape. The reason for that is simple: the conversion probability at the resonance does not depend on the density yet the photon energy at which effective conversion occurs does (note however that more complex behavior may occur in the strong non-resonance regime). An interesting case where the spectral feature remains narrow despite the spatial dispersion is when $\alpha=\beta/2=-1$. This case is clearly evident from equation 23 showing that, in this case, $\omega_0$ is essentially independent of location.

We have noted before that  individual narrow components which trace the overall envelope are unlikely to be observed since astrophysical objects do not have their properties maintain a constant value to a very high precision over time. The oscillations predicted here at the blue wings of the feature can be relatively broad suggesting that the system properties need to fluctuate by a factor of order unity to smear them out. As it is not clear by how much these properties fluctuate over time in realistic objects, it is not clear just how conspicuous these features might be in the spectra of real objects. That said, it is likely that objects with long dynamical or light crossing timescales (such as quasars) compared to the observations' duration would show such features. Time resolved spectroscopy of more compact regions may show them too.

To summarize, spatial dispersion in the properties of the system (plasma and magnetic field) results in a much richer variety of spectral conversion features than discussed in \S 3. In particular, the features may be considerably broadened spanning several orders of magnitude in energy and exhibiting a myriad of forms. A common feature in the case of small $\alpha,~\beta$ is the appearance of relatively broad oscillations occurring in the "optically thin" regime of the extended blue wing of the feature. 

\subsubsection{Comparison with atomic edges}

There is a concern that the broadened photon-particle conversion features may be similar to atomic ionization edges. There are several reasons to believe this may not be a substantial limiting factor in identifying such features: first and foremost, atomic edges have their optical depth, $\tau \propto (\omega/\omega_{\rm th})^{-\delta}$ where $-3<\delta<-2$ and $\omega_{\rm th}$ is the threshold for photo-absorption ($\omega>\omega_{\rm th}$). Therefore, the absorption probability is $\propto e^{-\tau}$ and depends exponentially on the energy. The resulting spectral features are considerably sharper than the photon-particle conversion features predicted here for a powerlaw behavior of the magnetic field and density on distance (see however \S5.1.1). Furthermore, oscillations in the optically thin part of the conversion feature are likely to show a unique oscillatory pattern which is not expected to occur in atomic absorption features nor in continuum emission features (such as black body or powerlaw emission). Lastly, in systems for which the dynamical timescale is short (e.g., of order the observation time span), temporal variability may be observed, depending on the evolution of $B$ and $\rho$ with time. As we shall later show (\S 5) such variability is unique and is not expected to occur for atomic features.

\subsection{Optically Thick Plasma}

To this point we have assumed that once photons are created they evolve according to equations 14, 38. This means that at no point along the photon path there is a "measurement" of the state of the photon-particle system where, by measurement, we mean an interaction which would result in the collapse of the system to one if its measurable states. In  particular, interaction with electrons (whether through Compton scattering or resonance absorption) would cause such an effect. In this case, the calculation must be restarted at each point where photons are scattered or emitted. Clearly, if the mean free path of photons is very short compared to the oscillation length, $\gamma_{\rm osc}$, photon-particle conversion will be significantly suppressed and its spectral features may only be observed at spectral regions where the gas is optically thin - such as line wings. Whether or not this will result in an observable effect of photon-particle conversion depends on the other parameters of the problem (e.g., the value of $g$) and the fraction of photons scattered into our line-of-sight.  

The complete treatment of such a problem requires full radiative transfer  while taking photon-particle oscillations into account at every point along the photon's trajectory. Such a calculation is challenging and sensitively depends on the properties of the gas in the medium which are unknown. We therefore do not treat these effects here but caution that when attempting to identify photon-particle oscillation features it would be best to avoid regions in the spectrum that are densely populated by optically thick atomic features whence simple superposition of the signals may lead to erroneous results. That said, regions which are optically thick to atomic transitions may not be co-spatial with the regions in which photon-particle conversion occurs. Indeed, it seems that for most cases of relevance, the plasma is thought to be highly ionized so that atomic features are scarce. Furthermore, if the predicted features  are broad then line-free zones may be identified. Narrow features in spectral regions scarcely populated with atomic lines may also be fairly easy to identify. 

Another type of resonance absorption of photons will occur in magnetized plasma regardless of its ionization state and is related to the presence of discrete Landau levels characterizing the allowed orbits of e.g., electrons in the plasma. Resonance transition between these levels is allowed and the cross-section could be high. These transitions are accounted for in our treatment of hot plasma. Also, the spectral regions where such an effect takes are narrow and do not generally overlap with spectral bands where photon-particle oscillations take place (see below).

\subsection{Polarization}

To this point we have neglected the effect of optical activity on the conversion probabilities and effectively assumed that the polarization of all the photons is fixed and is parallel to ${\bf e}_\|$. However, more realistically, astrophysical objects will not emit purely polarized radiation and even if so, the propagation of photons through active plasma may depolarize it. Additional processes that could change the photon polarization such as absorption and re-emission or scattering will not be considered here.

We set the system such that a fraction $x_{\|}$ of the photons are polarized along ${\bf e}_\|$ and consider two extreme cases: 1) Optical activity operates on scales much larger than $\gamma_{\rm osc}$,  in which case all the aforementioned probabilities which were derived for the case $x_\|=1$ still hold with the rescaling
\begin{equation}
P_{\gamma \rightarrow a}^{\rm obs}=x_{\|} P_{\gamma \rightarrow a}
\end{equation}
2) Optical activity operates on very short scales compared to the oscillation scale. This case is more difficult to estimate since the effect at different photon energies could be different so that the above equation does not strictly hold. Nevertheless, one expects the overall feature to be shallower by a factor $0.25-1$, depending on whether conversion occurs in the "optically-thin" or "optically-thick" limits. We do not treat the intermediate case where the oscillation length is of order the scale over which optical activity operates. A numerical treatment of the effect of polarization and optical activity will be carried out for specific cases in \S 5.

\section{Spectral Predictions for Astrophysical Objects}

In this section we consider specific astrophysical objects. We calculate more detailed spectral predictions taking into account the specific system in question and consider the particle mass and coupling constant  which may be probed by these systems. 

\subsection{Magnetars}

Magnetars are thought to be stellar remnants in the form of compact ($r_\star \sim 10$\,km; where $r_\star$ is the remnant's radius), highly magnetized ($B\sim 10^{15}$\,G) objects that rotate with a  period of $t_p\lesssim 10$\,sec (e.g., Harding et al. 2005). These objects are identified with soft $\gamma$-ray repeaters and anomalous X-ray pulsars. The physics of the plasma surrounding these compact remnants is believed to be similar to that of pulsars (see below) and follows the rationale laid out by Goldreich \& Julian (1969). In their model, the magnetic field co-rotates with the compact object up to the point where the edges of the field lines reach a rotational velocity equal to the speed of light (referred to as the light cylinder). Within this radius lies the magnetosphere whose name reflects the fact that the magnetic field dictates the plasma properties. In particular, the rapidly rotating field generates an electric field which is able to overcome the gravity of the central object and unbind the plasma which then moves out along magnetic field lines. A steady-state solution for the electron density (also termed the Goldreich-Julian density) is given by
\begin{equation}
n_e^{\rm GJ}=\frac{1}{t_p} \left \vert \frac{B}{ec} \right \vert \sim 7\times 10^{12} \left(  \frac{t_p}{10\,{\rm sec}} \right )^{-1} \frac{B}{10^{15}\,{\rm G}}\,{\rm cm^{-3}},
\label{gj}
\end{equation}
where $t_p$ is the magnetar's rotation period. By charge neutrality, the proton density is identical to the electron density. Nevertheless, in the Goldreich-Julian model, the charges could occupy somewhat separate regions of space so that the total particle density (protons and electrons) at any given volume is in the range $[n_e, 2n_e]$. There are some models suggesting that under certain circumstances, electron-positron pairs are created by the $\gamma$-ray radiation from accelerated electrons which result in a higher plasma density (e.g., Arons 2004 and references therein). We neglect these complications here. We also assume, for simplicity,  that the electrons and protons are well mixed. A case where the plasma density is considerably higher than the value considered here (e.g., Thompson et al. 2002) is treated below.

The photon emission mechanism from magnetars is poorly understood with different models attributing the emission to different parts of the magnetosphere, either close to the stellar surface or the separatix (e.g., Gruzinov 2007). In particular, the polarization of the emitted photon with respect to the magnetic field direction is  unknown and we shall assume the light is polarized in the direction of the magnetic field (non-polarized are considered in \S5.3.1).

We further simplify the problem by assuming azimuthal symmetry so that the plasma density and magnetic field along our line of sight do not change with time. This is clearly an over-simplification since  pulsations are observed indicating that azimuthal symmetry with respect to the observer is not obeyed. We briefly discuss the more general case below.

\subsubsection{Spectral predictions}

\begin{figure}
\plottwo{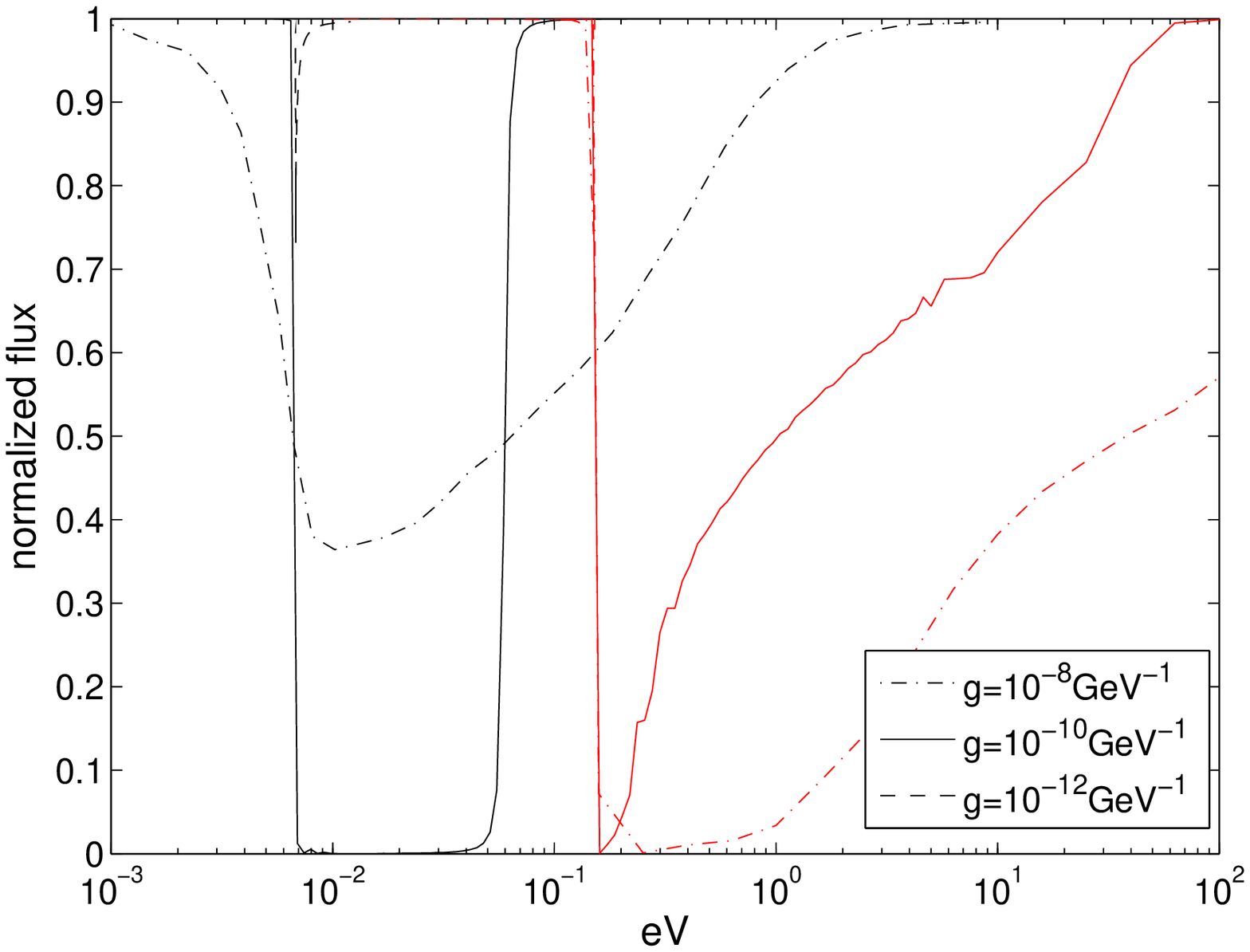}{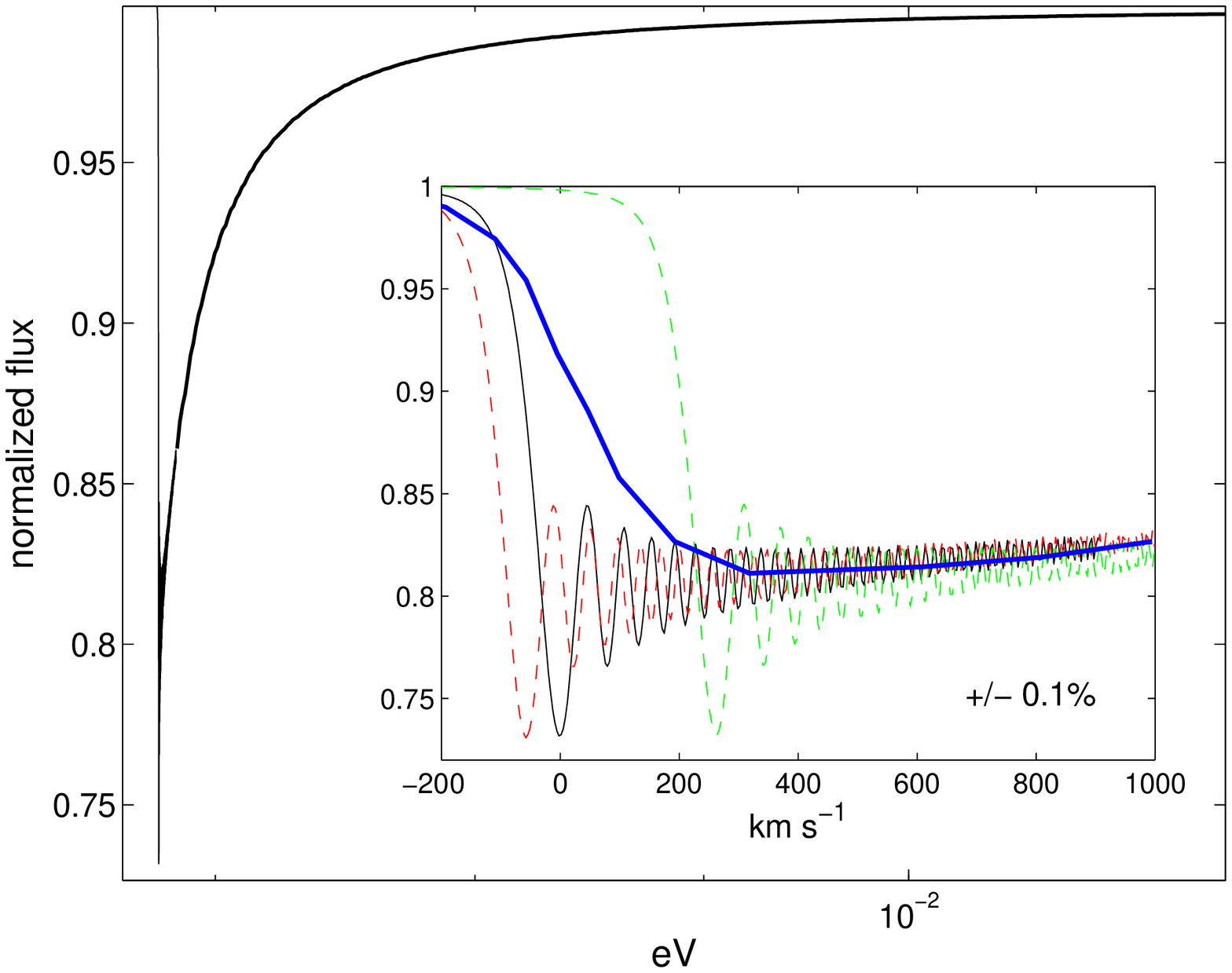}
\caption{{\it Left:} A few examples for the predicted conversion feature from a magnetar whose period is 10\,s with a magnetic field of $10^{15}$\,G, for several values of the coupling constant, $g$ ($m_a=10^{-5}$\,eV is assumed). Note that all features are situated around sub-mm wavelengths ($\sim 10^{-2}$\,eV) but span a large range in photon energy for large $g$ to the point that they resemble broad continuum absorption features. Red curves show similar predictions for the case in which the magnetosphere is $10^3$ denser than the Goldreich-Julian value (se text). {\it Right:} A close up on the feature obtained for $g=10^{-12}\,{\rm GeV}^{-1}$. The blue wing of the feature has an energy dependence which is initially much steeper than that which characterizes atomic edges  (with {\it roughly} $P(\omega)\propto {\rm exp}(-\omega^{-20}) $ compared to the ${\rm exp}(-\omega^{-3})$) but then declines much less rapidly than edges. {\it Inset:} Zooming in on the feature assuming the plasma and magnetic field values are constant in time shows rapid oscillations. Nevertheless, even small ($0.1$\%) variations in the initial values of the problem results in a considerable shift in the spectral pattern (see text). Realistically, a smooth feature may be expected for observations whose timescale is longer than the rotation period of the object (shown qualitatively as a blue line).}
\label{magnetar}
\end{figure}

We show a few examples for the predicted spectral conversion feature in the spectra of magnetars in figure \ref{magnetar}a spanning the more interesting range for the coupling constant which can be constrained by observations ($10^{-12}<g<10^{-8}\,{\rm GeV}^{-1}$). We assume fiducial numbers so that $B(r_\star)=10^{15}$\,G,$r_\star=10$\,km, and the Goldreich \& Julian density (eq. \ref{gj}). We also assume a dipole field so that both the plasma density and the magnetic field are $\propto r^{-3}$. We assume the plasma is quasi-neutral so that both protons and electrons are spatially mixed (see however Goldreich \& Julian 1969). We further assume that pair production is not important. The predicted feature in this case falls around $10^{-2}$\,eV; i.e., in the sub-mm wavelength range. Depending on $g$ the feature can be narrow (small $g$) or broad (large $g$) but can be observed even at modest spectral resolutions. Unlike absorption features where the depth of the feature is always an increasing function of the coupling constant, this is not the general case with conversion features (see above and figure \ref{magnetar}). For $g$ at the threshold of current terrestrial experiments for axion detection (and stellar ages constraints - $g=10^{-10}\,{\rm GeV}^{-1}$), the feature is broad spanning about a decade in energy.  Still larger coupling constants result in broader feature which could extend over several decades of energy. Interestingly, such efficient conversion of photons to axions may result in an object becoming undetectable in a specific band in magnitude-limited samples.  For densities higher than $n_e^{\rm GJ}$, the feature would be shifted to higher energies (see Fig. 10 and below).

It is worthwhile to note that a very efficient conversion is obtained even for very low values of $g$. In particular, using high-resolution, high signal-to-noise (S/N) spectra, coupling constant $<10^{-12}\,{\rm GeV}^{-1}$ may be probed (see next section). This comes about not only because the high conversion rates being $\propto B^2$ but more importantly, due to the dependence of vacuum birefringence term (in the direction parallel to the magnetic field projection, ${\bf e}_\|$) on $B$ which, for fields much stronger than the critical field, is $\propto B$. It can be easily shown that, in this limit $\omega_0 \propto \sqrt{B/\rho}=$const. and therefore a photon can transverse large distances while remaining in the strong resonance energy regime despite the radial variations in both $B$ and $\rho$ (provided $B\propto \rho$, as assumed here). That said, the feature is sensitive to small deviations in the $ \sqrt{B/\rho}$ ratio which still limits the sensitivity to low values of $g$ as compared to the case where constant density and magnetic field intensities are considered (see below).

Evidently, the predicted features are, to some degree, of oscillatory nature. In particular, in the "optically thin" limit these features are more pronounced and their wavelength period is wavelength dependent. Regions in which the plasma frequency is relatively high will result in more rapidly oscillating features in agreement with equation 35. We note, however, that, realistically, two factors limit our ability to detect such oscillations: 1) instrumental resolution and 2) time dependence. The instrumental resolution differs considerably across the electromagnetic spectrum and the detailed shape of the feature requires a convolution of the predicted terms with the instrumental kernel. This is beyond the scope of this work.  Time dependence of the system properties may also affect our ability to resolve individual oscillations in objects  where the observation timeline is of order or longer than the variability timescale since in these cases one integrates over many, potentially different, signals. The variability timescale may be related to the system's dynamical time scale (e.g., of a rotating pulsar magnetosphere; e.g., Lyutikov \& Thompson 2006) or viscous timescales (in the case of quasars).  We have attempted to estimate the magnitude of this effect for magnetars by considering small, $0.1\%$ variations in $B$  and show the predictions for the photon-particle oscillation feature in figure \ref{magnetar}. Clearly, even small changes can shift the feature over a few$\times 100~{\rm km~s^{-1}}$ (larger variations in $B$ or $\rho$ would result in larger shifts; c.f. Eq. 23). As such, it is likely that, realistically,  any finite duration observations of a realistic astrophysical object would result in a smeared feature. The exact shape depends on the exact variations in the properties of the system over time and space and high resolution predictions in this case are limited by our lack of knowledge. Figure \ref{magnetar} shows one possible such feature as would be observed with a high resolution spectrograph. We emphasize, however, that unless the variations in the system properties are of order unity (see below), the overall shape of the feature would be conserved.

\begin{figure}
\center{\includegraphics[width=8cm]{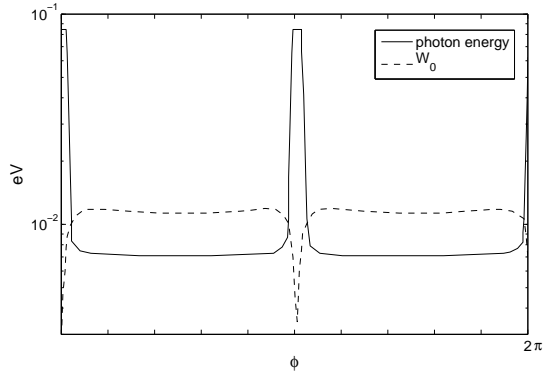}}
\caption{The dependence of the photon energy where conversion is maximal on the orbital phase of the magnetar, $\phi$ (in solid line) shows significant non-linearities and indicates that the conversion feature is highly time-dependent. Similar conclusions apply to the rest equivalent width of the feature, $W_0$ as a function of $\phi$ (dashed line). Variations can be easily detected and used to verify that the spectral features are indeed consistent with being photon-particle oscillation features.}
\label{magnetar2}
\end{figure}

Contrary to our assumption, magnetars are not spherically symmetric. In particular, their dipole fields rotate and are aimed at us only once every few seconds. Hence, one may expect the observed conversion feature to be time-dependent. The exact time-dependence and shape of the feature depend on the alignment of the magnetic field with our line-of-sight and a detailed calculation is beyond the scope of this paper. Nevertheless, assuming, for simplicity, that  $B \propto \rho \propto {\rm sin}(\phi)$ where the phase, $\phi=2\pi t/t_p$ (and that the same radial dependence applies), we calculated the observed features as a function of phase for the case of $g=10^{-11}\,{\rm GeV}^{-1}$ and show those in terms of 1) the photon energy at which maximum conversion occurs and 2) $W_0$ in figure \ref{magnetar2}. Non-linear effect are clearly seen as the shape of both quantities is non-sinusoidal. Both quantities which describe the spectral conversion feature vary with time in a well defined manner and, as such, can be used to verify that candidates for photon-particle conversion features are indeed behaving as expected. In particular, atomic lines and edges  are not expected to show a similar dependence unless very contrived scenarios are invoked.

Figure 12 shows the particle parameter space (spanned by the particle mass and coupling constant) which is covered by spectral observations of magnetars. When constructing the diagram we assume that we have perfect knowledge of the underlying continuum. This is especially important for large values of $g$ where the conversion feature is broad and can span several decades in energy and even be achromatic over a limited energy range. We have assumed that it is possible to detect the chromatic feature down to a 5\% level (this also applies to all diagrams of this sort presented below). As shown, magnetars provide excellent probes for a wide range of particle masses ($m_a<10^{-3}$\,eV) and down to $g\sim 10^{-12}\,{\rm GeV}^{-1}$.  We have also calculated the diagram for the case in which the plasma density is two orders of magnitude higher than the Goldreich \& Julian model (c.f., Lyutikov \& Thompson 2005) and find the results similar within the particle mass range plotted (though there is no drop in sensitivity for $m_a\sim 10^{-3}$\,eV; not shown). 

There have been claims (e.g., Rea et al. 2008 and references therein) that the plasma density in the magnetosphere of megnetars may be considerably higher (by roughly 3 orders of magnitude) than the Goldreich-Julian density due to pair production. In this case the results are similar to those obtained before. In particular, the spectral oscillation signatures are shown in figure 10a (red curves).  The features, in this case, are shifted  to higher, infrared (around $5\mu$m) photon energies and show more extended blue wings. The latter results from the fact that the plasma frequency even in regions far away from the star (where the magnetic field is also smaller) is still larger than the assumed axion mass ($m_a=10^{-5}$\,eV) so that resonance conversion is possible. This  is different than the lower density case considered above where the relatively sharp cutoff in the conversion probability at high photon energies results from the $m_a>\omega_p$ beyond some radius prohibiting resonance conversion.  The broader blue wings in this case for $g\sim 10^{-10}\,{\rm GeV}^{-1}$ may be detectable up to optical and UV energies. Due to the somewhat narrower individual oscillation features obtained for higher plasma densities (Eq. 35), features may be detected down to $g\sim 2\times10^{-12}\,{\rm GeV}^{-1}$ (see Fig. 12). That said, the higher plasma densities in this case allow the detection of more massive axions, up to $m_a\gtrsim 0.01$\,eV. We caution that in cases where $\omega_p \sim m_a$ an extended red-wing may be observed (Eq.  23) which could extend to photon energies $\omega_p$ hence $m_a$. In this case, the formalism outlined above does not hold and the predictions dubious. We do not treat such extreme cases in this work.

For comparison with the case of the Goldreich and Julian magnetosphere model, we show the particle parameter space which can be probed when there is no dependence of the magnetic field or the density on the location in the magnetosphere  ($\alpha=\beta=0$) and their values equal those at the stellar surface (see above). The size of the system is assumed to be the stellar size, i.e., 10\,km. This model is probably less realistic for the following reason: resonance features are narrow (with fractional widths of order $10^{-7}-10^{-6}$) and so the magnetic field  needs to be held at constant values with fractional variations of at most the same order. This is rather difficult to obtain for realistic  systems prone to all sorts of dynamical instabilities. That said, it is interesting to note that for high enough resolution power, one can probe oscillation features down to coupling constants as low as $10^{-15}~{\rm GeV}^{-1}$ which is about 5(!) orders of magnitude lower than can be probed by other means. 

We have not investigated the effects of hot relativistic plasma on the spectral predictions of our model. As we shall show below (\S5.3), these effects are small for the case of magnetars where the expected resonance conversion features lay below the electron cyclotron frequency. The effects of polarization and hot plasma will be discussed in \S5.4 in relation to pulsars.

\begin{figure}
\center \includegraphics[width=8cm]{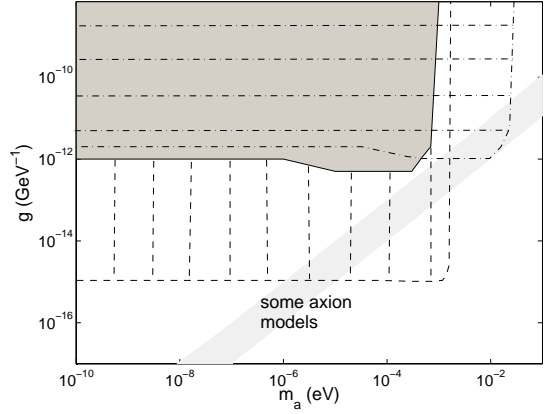}
\caption{The particle parameter range spanned by its mass $m_a$ and its coupling strength to the electromagnetic field, $g$, which can be probed by looking for photon-particle oscillation features in the spectra of magnetars. As shown, over the interesting mass range, magnetars can probe down to $g\sim 10^{-12}\,{\rm GeV}^{-1}$. The sensitivity considerably reduces once $m_a \gtrsim 10^{-3}$\,eV which is greater than the typical plasma frequency of the system (but not for the case in which the magnetosphere density is two orders of magnitude that of the Goldreich \& Julian value; see text). The trend extends to lower values of $m_a$ than shown. Denser magnetospheres ($10^3\times n_e^{\rm GJ}$) allow to probe more massive axions with little sacrifice in $g$-sensitivity (dash-dotted hatched region). We also show the case in which constant conditions exist around the star out to twice the stellar radius (see text). In this case much lower coupling constants may be probed (hatched region). Predictions of some axion models are shown as light grey area.}
\end{figure}

\subsection{Active Galactic Nuclei}

Active galactic nuclei (AGN) are among the brightest objects in the universe spanning a large redshift range. They are thought to be powered by gas accretion onto a supermassive black holes at the centers of galaxies. It seems that magnetic fields play a key role in advecting angular momentum away from the compact region thereby enabling efficient accretion and energy dissipation. In addition, magnetic fields may be crucial for launching the commonly observed gaseous outflows from these objects as well as jets (e.g., K\"onigl \& Kartje 1994 and references therein).  Blandford \& Payne (1982) presented a self-similar solution for hydromagnetic winds from accreting black holes. In their model, a large scale, ordered magnetic field configuration is obtained far from the black hole, and gas which is launched from the accretion disk is flung along magnetic field lines. This general framework was later expanded by  Emmering et al. (1992) and Contopoulos \& Lovelace (1994) who showed that a certain scaling must be obeyed by such self-similar solutions so that the density profile and the magnetic field radial dependence satisfy $\beta=(\alpha-1)/2$. K\"onigl \& Kartje (1994) have shown that a good fit to observations is obtained for $\alpha=-1$ hence $\beta=-1$ which correspond to a "minimum" energy solution (see their paper). In this work we consider those values as representatives of such accreting systems. 

A limitation of the Blandford \& Payne model is the fact that only scaling relations may be deduced while the amplitudes of the magnetic field and density cannot be derived from first principles (c.f. Goldreich \& Julian). To this end, we rely on some general characteristics of bright AGN: 1) the central black-hole mass is of order 
$10^9\,{\rm M_\odot}$  so that the last stable orbit of the accretion disk, $r_\star=10^{15}$\,cm. 2) Our view of the central engine of (type-I) AGN does not suffer from significant absorption hence the plasma in our line-of-sight to $r_\star$ is Compton thin. These two constraints set an upper limit on a volume-filling plasma  of $n(r_\star)\sim 10^9\,{\rm cm^{-3}}$ (while the radial integration of the density profile formally diverges at infinity, in reality all systems are finite and we take an outer radius to satisfy $r_{\rm out}=100r_\star$; this arbitrariness does not affect our spectral predictions for the relevant particle mass range since $\omega_p <m_a$ already on small scales).  We emphasize that the true value of the density in those objects may be much lower than this limit and in what follows we consider a broad range of densities. We neglect all possible contributions from electron-positron pairs to the plasma density  (associated with e.g., jets) and consider only electron-proton plasma.

We note that above the disk corona, the prevailing heating and cooling processes are thought to be mainly radiative hence the gas temperature is unlikely to exceed the Compton temperature which, for the radiation field typical of AGN, is in the range $10^6-10^8$\,K (Krolik et al. 1981).

The magnitude of the magnetic field can be estimated from equipartition arguments so that the magnetic energy and the gas kinetic energy (or the gravitational energy) in the accretion disk  are similar. Here we follow K\"onigl \& Kartje (1994) who estimated the minimum field intensity
\begin{equation}
B_{\rm min}(r_\star)=\left [ \eta^{-1} \frac{L}{c^2} \frac{ v_k(r_\star)}{3r_\star^2}   \right ]^{1/2} \sim 2\times 10^4 \left ( \frac{L}{10^{45}\,{\rm erg~s^{-1}} }\right )^\delta\,{\rm G}
\end{equation}
(c.f., Punsly 1991) where $v_k$ is the Keplerian speed and $\eta\sim 0.1$ is the efficiency by which mass is converted to radiation during the accretion process. There are many uncertainties associated with the black-hole mass--luminosity relation  in AGN (Kaspi et al. 2000) which may impact both the normalization and slope, $\delta$ of this relation. In what follows we take $B(r_\star) = 3 B_{\rm min}(r_\star)$ (c.f., K\"onigl \& Kartje 1994) and $\delta=0.1$ but note that, observationally,  $-0.1<\delta <0.1$ with a considerable scatter (Kaspi et al. 2000).

\subsubsection{Spectral Predictions}

\begin{figure*}
\plotone{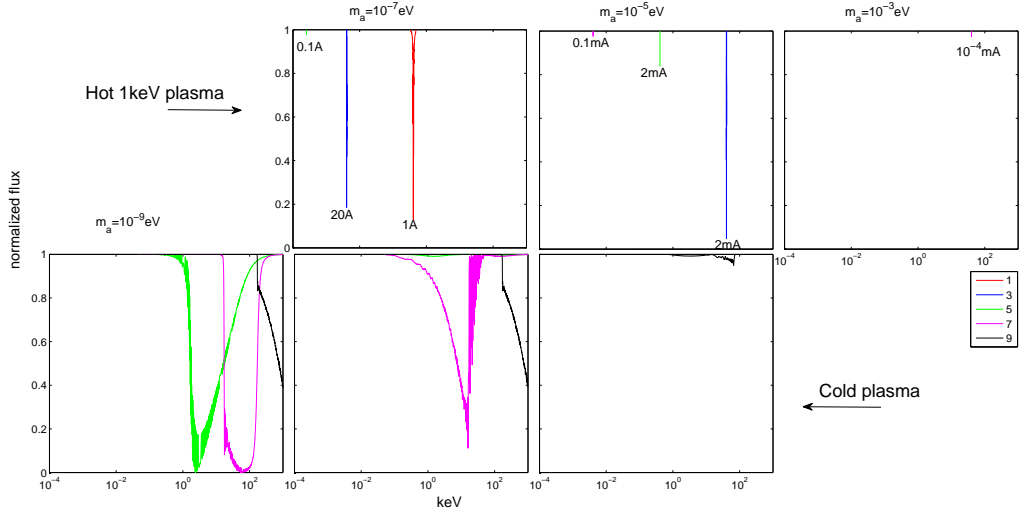}
\caption{The photon-particle spectral oscillation features predicted to occur in quasars for the case of hot, 1\,keV plasma (upper panel) and cold plasma (lower panel) for several particle masses and plasma densities ($\alpha=\beta=-1$ in all cases), and for $g=10^{-12}\,{\rm GeV}^{-1}$. Clearly, the expected features are very different in the case of cold and hot plasma yet in both cases observable features may be observed for the relevant range of particle masses.  For the case of hot plasma, the features are much narrower though generally broader compared to most atomic lines. Rest equivalent predictions are shown next to each narrow feature in the case of hot plasma (this is not shown for cold plasma on accounts of the large width of the features spanning, in some cases, more than one decade of energy). The shapes are diverse and may show red or blue wings. In particular, the red wings result from  $m_a\lesssim \omega_p$ so that $\omega_0$ extends to very low energy values (Eq. 23). For hot plasma, resonance occurs even at low plasma densities due to the higher (negative) refractive index at frequencies above the electron cyclotron frequency (see Fig. 5c).}
\end{figure*}

\begin{figure}
\plottwo{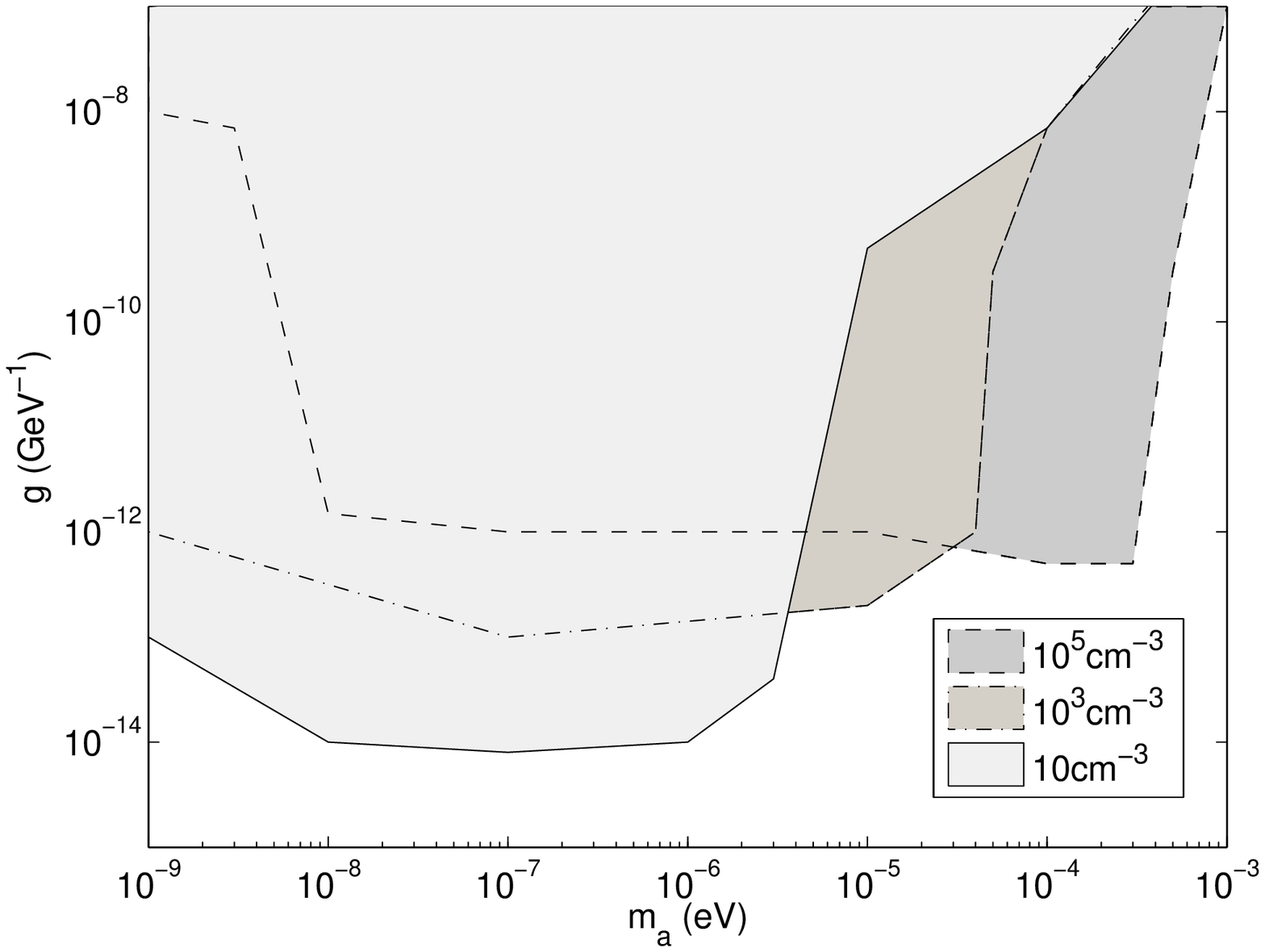}{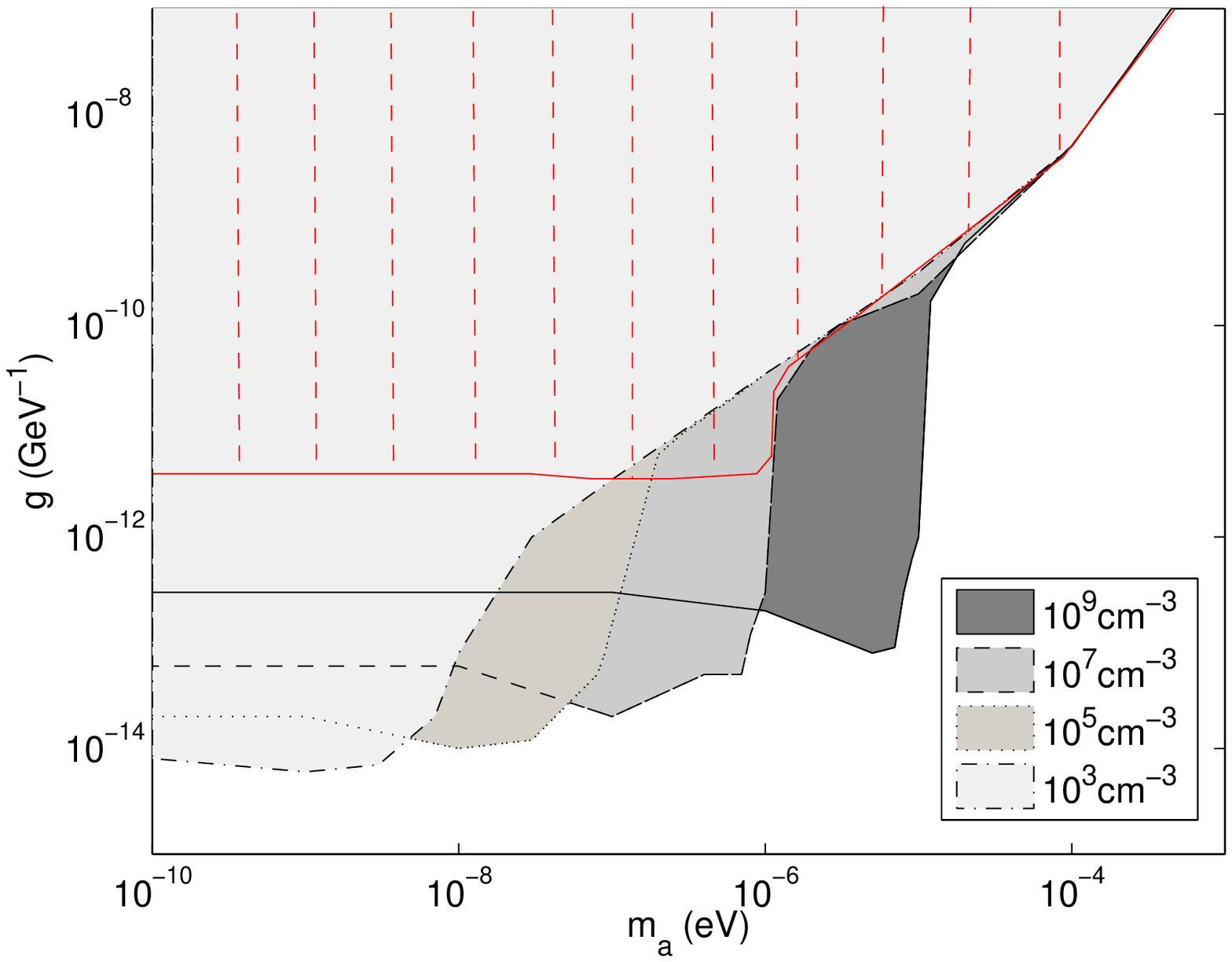}
\caption{The particle parameter space probed by looking for photon-particle conversion features in the spectra of quasars for hot 1\,keV plasma (left) and cold plasma (right) and for several values of the plasma density (see legend). Hot plasma allows to probe more massive particles owing to the higher value of the refractive index at high energies. Both hot and cold plasma models for the quasar plasma allow to probe a broad range of particle masses with $m_a<10^{-4}-10^{-3}$\,eV and down to coupling constants $g\gtrsim 10^{-14}\,{\rm GeV}^{-1}$; i.e., four orders of magnitude lower than those reachable by other current means. A sharp decrease is evident for $\omega_p\sim m_a$ (for cold plasma) or where the spectral feature falls at $>$MeV. At such high photon energies, the effect of pair production - not included in our calculations - becomes an issue.  Evidently, hot plasma allows to probe more massive particles on accounts of its higher (negative) refractive index at the same plasma density. For the case of cold plasma we also show the parameter range which can be probed by studying the spectra of faint AGN (hatched red region; see text).}
\label{rel}
\end{figure}

Figure 13 shows the expected spectral signature of photon-particle oscillation as a function of the particle mass and cold plasma density for a case with $g=10^{-12}\,{\rm GeV}^{-1}$. Clearly, a prominent feature is predicted in most cases. The feature is broad and spans at least one decade of energy. For more massive particles, a prominent oscillation feature is obtained only if the plasma density is high enough. In particular, for the particle mass range most relevant to dark matter physics, an oscillation feature is expected at a few$\times 100$\,keV energies. Overall, the relatively low plasma density of AGN magnetospheres make them less good probes of massive particles yet they are excellent probes for lower mass ones. As shown, prominent conversion features may be obtained for $m_a<10^{-5}$\,eV down to very low values of $g$. The features are broad with a shape which is distinct from atomic features such as lines and edges. That said, the width of the feature requires that the intrinsic continuum shape of the AGN be well known. Some of the spectral features show red wings which result from the fact that the particle mass is of order the plasma frequency. In this case the resonance frequency tends to extend to very low energies since $\lim_{m_a \to \omega_p} \omega_0=0$ (Eq. 23). 

The above spectral predictions do not use any smoothing of the signal (note the rapid variations over small energy scales in Fig. 13) nor does it account for time-dependent physics, as discussed for the magnetar case.  

The plasma temperature around AGN is thought to be weakly-relativistic, reaching up to the Compton temperature of $10^6-10^8$\,K (but could be lower than that limit). We have calculated the photon-particle conversion feature for 1\,keV ($\sim 10^7$\,K) plasma whose density is in the range $10-10^7\,{\rm cm^{-3}}$ and assuming $\theta=45^\circ$ for several particle masses. All other parameters are fixed at their standard values and the predicted spectral oscillation features  are shown in figure 13. The effect of finite plasma temperature on the spectral oscillation features is rather dramatic and results from the fact that  the relevant frequency range over which efficient photon-particle conversion is thought to occur is much higher than the typical electron cyclotron frequency in AGN (of order $10^{-3}$\,eV). In this case, the physics sensitively depends on the shape and value of  $F_{\| \|}(\omega)$. In particular, it is likely that $F_{\| \|}(\omega)$ is an increasing function of $\omega$ (unless $\theta=90^\circ$) which results in the refraction index being higher than in the cold plasma case. This in turn allows for resonance conversion to occur in lower density plasma, even at high photon energies (see Fig. 13) since $F_{\| \|}(\omega)>1$. Comparing the shape of the features to the case of cold plasma, it is evident that the spectral signatures from hot plasma are considerably narrower. This, however, is not a generic feature but results from our assumption that the density follows the magnetic field, i.e., $\alpha=\beta$. In this case the combination $\omega_p\sqrt{F_{\| \|}}/B\propto \omega_0=$const. (note that at high energies, $F_{\| \|} (\omega) \propto \omega$; Fig. 5) hence $\omega_0$=const (Eq. 23). The rest equivalent width, $W_0$ for the narrow features is denoted in figure 13 showing that some of those features may be easily detectable even in low resolution optical and UV spectra while others would probably require X-ray grating observations to be detected. 

Figure 14 shows the parameter space which can be probed by observations of AGN. Clearly, much lower values of the coupling constant may be probed which are inaccessible by other methods. Furthermore, under a wide range of densities and plasma temperatures, quasars are as good probes as magnetars. In particular, bright AGN are more sensitive probes than magnetars when low mass particles are concerned and can reach coupling constants which are $\sim 4$(!) orders of magnitude smaller than those probed by terrestrial experiments and inferred from indirect astrophysical arguments. 

We note that not all AGN are bright and that, in fact, most AGN in the local universe are relatively faint (so called Seyfert galaxies).  In these objects,  both $B(r_\star)$ and $r_\star$ are likely to be smaller than assumed above. In figure 14b  we show the parameter range spanned by faint AGN with a magnetic field of $10^4$\,G and $r_\star=10^{13}$\,cm (these correspond to object whose bolometric luminosity is of order $10^{45}\,{\rm erg~s^{-1}}$). We assume cold plasma with density of $10^7\,{\rm cm}^{-3}$ but note that such objects can, in principal, accommodate much higher densities before becoming Compton-thick. Clearly, the parameter range explored by observations of such objects is somewhat narrower than the case of bright quasars. Nevertheless, they can still probe down to $g\sim 3\times 10^{12}\,{\rm GeV}^{-1}$. Similar particle masses may be probed in this case since the limits are set primarily by the plasma density rather than the magnetic field strength and/or the system size.

\subsection{Pulsars}

The case for pulsars is quite similar to that of magnetars. These too are thought to be stellar remnants and are probably rapidly rotating neutron stars with large (dipolar) magnetic fields. These objects were first discovered in the radio, emitting short period pulses that were later shown  to span a very large energy range (up to X-ray energies in some cases).  These objects have been studied for several decades with more than 100 pulsars known to date. The physics of magnetars (described above) is thought to be analogous to that of pulsars and shall not be repeated here. Typical rotation periods, $t_p$, for pulsars are in the range $10^{-3}-1$\,s and their magnetic fields are lower compared to magnetar fields, $10^9<B<10^{14}$\,G (e.g., Harding et al. 2005). We use the following fiducial numbers in this work: $t_p=1$\,s, $B(r_\star)=3\times10^{12}$\,G, and $r_\star=10$\,km. Here too, a dipole radial dependence is assumed for the magnetic field and that the plasma density, $\rho \propto B$. 

The case of pulsars is qualitatively different than that of magnetars in the sense that the magnetic field is subcritical ($B<B_c$). The Goldreich \& Julian plasma density is expect to be of order  $2\times 10^{11}\,{\rm cm}^{-3}$. It has been suggested that an additional pair plasma component may be present, at least part of the time and at specific viewing angles, yet its properties such as density and temperature are poorly understood. Here we consider two extreme cases and investigate their effect on the resulting spectral photon-particle conversion feature: 1) hadronic plasma consisting of protons and electrons only  and 2) leptonic plasma consisting of electron-positron pairs (note that, in this case, the plasma is inactive). We assume that the density in both cases is that given by Goldreich \& Julian (1971).

\subsubsection{Spectral predictions}

\begin{figure}
\plottwo{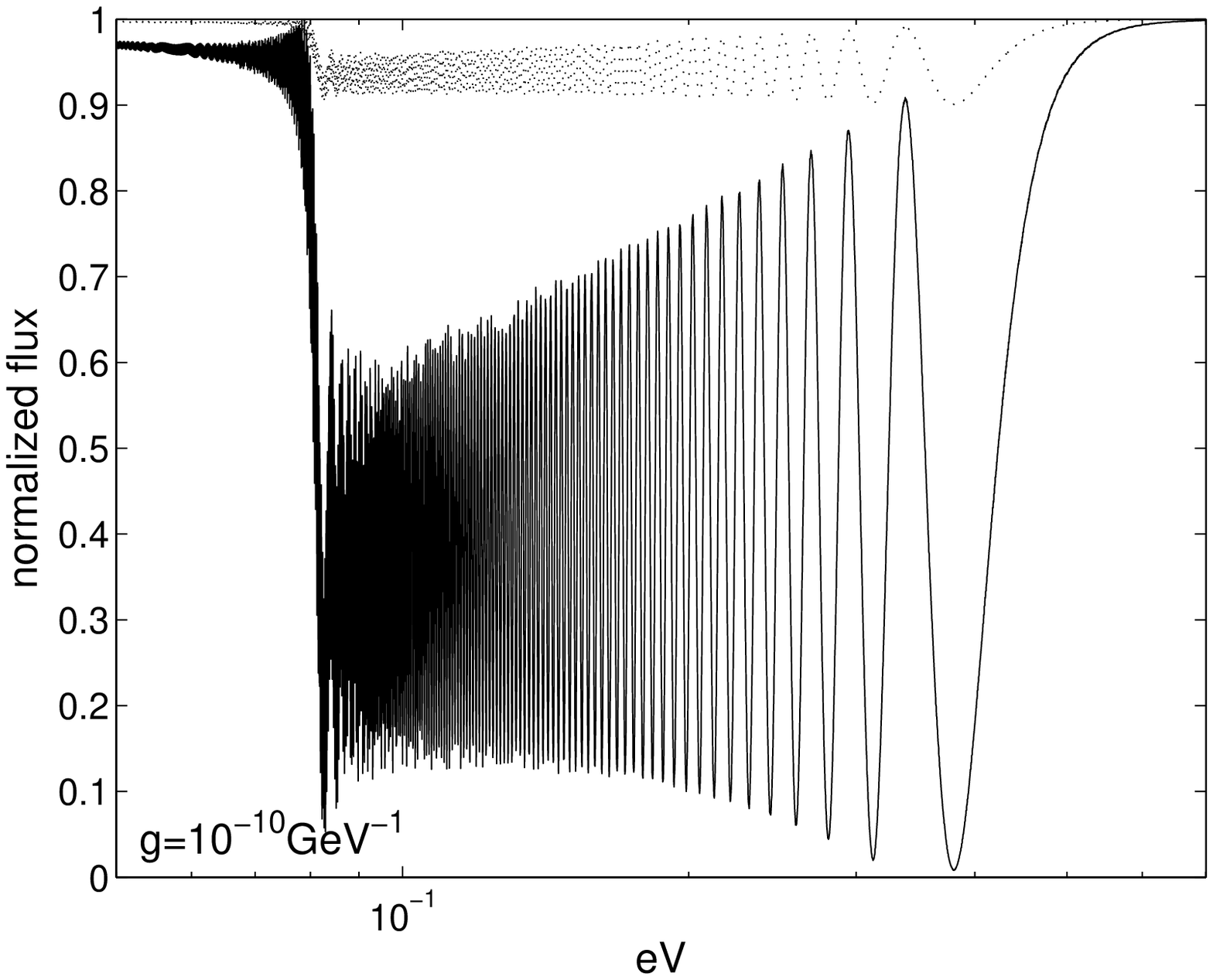}{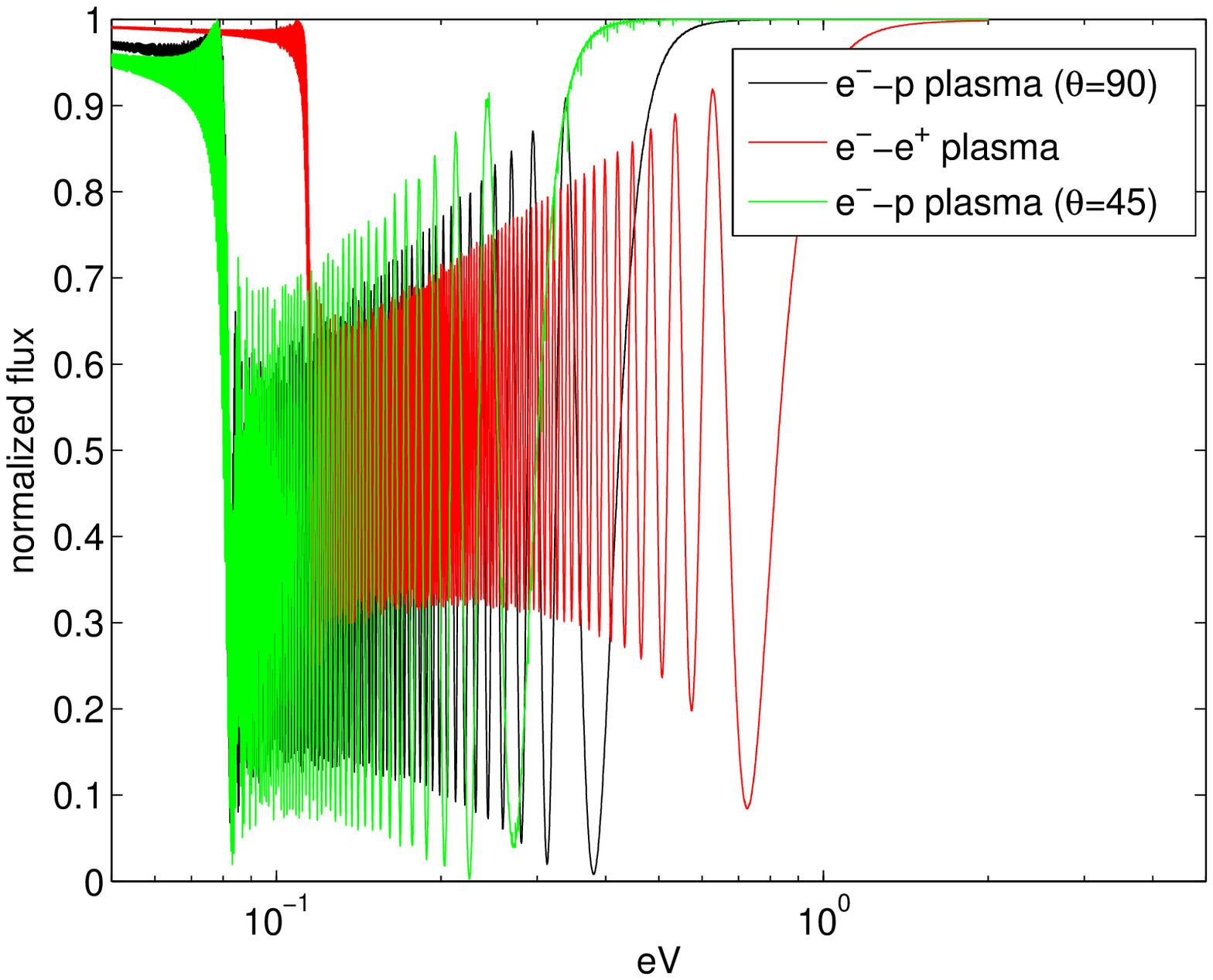}
\caption{{\it Left:} The predicted spectral oscillation feature for the standard pulsar model considered here (solid line). Unlike the case of magnetars, the feature is quite broad spanning almost a decade of energy in the sub-eV range. Also shown is the case for an object emitting non-polarized light. As expected, the feature is shallower on accounts of the smaller photon fraction that is able to undergo oscillations. In both models electron-proton plasma is assumed. {\it Right:} Spectral oscillation features for the standard pulsar case and for two plasma compositions. Electron-positron plasma shifts the feature to higher energies on accounts of the increase value of the refraction index at each energy. The effect of the photon angle with respect to the magnetic field is also shown. The feature (green line) is shifted to lower energies with respect to the standard model (black line) due to the proton-cyclotron line laying at sub-eV energies.}
\label{pulsar}
\end{figure}

The spectral conversion features for our canonical pulsar model are expected to fall around sub-eV energies that is, around 1\,$\mu$m (Eq. 23). The shape of the feature is somewhat different than that obtained for magnetars (at least for small values of the coupling constant, $g$) since, in this case, the vacuum birefringence term (being in the sub-critical magnetic field regime) has a non-trivial dependence on the radial coordinate via the $\sqrt{\rho}/B$ term (see Eq. 23), so that the conversion probability is more uniform as a function of photon energy and the resulting feature is quite broad. This results in a feature being observable for higher values of $g$ compared to the magnetar case. The shape is similar to the relevant canonical shape shown in figure 9, and is shown for the specific case considered here in figure \ref{pulsar}a. 

Until now we have assumed that all photons are created with their polarization vector being parallel to the projection of the magnetic field on the plane perpendicular to the photon propagation direction (i.e., ${\bf e}_\|$; see figure 1). However, the emission mechanism for continuum photons from the pulsar is not known, and a source emitting non-polarized light is a possibility. In figure \ref{pulsar}a we show the predicted signal for non-polarized light which undergoes photon-particle conversion. As expected, the conversion probability is lower  in this case and the feature shallower over the full spectral range. Optical activity has a negligible effect in mixing the photon polarizations in this case, and equation 39 holds. 

In addition to photon polarization, there is the effect of the angle $\theta$ between the propagating photon and the magnetic field. This can have a significant effect in cases where the oscillation feature falls just below (or around) electron/proton cyclotron line frequencies. In this case, the refractive index changes its value rapidly 
over narrow energy intervals and may even attain positive values due to cyclotron lines where resonance oscillation is forbidden (in this case the proton cyclotron frequency is at $\lesssim 2$\,eV). This is shown in figure 15b where the feature is moved to longer wavelengths as a result of this effect.

Plasma composition can also affect the spectral shape of oscillation features as is shown in figure \ref{pulsar}b. Evidently, the difference is mainly in the location of the feature in energy space while the overall shape is conserved. The feature shifts to higher energies for leptonic plasma since the contribution of positrons to the refraction index is much higher than the protons (on accounts of their smaller masses). We note that other processes, excluding composition,  may also lead to shifts in the position of the feature (e.g., changing of the magnetic field strength or plasma density).  Varying the plasma composition may  introduce different cyclotron lines which can significantly affect  the shape of conversion feature laying next to them. As protons have their cyclotron frequency at much lower energies than electrons, this becomes important in the case of pulsars as shown in figure \ref{pulsar}b. To demonstrate this effect we considered the case where the photon propagation direction is with an angle $\theta=45^\circ$ to the direction of the field (see figure 1). Had it not been for the presence of nearby proton cyclotron lines, the result would have been similar. However, the presence of the proton line at energies $< 2$\,eV results in a very different shape for the conversion feature exhibiting a much less pronounced extended blue wing at high energies. The explanation is simple: the plasma refraction term changes rapidly in this energy range spanning a large range of values and thereby allowing resonance conversion over a narrow photon energy range. This effect is less pronounced for $\theta=90^\circ$ since the plasma term is more localized in energy range (see e.g., Fig. 3b).

The effect of plasma temperature on the spectral signature of the photon-particle oscillation process is studied here for the somewhat extreme case (see e.g., Araya \& Harding 1999 and references therein) of a sub-relativistic electron plasma with $T=100$\,keV. Unlike the case of AGN and X-ray binaries (see below), a resonance feature is expected to fall below the electron cyclotron line so that the effect of the finite plasma temperature is negligible (and that of the angle $\theta$ is also small). This is demonstrated in figure 16 where the solution to equation 23 is plotted (for simplicity we assume uniform conditions in the pulsar magnetosphere in this case). Clearly, the resonance at low energies is unaffected by the increased value of the refraction index at higher energies and its shape and location remain unaltered. However, an additional solution appears at high energies where a second resonance exists (a third resonance also exists in this case but is at $>1$\,MeV energies). Such a resonance may be observed, in principal, under favorable conditions. In reality, however, the much higher values of the refractive index results in the feature being rather narrow. 

\begin{figure}
\plottwo{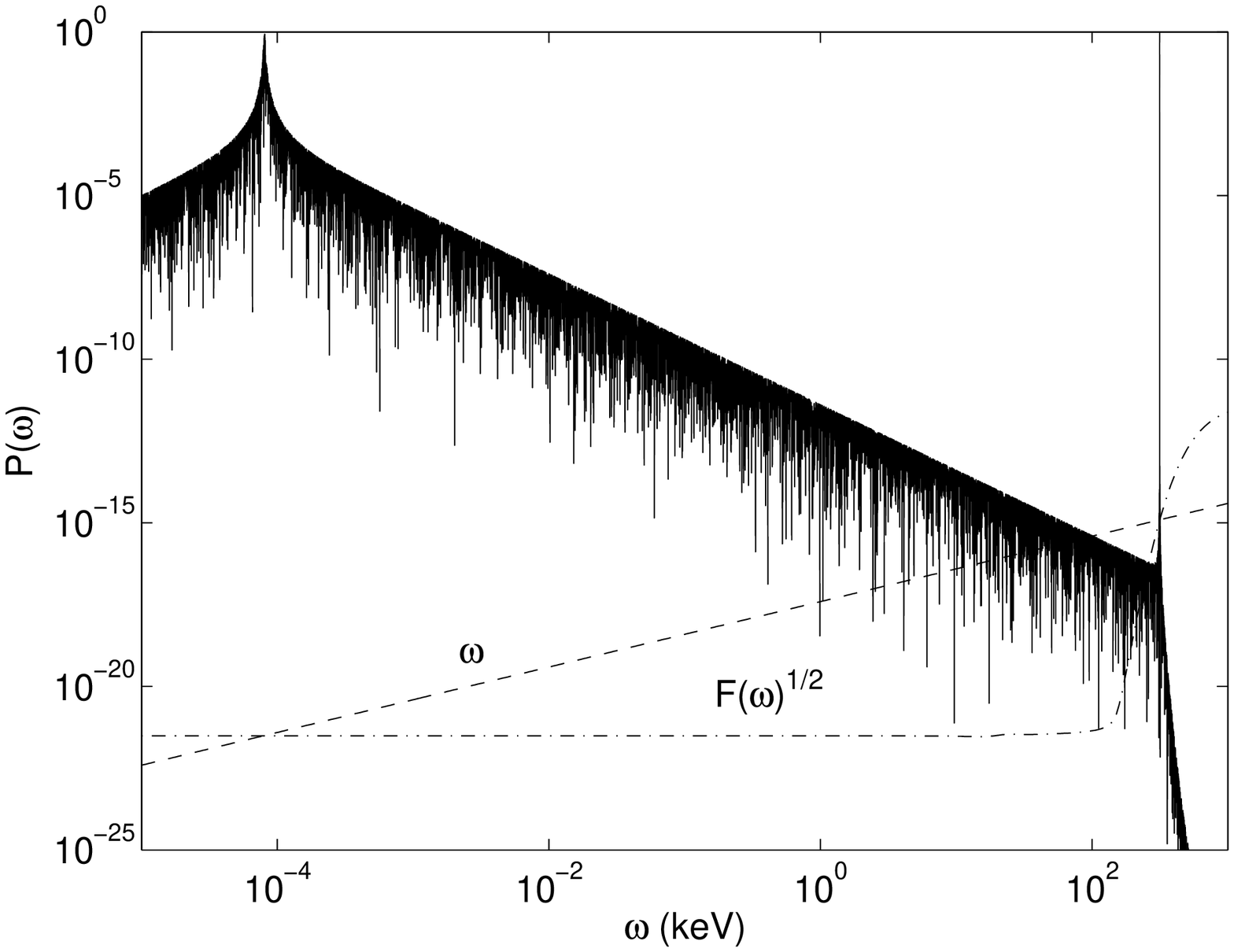}{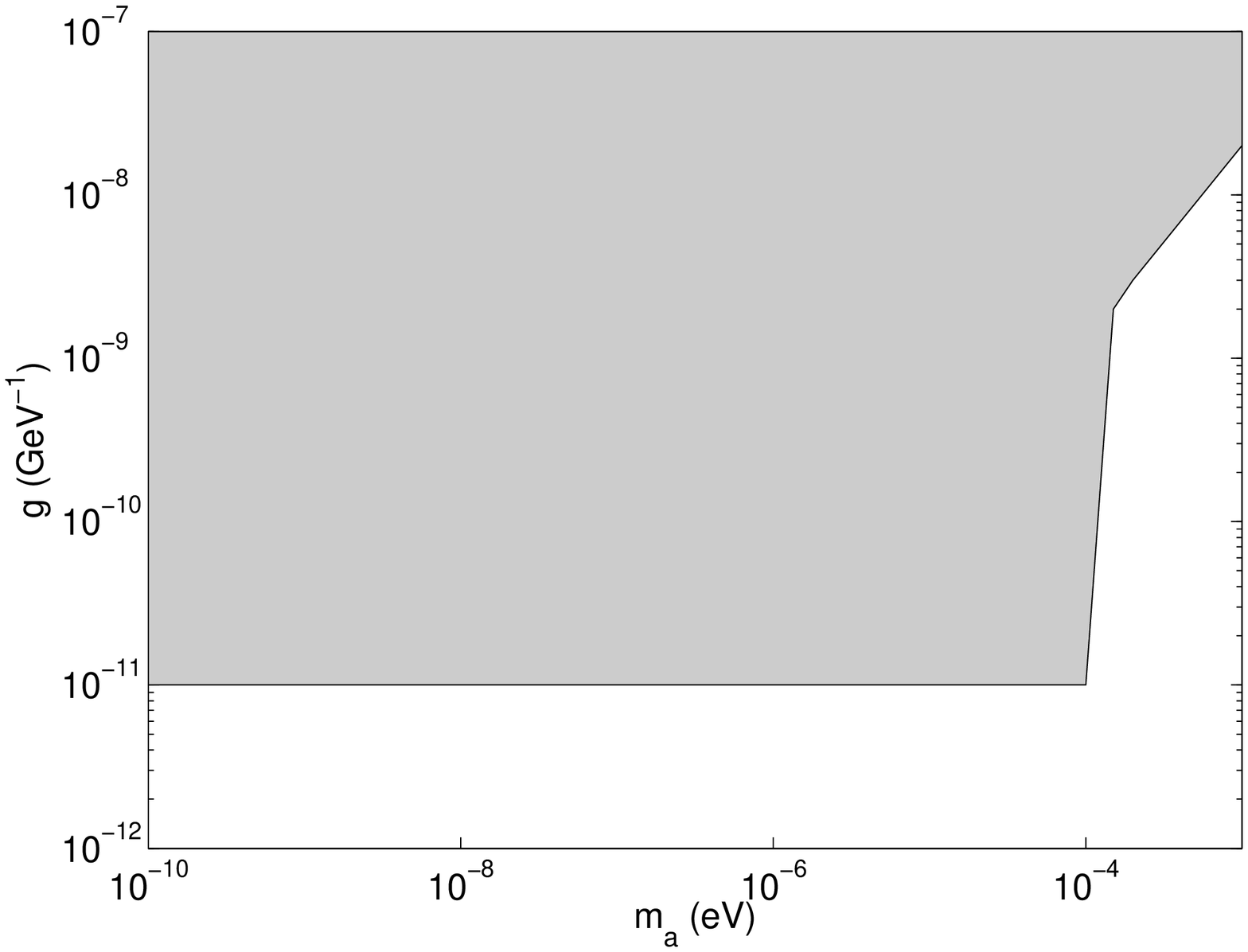}
\caption{{\it Left:} The effect of relativistic plasma on the photon-particle resonance features for a uniform pulsar magnetosphere ($\alpha=\beta=0$). A numerical solution to equation 16 is shown where the location of resonances for the case of 100\,keV plasma occurs at the intersections of the $\omega$ (dashed) line and $\sqrt{F(\omega)}$ (dash-dotted) curve; arbitrary units are used. While the location of the resonance below the electron cyclotron frequency is unchanged, a second solution appears at higher energies. The photon-particle conversion probability for the standard pulsar with $m_a=10^{-5}$ and $g=10^{-10}$.}
\caption{{\it Right:} The parameter space spanned by the particle mass, $m_a$, and coupling constant, $g$, which is probed by pulsar spectra assuming a detection threshold of 5\% with sufficient resolution in the sub-mm band. As shown, photon-particle oscillations can be probed over a large range of particle masses (extending to $m_a<10^{-10}$\,eV; not shown) and down to very low values of $g$. For $m_a\gtrsim 10^{-4}$, $m_a>\omega_p$ and the non-resonance regime begins and the sensitivity decreases sharply for more massive particles.}
\end{figure}

As in the magnetar case, time dependent effects are likely to be important and the observed trends qualitatively similar to those discussed in \S5.1.1.

The phase space covered by pulsar observations is shown in Fig 17. Clearly, pulsars are less sensitive to small values of $g$ on accounts of their lower magnetic fields. In addition, they probe a somewhat more restricted range of particle masses due to their lower plasma densities. That said, pulsars may still probe coupling constants which are an order of magnitude lower than those reached by e.g., terrestrial experiments.

\subsection{X-ray Binaries}

\begin{figure}
\plottwo{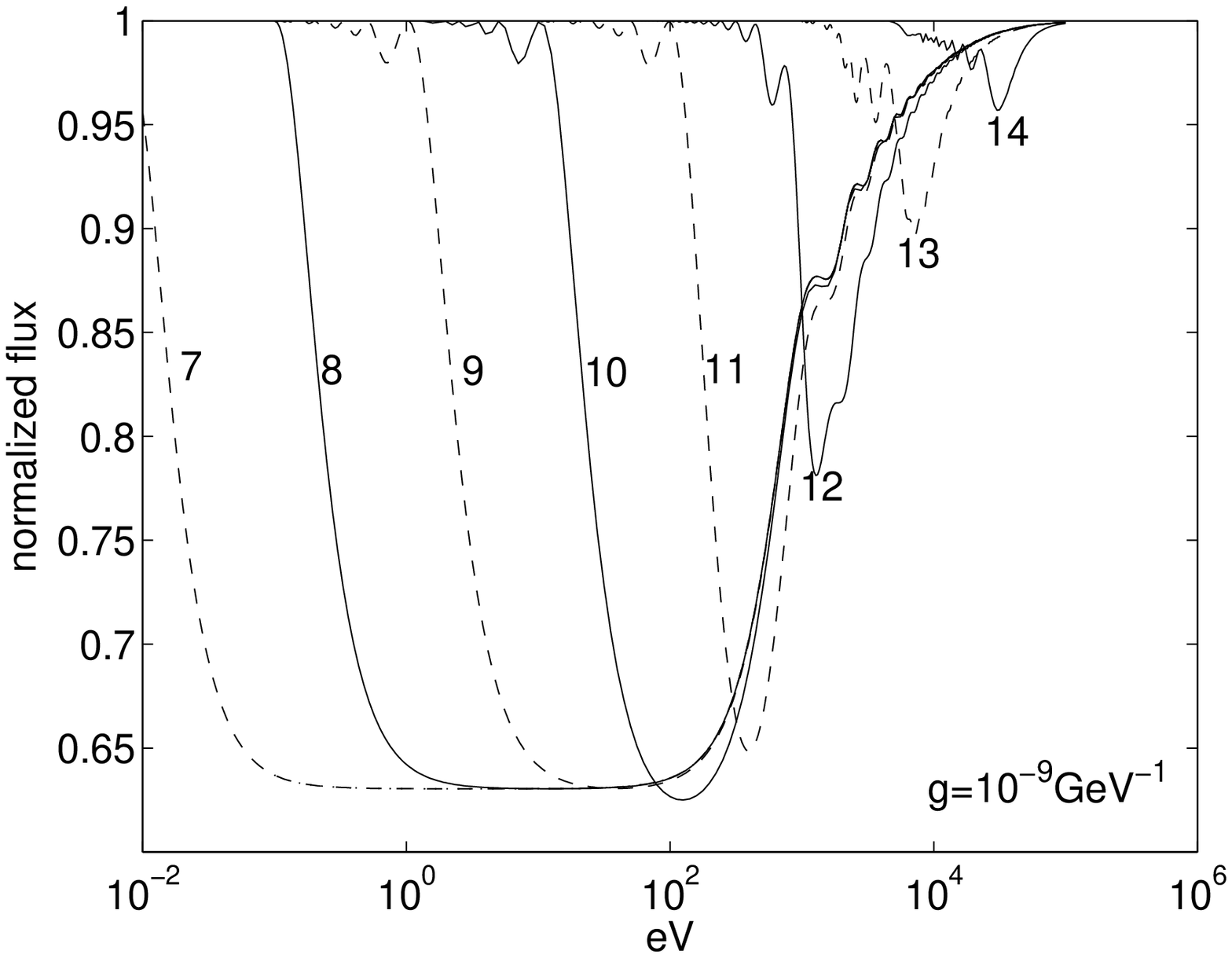}{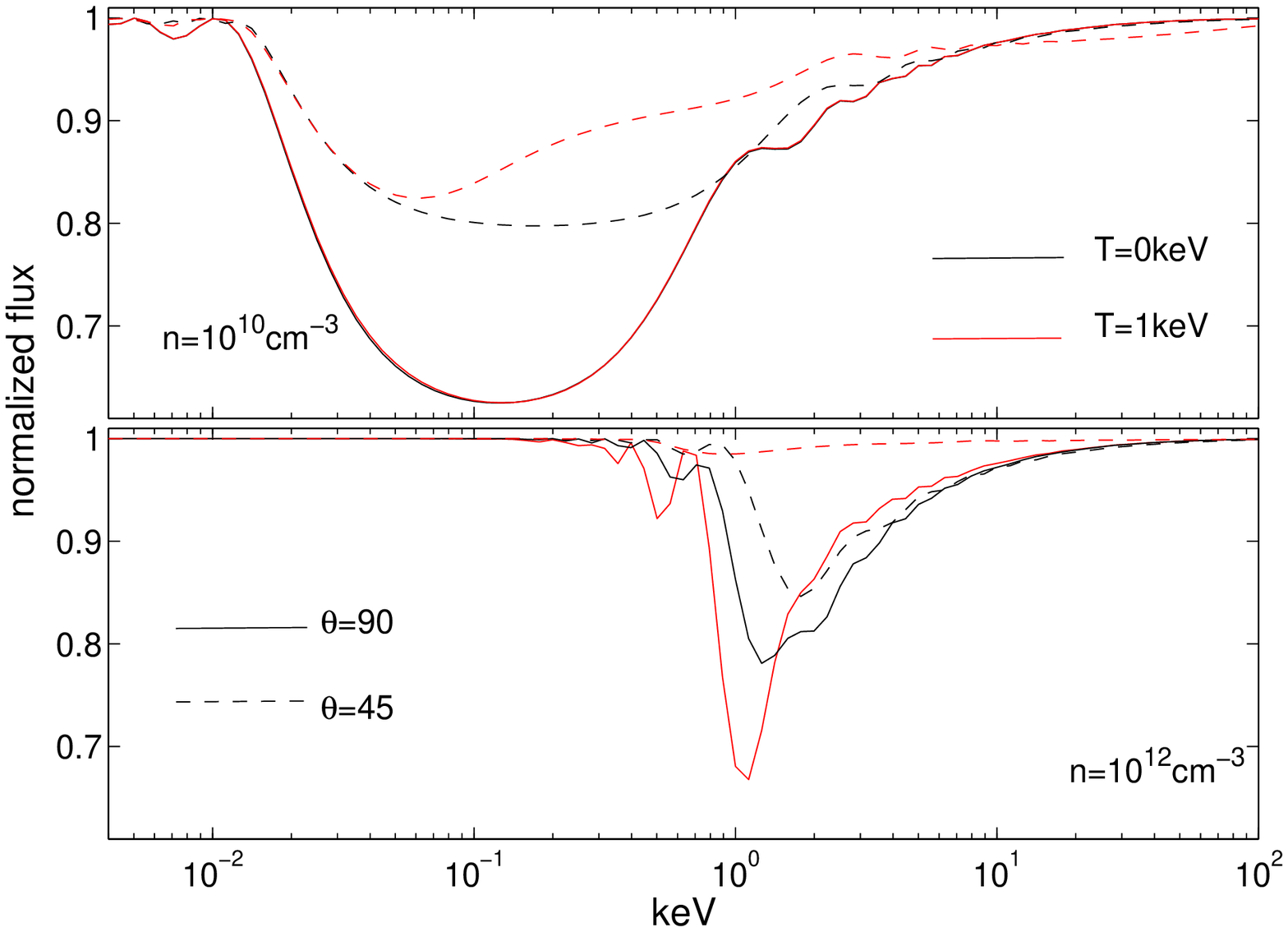}
\caption{{\it Left:} A few examples for the predicted spectral feature from X-ray binaries assuming $m_a=10^{-5}$\,eV and $g=10^{-9}\,{\rm GeV}^{-1}$, and for several values of the density denoted next to each curve. Clearly, an observable feature is expected which may be narrow (in case the particle density is high) or broad (for low plasma densities).  {\it Right:} The spectral predictions assuming cold (black curves) or hot, 1\,keV plasma (red line), and for two values for $\theta$ (see legend). Due to the higher magnetic field and density values compared to the quasar case, the features are not as strongly affected by the plasma temperature as in the quasar case. That said, higher densities are more sensitive to temperature and photon angles since the resonance feature in this case is pushed to higher energies where these effects are more pronounced (see Fig.  5c).}
\label{xrb}
\end{figure}

X-ray binaries are believed to be either a neutron star or a black hole accreting material shed by a less evolved donor star which fills its Roche Lobe and supplies gas to an accretion disk about the compact object. The magnetic field strength is estimated, as for the case of AGN, from equipartition arguments to be of order $10^9$\,G with $r_\star\sim 10^6$\,cm. The plasma density and radial dependence, as for the case of active galactic nuclei is less certain. We assume complete analogy with AGN so that the radial dependence of the plasma and the magnetic field satisfy $\alpha=\beta=-1$ (K\"onigl \& Kartje 1994). The normalization, $n(r_\star)$ is not well known and we have experimented with a broad range of densities. Some of the predicted spectral feature, for a case with  $g=10^{-9}\,{\rm GeV}^{-1}$ and $m_a=10^{-7}$\,eV, are shown in figure \ref{xrb}. Evidently, a broad spectral feature may be observed extended up to 100\,keV for plasma densities $<10^{15}\,{\rm cm^{-3}}$. In particular, for low plasma densities, the feature may span 4 decades in energy, from the sub-mm range and up to soft X-ray energies. For denser plasma the feature shifts beyond MeV energies where other opacity mechanism may become important (e.g., pair production).  Clearly, such broad features may be easily detected even with broad band photometry yet their identification requires  good understanding of the continuum emission mechanisms in those objects. 

As for the case of AGN, the electron cyclotron frequency (here around 10\,eV) is lower than the typical frequency range over which efficient conversion is expected to occur, at least for dense plasma. This results in the resonance feature being quite sensitively dependent on the temperature as well as on the angle between the propagating photon and the magnetic field, $\theta$ (c.f. \S5.2). In figure \ref{xrb}b we show a few examples of this effect comparing the spectral features in cold and hot, 1\,keV, plasma assuming two models for the density (with $n(r_\star)=10^{10},~10^{12}\,{\rm cm^{-3}}$) and two extreme values for the angle $\theta$. As expected, the changes between hot and cold plasma are more pronounced for the denser model. Also, the hotter plasma models are more sensitive to the value of $\theta$ which sets the value of the refraction index above the cyclotron lines. Clearly, there is a rich spectrum of possibilities once various temperatures and photon-magnetic field angles are included.

To conclude, we find that X-ray binaries can be used as probes for pseudo-scalar particles down to coupling constant values that  are comparable to those probed by current terrestrial experiments and inferred from the thermal properties of stars in globular clusters. Typically, the threshold for a few  per-cent detection occurs at $g>10^{-10}\,{\rm GeV^{-1}}$. For this reason, searching the spectra of X-ray binaries for spectral oscillation features is important for corroborating results by CAST as well as other astrophysical constraints, although these objects are unlikely to considerably extend the phase space available to us.

\subsection{Magnetic stars}

While all stars have finite magnetic fields, some stars, known as Ap stars, have intense magnetic fields approaching $10^4$\,G near their surface with approximately dipolar configurations (see however Breithwhite 2008 for possible deviations from this simplified picture), implying considerable magnetic energy stored in the volume around the star. Typical sizes of such stars are typical of ordinary A stars; i.e., of order $10^{11}$\,cm.  
The plasma density around those stars is however less secure: it is known that such stars shed winds extending to large radii yet the mass loss rate is somewhat uncertain. Here we take a fiducial value for the mass loss rate of $10^{-7}\,{\rm M_\odot~yr^{-1}}$ which translates to a mean particle volume density of $\sim 10^{12}\,{\rm cm^{-3}}$ (we neglect the possibility of clumpy plasma in this analysis). We have experimented with several density profiles for which $-3<\beta<-2$ and the results remain qualitatively similar: Ap stars cannot be used to probe photon-particle oscillations if the coupling constant $g<10^{-7}\,{\rm GeV}^{-1}$. (A small fraction of Ap stars exhibits magnetic fields of order $10^5$\,G and allow to probe down to $g\sim 10^{-8}\,{\rm GeV}^{-1}$.) Furthermore, due to the relatively high plasma densities in the stellar wind, the feature is expected to fall in the hard X-ray band, or even in the $\gamma$-ray band, where the flux is negligible. Finite plasma temperatures only shift the feature to even higher energies making the observations  more challenging. Hence, limits on $g$ obtained in this way are relatively uninteresting given the phase space already probed by other means such as  stellar ages and terrestrial experiments. We note that corroborative information may still be valuable yet we choose to not explore this avenue any further in this work. Clearly, non-magnetic stars have much lower magnetic fields resulting in even smaller conversion probabilities hence do not provide any advantage with respect to currently available limits. We note that the limit of $g=10^{-10}\,{\rm GeV}^{-1}$ obtained from cooling timescale considerations for stars in globular clusters are better than the spectroscopic limits since  higher magnetic fields are encountered in the interiors of stars compared to their surfaces.

\subsection{White Dwarfs and Magnetic Cataclysmic Variables}

White dwarfs  (WD) are the final evolutionary phase of most (not too massive) stars. Their  size is of order $10^9$\,cm  and most objects have magnetic field intensities  in the range $10^3-10^6$\,G (e.g., Putney 1999  and references therein). A sub-population of white dwarfs, called magnetic white dwarfs (mWD), has field intensities as high as $10^6-10^8$\,G (e.g., Kemp et al. 1970) and are thought to evolve from Ap stars. Similar to Ap stars, the field configuration in those objects  is thought to be dipolar.  Magnetic cataclysmic variables, (mCV),  are systems in which the magnetic white dwarf is thought to be a part of a binary system, and also exhibit magnetic fields of order $10^8$\,G (e.g., Tapia 1977). The density of the plasma in the magnetospheres of these objects and its radial dependence are poorly known and we shall assume, as before, that the plasma follows the magnetic field.

If mCVs and mWDs are descendent of Ap stars then they provide, in principal (assuming $\omega_p>m_a$), better probes than Ap stars since the product of the magnetic field and system size is larger (see Eq. 33). In particular, assuming  $B=10^8$\,G, $r_\star=10^9$\,cm, $\alpha=\beta=-3$, and a particle density of $10^{12}\,{\rm cm^{-3}}$ at $r_\star$ (as for Ap stars), then a conversion feature may be observed down to $g\sim 10^{-10}\,{\rm GeV}^{-1}$ at hard X-ray energies (assuming $m_a\sim 10^{-5}$\,eV and cold plasma). For much lower plasma densities of $\sim 10^8\,{\rm cm^{-3}}$  at $r_\star$, low mass particles may be detected down to $g\sim 3\times 10^{-11}\,{\rm GeV}^{-1}$ in the soft X-ray band (around 0.1\,keV energies). UV observations may place interesting limits on light axions ($m_a<10^{-6}$\,eV)  if the plasma densities in the objects' magnetospheres is still lower, of order $10^6\,{\rm cm}^{-3}$ at $r_\star$. As in the X-ray binary case, the interesting resonance features are likely to fall above the electron cyclotron frequencies making the predictions rather sensitive to the (unknown) plasma temperature. We do not consider this specific case here. To conclude, this class of objects may, under favorable conditions, extend the particle phase space accessible to us down to $g\sim 3\times 10^{-11}\,{\rm GeV}^{-1}$. In cases where $\theta \ll 90^\circ$ and/or polarization measurements are impossible, $g$-limits may be even higher and so the parameter space explored comparable to that accessible to CAST.

\subsection{Young Stellar Objects}

Young stellar objects (YSOs) such as T-Tauri stars are stellar objects in the making. They are characterized by a protostar whose surface magnetic field is of order $10^3$\,G and whose magnitude drops with distance similar to or faster than a dipole (i.e., $r^{-3}$). The central object is enshrouded by gas which accretes onto it and whose density is roughly of order $10^{14}\,{\rm cm^{-3}}$ with a $r^{-5/2}$ radial dependence (K\"onigl 1991). The relatively low magnetic fields augmented by the small sizes of the system result in such objects being relatively poor probes for photon-particle oscillations. In particular, for the parameter space explored in this work, we obtain that photon conversion may be observed only for relatively large values of the coupling constant ($g> 10^{-8}$) and even that only at very high energies (unless the density is considerable lower than assumed). Given that better constraints are obtained from stellar ages and terrestrial experiments, there is little motivation to further explore the spectral predictions for photon-particle oscillations in this work.

\section{Discussion}

\begin{figure*}
\plotone{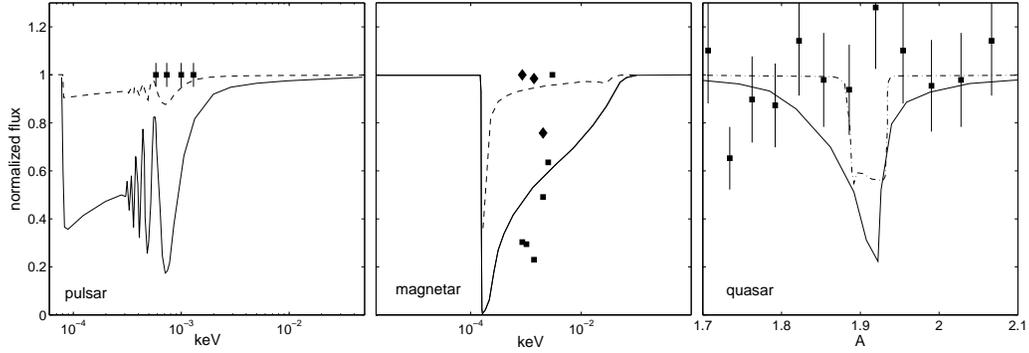}
\caption{Spectral oscillation features for $m_a=4\times 10^{-6}$\,eV particles with $g=10^{-10}\,{\rm GeV}^{-1}$ for pulsars (left panel), magnetars (middle panel) and quasars (right panel). Over-plotted are the slope-normalized data for pulsars (Sollerman 2003) and for quasars (Mkn\,279; Yaqoob \& Padamabhan 2004). For magnetars we show the normalized and de-reddened broad band photometric data  for two objects (circles and diamonds) taken from Durant \& van Kerkwijk (2005; error bars are small and cannot be seen and we have not included the marginal $B$-band data). Also shown in dashed line is a model for photon-particle conversion in pulsars/magnetars with $g=3\times 10^{-11}\,{\rm GeV}^{-1}$. The dot-dashed line is a model for photon-particle conversion in quasars with $g=5\times 10^{-12}\,{\rm GeV}^{-1}$. Note that different x-axis scale in each panel.}
\label{g10}
\end{figure*}

We have given detailed predictions for the spectral signatures of photon-particle oscillations in astrophysical objects. While these features appear as absorption features in the spectrum of an object, the physics behind them is {\it very} different: photons are not a well determined state in highly compact and magnetized environments filled with dilute plasma and a more general definition is that of a photon-particle state where, under conditions prevailing in some objects, could manifest itself as a particle state showing an effective deficit of photons at particular energies across the electromagnetic spectrum. 

We have demonstrated that spectroscopic observations of magnetars, pulsars, and quasars can be 2-4 orders of magnitude more sensitive to small values of the coupling constant between particles and the electromagnetic field leading to detectable photon-particle oscillations in cases where little or no signal is expected for current terrestrial experiments. This approach is also much more sensitive than indirect astrophysical constraints, such as cooling of old stars in globular clusters, providing limits on the coupling constant, $g$. In addition, the method presented here, probes particles in a cosmologically relevant mass range where they serve as viable dark matter candidates, and where other constraints (e.g., those related to optical diffuse light or the closure of the universe) are limited. 

Unlike terrestrial experiments whose set up may be controlled and their results checked for consistency (for which the PVLAS experiment is an excellent example), astrophysical sources cannot be controlled and their parameters cannot be tweaked by an observer. Moreover, by using compact objects as effective laboratories, the experimental setup is only qualitatively understood. Nevertheless, several lines of reasoning suggest that despite these uncertainties and limitations, the approach explored here may prove very useful: first, the parameter space range probed by different astrophysical environments overlap providing further corroboration for the findings. In addition, the increased sensitivity is not by a factor of a few but rather by several orders of magnitude compared to current methods, hence less susceptible to the details of the system in question (clearly, if our understanding is not even in the order of magnitude range than considerable deviations from the above predictions may occur).

It is {\it not} the purpose of this paper to undertake an exhaustive search of the parameter space in each class of objects by fitting a large range of spectral models to observations (while marginalizing over model uncertainties in each case and taking into account possible time-dependence of the spectral features). We do, however, wish to demonstrate the feasibility of this approach by showing spectral predictions for each class of objects using the current lowest limits on the coupling constant from CAST and indirect astrophysical arguments of $g=10^{-10}\,{\rm GeV}^{-1}$. To this end, we shall assume the aforementioned fiducial parameters for each class of systems considered above, and check the observational implications for the case of $m_a=4\times 10^{-6}$\,eV which remains  in the unexplored mass range of the microwave galactic haloscope experiments (whose limits are meaningful only if axions are numerous enough so that they constitute most of the dark matter in our galaxy). We emphasize that one should not draw any conclusions on the presence or absence of photon-particle oscillation features from the following demonstration of feasibility as only a very specific model is considered and we do not marginalize over model uncertainties. 

Clearly, the spectral features are pronounced (see Fig. \ref{g10}) and may be clearly detected even at low resolution and modest signal-to-noise conditions (under our set of assumptions concerning the physical properties of the objects).  A straightforward constraint on the axion properties can from pulsars where a very broad spectral feature is predicted yet is not seen in the data.  Magnetars can, in principal, provide similar constraints given if the densities in their magnetosphere is higher than the Goldreich-Julian value by several orders of magnitude. In this case, the broad features may extend to optical and UV energies (Fig. 19) were data for a few objects are available. Nevertheless, our current understanding of the various emission mechanisms contributing to the emission in these wavebands is at its infancy and different magnetars seem to have very different spectral behaviors (compare the two data sets in Fig. 10). These issues are likely to pose considerable difficulties when interpreting the spectra and attempting to draw robust conclusions of any kind.  At face value, the spectral energy distribution of both magnetars shown is inconsistent with the specific oscillation feature considered here.

For quasar, a broad X-ray feature is predicted yet is not seen in the data (Fig. 10). Interestingly, the oscillation feature, in this case, lies in the part of the spectrum close to the iron $K\alpha$ line and a more detailed analysis including the effect of atomic features is in order. This is, however, beyond the scope of this paper. If the lack of discernible features in the spectrum is to be taken seriously, then, given the current quality of the data and given our restrictive set of model assumptions, a tentative limit (not marginalizing over model uncertainties!) on the coupling constant of $g<3\times 10^{-11}\,{\rm GeV}^{-1}~(<5\times 10^{-12}\,{\rm GeV}^{-1})$ may be obtained for pulsars and quasars, respectively.  

We emphasize that these observations were not conducted to maximize the efficiency for the detection of photon-particle oscillations in those objects and that, in principal, much better data and analysis are required to reach meaningful limits. Specifically, high quality and high resolution X-ray data for quasars as well better understanding of the infrared to optical spectral energy distribution (via photometry and spectra) of magnetars may yield considerably better  limits in this case. We re-emphasize that the above limit on $g$ is given here {\it only} as a proof-of-concept and applies {\it only} within our restrictive set of assumptions concerning the physics of the relevant astrophysical objects. 

Thus far we have considered pseudo-scalar particles such as the axion. The case of scalar particles is completely analogous to the one considered here with the interchange of ${\bf e}_\|$ and ${\bf e}_\bot$. By symmetry, all the predictions given here remain valid with the proper transformation. Naturally, the limits which can be obtained on such a class of particles are identical to the case of pseudo-scalar particles.

\begin{figure*}
\plotone{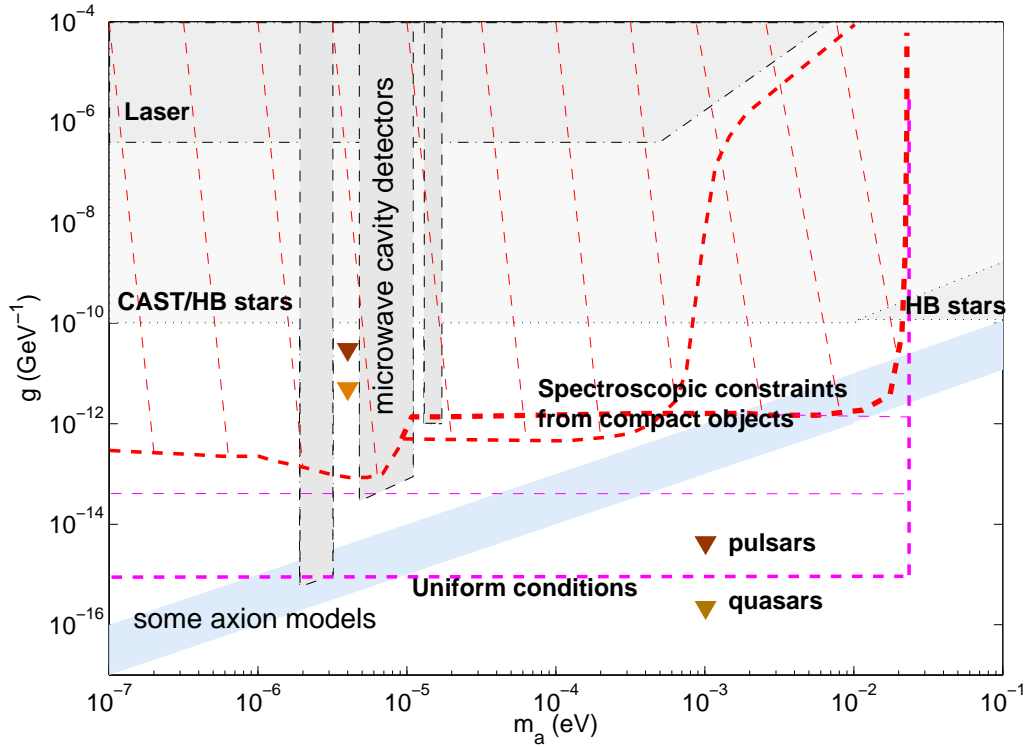}
\caption{The particle parameter space (spanned by mass and coupling constant) which is probed by the spectroscopic constraints discussed in this work (hatched red surfaces whose difference is related to the assumptions concerning the density of the magnetosphere in magnetars; see \S5.1.1) as compared to other currently used methods such as laser experiments, microwave experiments,  solar axion telescopes, and indirect astrophysical considerations. Also shown (hatched magenta region) is the range probed by compact objects under the assumption of uniform conditions (see text). Clearly, the method described here can {\it directly} probe a considerably larger parameter range than is accessible by other methods. The proof-of-concept limits obtained for quasars and pulsars are also shown (for the case of cold plasma; see text).}
\label{total}
\end{figure*}

The higher sensitivity (assuming 5\% detection threshold) of compact astrophysical objects for probing photon-particle oscillations over an interesting range of particle masses is summarized in figure \ref{total} and is compared to the regions that can now be probed by other means (CAST, microwave resonance haloscopes, and laser experiments). Also shown is the sensitivity assuming uniform magnetic field and density conditions over a length scale $r_\star$ across. Overall, significantly larger phase space may be probed by studying the spectra of compact objects which is unreachable by laboratory means. As such, the approach proposed here may allow us to {\it directly} detect the long sought axion (and/or scalar particles) which provide perhaps the best solution to the strong-CP problem and may also solve the dark matter mystery. It might even be possible to probe the phase space relevant to some axion models.  We have shown that, given current limits on $g$, oscillation features in some objects may be rather broad and could, in principal, be detected by broad band photometry. Reaching down to low values of $g$ may require medium and high resolution spectroscopic observations of compact objects.

Apart from particle physics, there are also implications for the physics of compact objects: should light (pseudo-)scalar particles be detected via the route suggested here or by some other means, the spectral signature (or lack of) would teach us about the physics of inner engines of these objects. In particular, it may teach us about the large scale magnetic field configuration in those objects, the plasma properties in such extreme environments, as well as about their emission mechanism. It may also shed light on their time-dependent properties and could provide invaluable information on their evolution. This provides an additional link between particle physics and astrophysics.

\subsection{Q\&As}

Here we attempt to summarize a few of the topics related to the identification and interpretation of photon-particle oscillation features:
\begin{itemize}
\item
{\bf How can one tell photon-particle oscillation features from other spectral features?}  By their shape and variability. Calculations presented here show that the shapes of photon-particle oscillation features may be as narrow as atomic lines or broad as atomic edges or even continuum features. Nevertheless, the shape is, generally, very different from those produced by atomic physics. Nevertheless, secure identification requires good knowledge of the emission and absorption mechanism in the particular object probed. For systems with short variability timescales, distinct energy shifts and shape changes of the feature may be observed. Such variations would be set by the time-dependence of the system properties and are not expected to naturally occur for atomic features.
\item
{\bf What determines the properties of photon-particle spectral features?}
Unlike atomic lines and edges, the shape of photon-particle conversion features depends rather weakly on the plasma temperatures for highly magnetized objects (but not for weakly magnetized systems such as quasars). The width of the feature is primarily determined by the stratification of the magnetic field and plasma density in the astrophysical object probed such that more stratified objects are likely to show broader spectral features. The depth of the feature (or its rest equivalent width, $W_0$) depends on the polarization of the emitted radiation with polarized sources potentially showing more extreme cases of conversion (from no conversion at all to conversion of almost 100\% of the radiation leading to zero flux at certain frequencies). There could be dramatic temperature-induced variations in the properties of the feature  in objects where resonance conversion occurs above the electron cyclotron frequency and when the plasma is hot. An oscillatory pattern extending to higher photon energies can be observed in many cases when high resolution spectroscopy (and short time-frames in varying objects) is used.
\item
{\bf Which wavelengths should be searched?} Our calculations show that the feature may be observed over a very broad energy range. Generally, the relevant wavelength range is determined by the plasma density and magnetic field intensity. For more magnetized objects, the feature will be shifted to longer wavelengths where the refractive index is higher. We emphasize that several spectral features may be observed depending mainly on the composition and temperature of the plasma. In particular, while relatively broad features are expected off-cyclotron resonances, additional narrower features may overlap with electron and proton cyclotron lines. Generally, no features will be observed at photon energies below the plasma frequency where the object is optically thick. Care should be taken when attempting to identify oscillations in regions occupied by atomic features such as lines and edges.  Note that, unlike atomic features, the oscillation feature will shift in energy in objects whose magnetic field strength and density vary with time. This may prove crucial for correctly identifying and resolving oscillation features in compact objects.
\item
{\bf What kind of observations are needed?} This depends on how low a coupling constant, $g$, one wishes to be able to probe and how rapidly varying the astrophysical object probed is. We have shown that, for $g$-values of order the CAST limits, broad band photometry of magnetars and pulsars and high signal-to-noise non-grating X-ray spectra could suffice. Nevertheless, to be able to search for particles (axions) in a previously unexplored regions of the parameter space down to low $g$-values, medium to high resolution spectroscopy may be  required. This is particularly true for the case of magnetars in which a narrow feature is predicted, and for quasars, for which the feature is expected to fall in the (hard) X-ray band. As the X-ray ($<$10\,keV) spectrum of low luminosity quasars shows various lines and edges (e.g., Chelouche \& Netzer 2005), high resolution spectra are of paramount importance. As noted before, oscillation features may shift with time, especially in compact objects whose relevant timescales (e.g., dynamical ones) are short, and so multi-epoch short-exposure observations or time-tagging of individual photons may prove helpful in correctly identifying them and characterizing their shapes and positions.
\item
{\bf Can the particle parameters be directly inferred?} Unfortunately, this task is somewhat complicated since, unless one has a good understanding of the object in which efficient photon-particle conversion takes place, degeneracy between the particle parameters and the object parameters is a limitation. Some of the degeneracies may be alleviated if good S/N data are available in which case fitting of model templates to the data may reveal unique features (such as extended broad wings, oscillatory pattern).  It is particularly difficult to constrain the properties of  light particles  whose mass is considerably smaller than the plasma frequency. These very important issues, of extracting reliable particle properties (or limits to them) while marginalizing over model uncertainties, is beyond the scope of this introductory work and will be dealt with in a forthcoming publication.
\item
{\bf What can we learn about compact objects?} In the event that particles, such as axions, are detected by other means (e.g., using terrestrial experiments or in the spectra of several types of objects), then the presence or absence of an oscillation feature may teach us about the interiors of compact objects. As discussed here, the observed properties of photon-particle conversion features are sensitive to the value of the magnetic field, its spatial configuration (hence the effective size of the system), the photon propagation direction and polarization, and to the plasma properties. Therefore, studying those features in the spectra of celestial objects may shed light on the properties of astrophysical magnetospheres and enhance our understanding of such systems. 
\end{itemize}

\section{Summary}

We construct a general formalism to calculate the spectral signatures of photon-particle oscillations in several classes of astrophysical objects. We include QED effects as well as processes characterizing the interaction of photons with active relativistic plasma. We focus on the observational manifestation of the photon-particle oscillation phenomenon and emphasize the differences between the spectral features associated with it and the more familiar atomic features such as line, edges, and continuum emission features. We give detailed spectral predictions for several types of compact objects, among which are magnetars, pulsars, and quasars. We demonstrate the the photon-particle oscillations features may occur over a very broad spectral range, from sub-mm wavelengths and up to $\gamma$-ray energies, depending on the properties of the compact object as well as the particle mass. We characterize the shapes and strengths of the expected spectral features using spectroscopic nomenclature and argue that such features can be detected using low to medium resolution spectra if the coupling constant of the (pseudo-)scalar particles to the electromagnetic field is $>10^{-13}\,{\rm GeV}^{-1}$ and their mass is $<10^{-2}$\,eV (assuming our understanding of compact astrophysical objects is qualitatively correct). Such detection limits in terms of the coupling constant are about three orders of magnitude better than those achievable by current terrestrial experiments and indirectly deduced using additional astrophysical arguments. We conclude that, by studying the spectra of magnetars and quasars, one can increase current sensitivities for axion detection by several orders of magnitude.  In particular, using current  photometric and spectroscopic data for these objects one may already probe an interesting, and relatively unexplored, range in the parameter space. The detection of (pseudo-)scalar particles can have major implications for fundamental physics among which strong interactions, string theory, and also cosmology. It may further enhance our understanding of compact astrophysical objects and provide us with information which is complementary to that  gained by other spectroscopic lines of investigation.

 \acknowledgements

It is a pleasure to thank Stephen Adler, Edward Witten, Pierre Sikivie, Matt Kleban, Andrew MacFadyen, Carlos Pe\~na-Garay, and Andrew Youdin for many illuminating discussions. Special thanks go to the organizers and participants of the 17th Kingston meeting on compact stars for  a wonderful learning experience. While at the Institute for Advanced Study, D.C. was supported by NASA through a Chandra Postdoctoral Fellowship award PF4-50033. D.~C. is grateful to David Bowen for continuous support, and to Princeton's Small World's Coffee  for daily encouragements.

\appendix

\section{The Dielectric Tensor of Hot and Active Plasma}

Using the properties of the modified Bessel functions and the exact PDFs one can recast the relativistic tensor by Trubnikov (1959) in the form
\begin{equation}
\epsilon_{ij}(k,\omega,\mu )=\delta_{ij}-\mu \left ( \frac{\omega_p}{\omega} \right )^2F_{ij}
\end{equation}
where
\begin{equation}
\begin{array}{l}
\displaystyle F_{11}=\lambda^{-1} \sum_{n=-\infty}^{\infty} n^2 \sum_{k=0}^\infty A_n^{\vert n \vert +k} Z_{\vert n\vert+k+3/2 }(a,z_n,\mu ) \\
\displaystyle  F_{12}=-F_{21}=-i\sum_{n=-\infty}^\infty n \sum_{k=0}^\infty \frac{dA^{\vert n \vert +k}}{d\lambda} Z_{\vert n \vert +k+3/2} (a,z_n,\mu ) \\
\displaystyle F_{22}=F_{11}-2\lambda^{-1} \sum_{n=-\infty}^\infty \sum_{k=0}^\infty \left ( \lambda^2 \frac{dA}{d\lambda} \right )^{\vert n \vert +k} Z_{\vert n\vert+k+3/2 }(a,z_n,\mu ) \\ 
\displaystyle F_{13}=F_{31}= -\frac{1}{\sqrt{2\lambda}} \sum_{n=-\infty}^\infty n \sum_{k=0}^\infty A_n^{\vert n \vert +k} \frac{\partial  Z _{\vert n \vert +k+5/2}(a,z_n,\mu )}{\partial z_n} \\
\displaystyle F_{23}=-F_{32}=-\frac{i}{\sqrt{2 \lambda}} \sum_{n=-\infty}^\infty \sum_{k=0}^\infty  \left ( \lambda \frac{dA_n}{d\lambda} \right )^{\vert n \vert +k} \frac{\partial Z_{\vert n \vert +k +5/2}(a,z_n,\mu )}{\partial z_n} \\ 
\displaystyle F_{33}=\sum_{n=-\infty}^\infty \sum_{k=0}^\infty \left [  A_n^{\vert n\vert +k} Z_{\vert n \vert +k +5/2} (a,z_n,\mu ) + \frac{A_n^{\vert n \vert +k}}{2} \frac{\partial^2 Z_{\vert n \vert +k +7/2}(a,z_n,\mu )}{\partial z_n^2}     \right ]
\end{array}
\end{equation}

The exact PDFs of half -integer index, $q+3/2$ where  the harmonic number $q=0,1,...$, are defined by Cauchy-type integrals defined on the real axis and approaching 0 at infinity (see Eqs. 19,20 in Castej\'on \& Pavlov 2006). They are given by
\begin{equation}
\begin{array}{ll}
\displaystyle Z_{q+3/2}(a,z,\mu )=-\sqrt{ \frac{-\beta}{2\pi \mu }} \frac{e^{-2\beta a -\mu }}{K_2(\mu ) a^{(q+1/2)/2}} \int_{-\infty}^\infty \frac{dt e^{\beta t}}{t-z}  \left ( a-t+\frac{t^2}{2\mu } \right )^{(q+1/2)/2} K_{q+1/2} \left ( -2\beta \sqrt{a^2-ta+\frac{t^2a}{a\mu }}  \right ); & \displaystyle \vert N_{3} \vert >1 \\

\displaystyle Z_{q+3/2}(a,z,\mu )=\sqrt{\frac{\pi \beta}{2\mu }} \frac{e^{-\mu \sqrt{\beta}}}{K_2(\mu ) a^{(q+1/2)/2}} \int_{-\infty}^\infty \frac{du e^{-\beta u}}{u-a_r+z} \left ( \frac{u^2}{2\mu } + \frac{u}{\sqrt{\beta}}  \right )^{(q+1/2)/2} I_{q+1/2} \left ( 2\beta \sqrt{ \frac{au^2}{2\mu} +\frac{a u}{\beta}} \right ); & \displaystyle 0<\vert  N_{3} \vert <1
\end{array}
\end{equation}
where $a=\mu N_{3}^2/2,~z=\mu (1-\Omega_e/\omega),~\mu =mc^2/k_B T$ and $\beta=1/(1-N_{3}^2 )$. $a_r=\mu (1-\sqrt{1-N_{3}^2})$ is the relativistic counterpart of $a$. $N_{3}=k_{3} c / \omega$. $I_{q+1/2},~K_{q+1/2}$ are the modified Bessel and Mac Donald functions respectively where the square root of the argument refers to the positive branch. Contour integration is chosen to pass below the pole for $N_{3}>1$ and above the pole for $0<N_{3}<1$. There are easily calculable non-singular integral forms for the above expression in the limit of very small (see equations 43,44 in Castej\'on \& Pavlov 2006) and very large $z$ (see equations 28,29 in Castej\'on \& Pavlov 2006).

\end{document}